\documentclass{aa}  
\usepackage{natbib,epsfig,txfonts,graphicx}
\bibpunct{(}{)}{,}{a}{}{,}

\newcommand{\msun}{\ensuremath{M_{\odot}}}
\newcommand{\lsun}{\ensuremath{L_{\odot}}}
\newcommand{\rsun}{\ensuremath{R_{\odot}}}
\newcommand{\Zsun}{\ensuremath{Z_{\odot}}}
\newcommand{\Teff}{\ensuremath{T_{\rm eff}}}
\newcommand{\vinf}{\ensuremath{v_{\infty}}}
\newcommand{\mdot}{\ensuremath{\dot{M}}}

\newcommand{\msunyr}{\ensuremath{M_{\odot} {\rm yr}^{-1}}}

\newcommand{\mdu}{\ensuremath{10^{-6}\,M_{\odot} {\rm yr}^{-1}}}

\newcommand{\beq}{\begin{equation}}
\newcommand{\eeq}{\end{equation}}
\newcommand{\beqa}{\begin{eqnarray}}
\newcommand{\eeqa}{\end{eqnarray}}

\newcommand{\nbeq}{\begin{equation*}}
\newcommand{\neeq}{\end{equation*}}

\newcommand{\kms}{\ensuremath{{\rm km}\,{\rm s}^{-1}}}

\newcommand{\HII} {H\,{\sc ii}}
\newcommand{\HeI} {He\,{\sc i}}
\newcommand{\HeII}{He\,{\sc ii}}

\newcommand\NIII{N\,{\sc iii}}
\newcommand\NIV{N\,{\sc iv}}
\newcommand\NV{N\,{\sc v}}

\newcommand\SiIII{Si\,{\sc iii}}
\newcommand\SiIV{Si\,{\sc iv}}

\newcommand{\Hep}{H$_{\epsilon}$}
\newcommand{\Hd} {H$_{\rm \delta}$}
\newcommand{\Hg} {H$_{\rm \gamma}$}
\newcommand{\Hb} {H$_{\rm \beta}$}
\newcommand{\Ha} {H$_{\rm \alpha}$}

\newcommand{\Bra}{Br$_{\rm \alpha}$}

\newcommand{\Rstar}{\ensuremath{R_{\ast}}}

\newcommand{\Lstar}{\ensuremath{L_{\ast}}}

\newcommand{\logg}{\ensuremath{\log g}}
\newcommand{\YHe}{\ensuremath{Y_{\rm He}}}

\newcommand{\vesc}{\ensuremath{v_{\rm esc}}}

\newcommand{\vsini}{\ensuremath{v{\thinspace}\sin{\thinspace}i}}
\newcommand{\vturb}{\ensuremath{v_{\rm turb}}}
\newcommand{\vmic}{\ensuremath{v_{\rm mic}}}
\newcommand{\vmac}{\ensuremath{v_{\rm mac}}}

\newcommand{\trip}{\ensuremath{\lambda\lambda4634-4640-4642}}
\newcommand{\qua}{\ensuremath{\lambda\lambda4510-4514-4518}}

\newcommand{\nivem}{\ensuremath{\lambda4058}}
\newcommand{\nivab}{\ensuremath{\lambda6380}}
\newcommand{\niiir}{\ensuremath{\lambda4640}}

\newcommand\eg{\hbox{e.g.,}}
\begin{document}
\title{Nitrogen line spectroscopy in O-stars}
\subtitle{III. The earliest O-stars
\thanks{Based on (i) observations
collected at the European Southern Observatory Very Large
Telescope, under programmes 68.D-0369, 171.D-0237 (FLAMES), and
67.D-0238, 70.D-0164, 074.D-0109 (UVES); (ii) observations made with the
NASA/ESA {\it Hubble Space Telescope}, obtained from the Data Archive at the
Space Telescope Science Institute, which is operated by the Association of
Universities for Research in Astronomy, Inc., under NASA contract NAS
5-26555. These observations are associated with programmes 6417, 7739,
and 9412; and (iii) observations gathered with
the 6.5m Magellan telescopes at the Las Campanas Observatory, Chile.}
\fnmsep
\thanks{Appendix A and B are only available in electronic form 
at http://www.edpsciences.org}
}

\author{J.G. Rivero Gonz\'alez\inst{1}, J. Puls\inst{1},
        P. Massey\inst{2}\thanks{Visiting astronomer, Cerro Tololo International
Observatory, National Optical Astronomy Observatory, which is operated by the
Association of Universities for Research in Astronomy (AURA) under cooperative
agreement with the National Science Foundation.}, \and F.
Najarro\inst{3}}

\institute{Universit\"atssternwarte M\"unchen, Scheinerstr. 1, 81679 M\"unchen, 
           Germany, \email{jorge@usm.uni-muenchen.de} 
           \and
           Lowell Observatory, 1400 West Mars Hill Road, Flagstaff, AZ 86001,
           USA
	   \and 
           Centro de Astrobiolog\'{\i}a, (CSIC-INTA), 
	   Ctra. Torrej\'on a Ajalvir km 4,
	   28850 Torrej\'on de Ardoz, Spain}

\date{Received; Accepted}

\abstract 
{The classification scheme proposed by Walborn et al. (2002, AJ, 123,
2754), based primarily on the relative strengths of the \NIV\nivem\ and
\NIII$\lambda$4640 emission lines, has been used in a variety of
studies to spectroscopically classify early O-type stars. Owing to the
lack of a solid theoretical basis, this scheme has not yet been
universally accepted though.}
{We provide first theoretical predictions for the
\NIV\nivem/\NIII$\lambda$4640 emission line ratio in dependence of
various parameters, and confront these predictions with results from
the analysis of a sample of early-type LMC/SMC O-stars.}
{Stellar and wind parameters of our sample stars are determined by
line profile fitting of hydrogen, helium and nitrogen lines,
exploiting the helium and nitrogen ionization balance. Corresponding
synthetic spectra are calculated by means of the NLTE
atmosphere/spectrum synthesis code {\sc fastwind}.}
{Though there is a monotonic relationship between the \NIV/\NIII\
emission line ratio and the effective temperature, all other
parameters being equal, theoretical predictions indicate additional
dependencies on surface gravity, mass-loss, metallicity, and,
particularly, {\it nitrogen abundance}. For a given line ratio (i.e.,
spectral type), more enriched objects should be typically hotter.
These basic predictions are confirmed by results from the alternative
model atmosphere code {\sc cmfgen}. 

The effective temperatures for the earliest O-stars, inferred from the
nitrogen ionization balance, are partly considerably hotter than
indicated by previous studies. Consistent with earlier results,
effective temperatures increase from supergiants to dwarfs for all spectral
types in the LMC. The relation between {\it observed}\
\NIV\nivem/\NIII$\lambda$4640 emission line ratio and effective
temperature, for a given luminosity class, turned out to be quite
monotonic for our sample stars, and to be fairly consistent with our
model predictions. The scatter within a spectral sub-type is mainly
produced by abundance effects.}
{Our findings suggest that the Walborn et al. (2002) classification
scheme is able to provide a meaningful relation between spectral type
and effective temperature, as long as it is possible to discriminate
for the luminosity class. In terms of spectral morphology, this might
be difficult to achieve in low-$Z$ environments such as the SMC, owing
to rather low wind-strengths. According to our predictions, the major
bias of the classification scheme is due to nitrogen content, and
the {{\it overall}} spectral type-\Teff\ relation for low-metallicity
(e.g., SMC) O-stars might be non-monotonic around O3.5/O4.}


\titlerunning{The earliest O-stars}
\authorrunning{J.G. Rivero Gonz\'alez et al.}

\maketitle
%

\section{Introduction} 
\label{Introduction}

Though important for the evolution of the early Universe and the
present cosmos, massive stars are not as thoroughly understood as
desirable to safely infer or predict their interaction with their
surroundings, e.g., their input of (ionizing) radiation, wind-energy
and momentum, and nuclear processed material.

Particularly uncertain is the situation for the earliest O-stars
(earlier than O4, see below), both with respect to their physical
parameters, and the relation between their spectroscopic definition
and these parameters. Since the earliest O-type stars include the most
massive stars in the Universe, i.e., the top end of the stellar
initial mass function, such a lack of knowledge is intolerable. As
pointed out by \citet{massey05}, already a 10\% uncertainty in the
effective temperature, \Teff\ (which is still possible for the hottest
stars to present date), results in a factor of two or more uncertainty
in the Lyman flux, affecting our understanding of the ionization
balance of \HII\ regions and, e.g., the porosity of the interstellar
medium (e.g., \citealt{Oey97, Oey06}).

This large uncertainty is produced because the standard approach to derive
\Teff, exploiting the \HeI/\HeII\ ionization equilibrium, fails in
the earliest O-star regime. This is the result of rather insensitive \HeII\
lines to changes in \Teff, and quite weak strategic \HeI\ lines,
with their equivalent widths falling below 200 m\AA, about the limit achievable
with normal photographic emulsions in the 1970s. The apparent absence of \HeI\
was used by \citet{walborn71} to extend 
previous classification criteria to the O3 spectral type, which then
displayed a degeneracy with respect to effective temperature. This
degeneracy was partially broken by \citet{kudritzki80} and
\citet{simon83} who used better photographic data to detect \HeI\ in
the spectra of some O3 stars, but even today with modern CCDs and high
S/N the detection of \HeI\ is not always achievable. To circumvent
this problem \citet{walborn02b} suggested to use the
\NIV\nivem/\NIII\niiir\footnote{The individual components
NIII$\lambda\lambda$4640.64-4641.85 become blended for a projected
rotational velocity \vsini\ $\geq$ 40~\kms.} (hereafter \NIV/\NIII)
emission line ratio as the primary classification criterion for the
earliest spectral types, instead of the \HeI/\HeII\ absorption line
ratio. By means of this criterion, the former O3 class was split into
three different types O2, O3, and O3.5, being O2 the degenerate one
instead.

This scheme has been used in a variety of studies during the past years to
classify spectra of early O-stars in the Large and Small Magellanic Clouds
(LMC and SMC) (\eg\ \citealt{walborn04}, \citealt{massey04, massey05,
massey09}, \citealt{evans06}) and in the Milky-Way, \eg\ \citet{sota11}.
However, these additional sub-types are not (yet) universally accepted. In
particular \citet{massey04, massey05} criticized that relying on the relative
strengths of the optical nitrogen emission lines lacks a solid theoretical
basis, since the (photospheric) line emission is the result of complex NLTE
processes. 

To provide more insight into this and related matter, we started a series of
publications dealing with nitrogen spectroscopy in O-type stars. In the first
paper of this series (\citealt[hereafter Paper~I]{rivero11}), we concentrated
on the formation of the emission at \NIII\trip.
In the follow-up paper (\citealt[hereafter Paper~II]{rivero12}),
we investigated the \NIV\nivem\ emission line formation, and applied our
accumulated knowledge to derive nitrogen abundances for an LMC O-star sample.

The primary goal of the present paper is a quantitative study of the
atmospheric parameters of the earliest O-stars, by means of nitrogen line
spectroscopy and building on the results from our previous work within this
series. Particularly, we concentrate on testing the \citet{walborn02b}
classification scheme on its capability of providing a reasonable relation
between spectral types and effective temperatures. To this end, we investigate
the theoretical \NIV/\NIII\ emission line ratio, and the impact of different
parameters on this ratio. Subsequently, our predictions are confronted with
corresponding observational results, derived from an analysis of an early O-type
sample of LMC/SMC stars.

This paper is organized as follows. In Sect.~\ref{model grid}, we describe the
tools used within this work, the atmospheric code {\sc fastwind} and a
suitable model grid. Open questions from our previous studies regarding the
formation of the \NIII\ and \NIV\ emission lines are addressed in
Sect.~\ref{theory}. Section~\ref{em_ratio} presents first theoretical
predictions on the \NIV/\NIII\ emission line ratio. In
Sect.~\ref{comp_cmfgen}, we compare our results and predictions with
corresponding ones from the alternative atmospheric code {\sc cmfgen}
(\citealt{hilliermiller98}). The stellar sample and the observations used
within this study as well as the procedure to determine stellar/wind
parameters together with nitrogen abundances are presented in 
Sect.~\ref{ana_obs}. Finally, we provide a discussion of our results in
Sect.~\ref{discussion}, and summarize our findings and conclusions in
Sect.~\ref{summary}.

\section{Code and model grid} 
\label{model grid} 
\begin{table}
\caption{Model-grid used within this work: coverage of fundamental
parameters. For \Teff\ and \logg\ coverage, see also asterisks in
Fig.~\ref{iso_ratio_z}.} \label{tab_grid}
\begin{tabular}{lcc}
\hline
\hline
Parameter       & Range                        & Typical step size \\
\hline
\Teff\ (kK)     & 35.0 - 55.0                  & 1.0 \\
\logg\ (cgs)    & 3.2, 3.5 - 4.2, 4.5          & 0.1 \\
$\log Q$\tablefootmark{a}/series & $-$14.00/A, $-$13.50/B, $-$13.15/C  & 0.35\\
                & $-$12.80/D, $-$12.45/E, $-$12.10/F & \\
\YHe\           & 0.08, 0.10 - 0.20            & 0.05 \\
\hline
&  Galactic & \\
\hline
$Z$             & \Zsun                        & - \\
$[{\rm N}]$     & 7.64, 7.78 - 8.98            & 0.2 \\
\hline
&  LMC & \\
\hline
$Z$             & 0.5\Zsun                     & - \\
$[{\rm N}]$     & 6.90, 6.98 - 8.58            & 0.2 \\
\hline
&  SMC & \\
\hline
$Z$             & 0.2\Zsun                     & - \\
$[{\rm N}]$     & 6.50, 6.78 - 8.38            & 0.2 \\
\hline
\end{tabular}
\tablefoottext{a}{in units of \msunyr/(\kms \rsun)$^{1.5}$.}
\tablefoot{Other model parameters adopted as follows: wind terminal
velocity, \vinf, as a function of the photospheric escape velocity,
\vesc\ (see \citealt{KP00}); stellar radius, \Rstar, as a function of
spectral type and luminosity class, corresponding to prototypical
values; velocity field exponent, $\beta = 0.8$; and micro-turbulence,
$\vmic$ = 10~\kms. Nitrogen baselines abundances (first entry in [N]
range) have been drawn from~\citet{hunter07}.}
\end{table}

\begin{figure*}
\center
{\includegraphics[width=165mm]{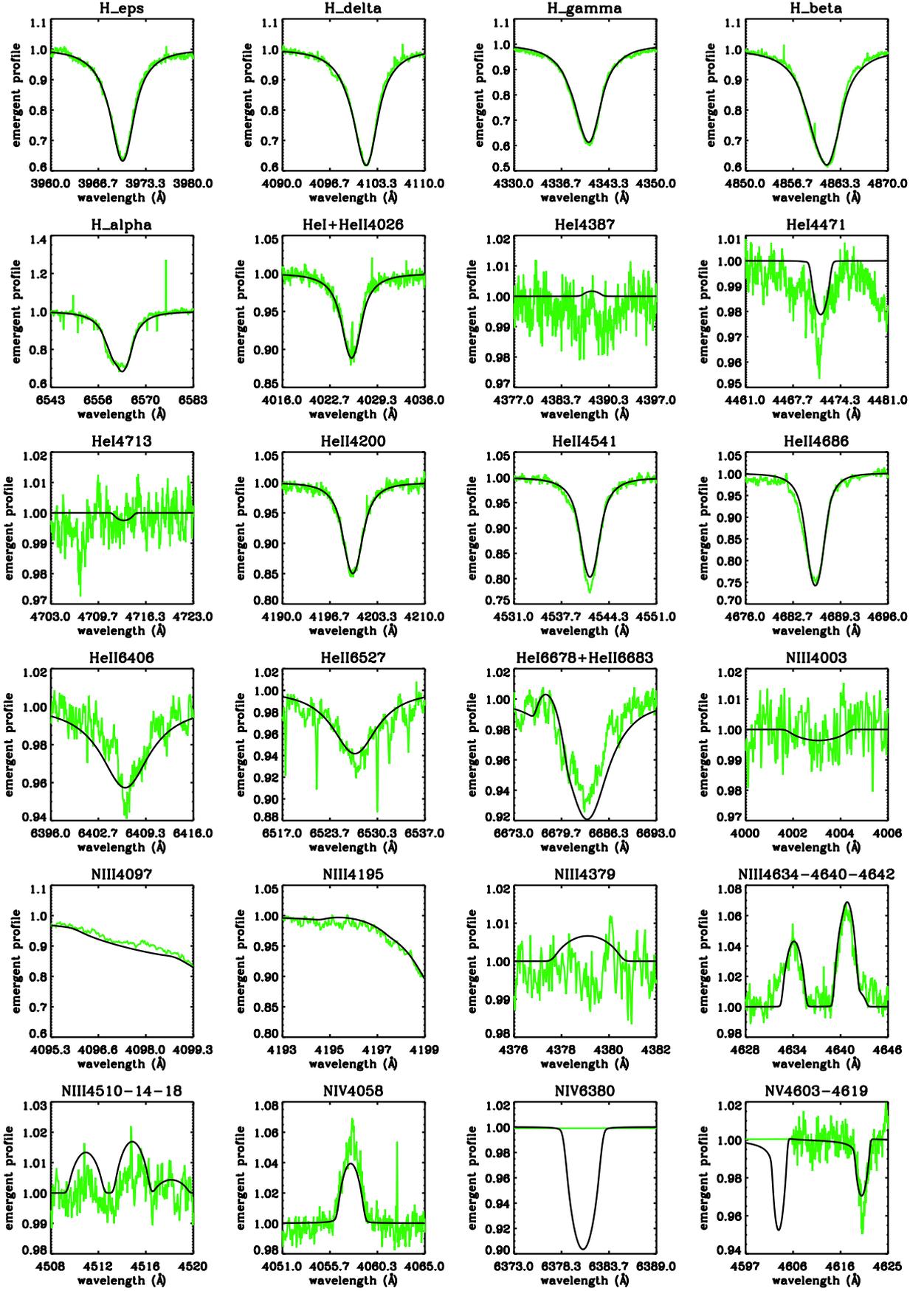}}
\caption{High resolution optical spectrum of the Galactic O3 V((f$^*$))
star HD\,64568  (green), compared with the best-fitting synthetic 
H/He/N spectrum from our grid (black, convolved with \vsini\ = 100
\kms, see \citealt{markova11}). \NIV\nivab\ and \NV$\lambda$4603 not
observed.
Grid parameters: \Teff\ = 48~kK, \logg\ = 4.0, $\log Q = -12.8$
(series `D'), \YHe=0.1, [N] = 8.38. Fine tuning of the parameters can
improve the fit. } 
\label{hd64568}
\end{figure*}

All calculations within this work were performed by means of the NLTE
atmosphere/spectrum synthesis code {\sc fastwind}
(\citealt{santo97,Puls05}), using the recently updated version v10.1
(see Paper~II). Most results presented here (except for the fine-tuned
fits in Sect.~\ref{ana_obs}) are based on a model-grid, with H, He,
and N as `explicit' elements.  Corresponding model atoms have been
described in \citet{Puls05} (H/He) and Papers~I/II (N), and
Table~\ref{tab_grid} provides the coverage of important grid
parameters. Details of the basic grid were already provided in
Paper~II, and only its `hot' range (\Teff\ $\ge$ 35kK) was used for our
current analysis. This subgrid has been extended to cover a broader
range in background metallicity, $Z$~=~1, 0.5, 0.2 \Zsun, associated
with the Milky Way (MW), Large Magellanic Cloud (LMC), and Small
Magellanic Cloud (SMC), respectively.  Moreover, we increased the
sampling with respect to \logg, and also the coverage of the
wind-strength parameter (or optical depth invariant),
Q~=~\mdot/(\vinf\Rstar)$^{1.5}$, towards larger values, resulting in 6
model series of different wind-strength (from series `A' with $\log
Q$~=~-14.00 to series `F' with $\log Q$~=~-12.10, for details see
Table~\ref{tab_grid}). Finally, the sampling in helium content, \YHe\
= $N_{\rm He}/N_{\rm H}$, and nitrogen abundance, [N] = $\log_{\rm 10}
(N_{\rm N}/N_{\rm H}) + 12$, has been improved, for studying the
reaction of important nitrogen lines on extreme changes in [N]
(Sects.~\ref{ab_niv} and~\ref{nitro_ratio}), and for better
constraining abundances in our analysis of LMC/SMC stars
(Sect.~\ref{ana_obs}). The current grid accounts for a total of
104,000 models.

The major potential of such a model-grid (besides theoretical
investigations) is the possibility of obtaining rather precise
estimates of stellar parameters within a reasonable amount of time
(few minutes). With the advent of large stellar surveys, such as the
VLT-FLAMES Tarantula survey (VTFS, \citealt{evans11}) or the IACOB
project (\citealt{simon-diaz11b}), this will turn out to become a very
useful tool. As an example for the present quality of our models, we
show in Fig.~\ref{hd64568} the comparison of a high resolution
(resolving power $R$ = 48,000), high S/N ($\ge$ 150) spectrum of the
Galactic O3 V((f$^*$)) star HD\,64568 (for details, see
\citealt{markova11}) with the best fitting synthetic spectrum from our
grid models. Obviously, the grid resolution is sufficient to achieve
quite a fair representation of the observed spectrum, both with
respect to H/He and N (note that lines from \NIII, \NIV, and \NV\ are
present in parallel). Further analyses of Galactic objects will be
performed in a forthcoming paper.


\section{\NIII/\NIV\ emission lines -- parameter-dependence}
\label{theory}

Our previous studies on the formation of \NIII\trip\ and \NIV\nivem\
prompted a number of questions, which are investigated in the
following. First we concentrate on the \NIII\ triplet emission in
low-$Z$ environments, to discriminate the counteracting effects of
line blocking (less blocking -- more emission) and mass-loss (less
mass-loss -- less emission) under these conditions.
Subsequently, we study the impact of $Z$ and [N] on \NIV\nivem.\footnote{For a
detailed description of the formation process of \NIII\ and \NIV\ 
emission lines, we refer to Paper~I and Paper~II, respectively.}

\subsection{\NIII\ emission line formation: EUV line-blocking vs.
wind-strength}
\label{niii_block_wind}
In Paper~I we argued that, for Galactic conditions, the canonical
explanation for the presence of emission at \NIII\trip\ (related to
dielectronic recombination, \citealt{mihalas73}) no longer or only
partly applies if one accounts for the presence of
line-blocking/blanketing and winds. The key role is now played by the
stellar wind, as long as the wind-strength is large enough to enable a
significantly accelerating velocity field already in the photospheric
formation region of the resonance line(s) connected to the upper level
of the involved transition. Furthermore, our study implied that 
particularly the efficiency of the `two electron' drain (depopulating
the lower level, see Paper~I) is strongly dependent on the degree of
EUV line-blocking, i.e., on $Z$. For a given wind-strength and
nitrogen abundance, the emission should become stronger in low-$Z$
environments, because of less blocking (see Fig~16 in Paper~I).
Nevertheless, this comparison might be unrealistic, since less
blocking goes hand in hand with a lower wind-strength, which might
(over-) compensate the discussed effect. Moreover, one might need to
consider a lower nitrogen content, owing to a lower baseline
abundance.

Here we investigate the combined effect. First, we compare the
behavior of the \NIII\ emission lines for two different $Z$
environments (MW and SMC), applying a consistent scaling of $\log Q$,
via \mdot\ $\propto (Z/Z_\odot)^{0.72}$ (clumping corrected,
\citealt{mokiem07b}) and \vinf\ $\propto~ (Z/Z_\odot)^{0.13}$
\citep{Leitherer92}. For the SMC with $Z/\Zsun = 0.2$, this yields a
reduction of $\log Q$ by $-0.37$~dex compared to the Galactic case,
and corresponds well to the step size of $\Delta \log Q = 0.35$ used
within our model grid. Thus, Galactic models from, e.g., series `E'
need to be compared with SMC models from series `D'.

\begin{figure}
\resizebox{\hsize}{!} {\includegraphics{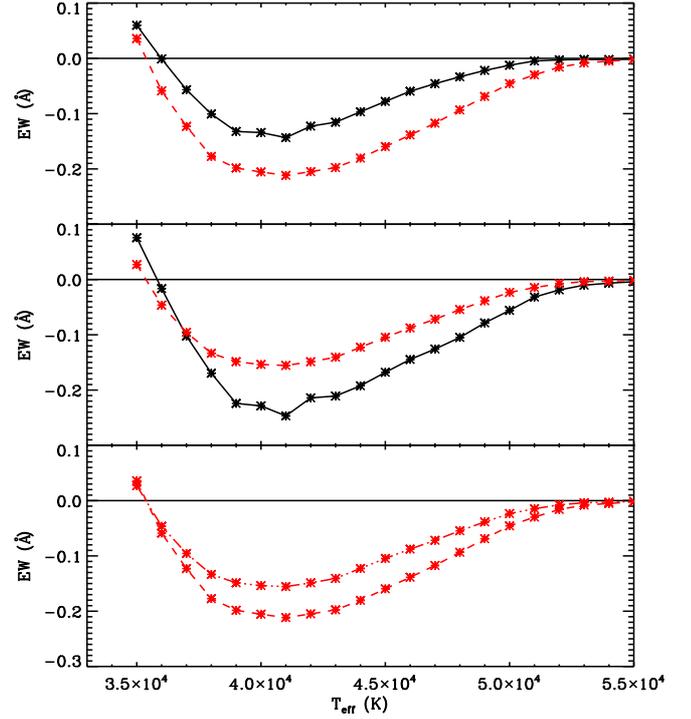}} 
\caption{Equivalent width of \NIII\niiir\ as a function of \Teff, for
MW and SMC models at \logg = 4.0 and mass-loss rates scaled according
to $Z$ (see text). Solid/black and dashed(-dotted)/red curves
refer to MW series `E' and SMC series `D' models, respectively
($\Delta \log Q = -0.35$~dex). Upper panel: [N]~=~7.78 for both
Galactic and SMC models. Middle panel: comparison accounting for the
theoretically expected maximum [N] enrichment, drawn from tailored
evolutionary calculations by~\citet{brott11a}: MW models with
[N]~=~8.18 and SMC models with [N]~=~7.58. Lower panel: SMC
series `D' models, with [N]~=~7.58 (dashed-dotted) and [N]~=~7.78
(dashed). The line emission increases with increasing [N].} 
\label{gal_smc_4640}
\end{figure}

In Fig.~\ref{gal_smc_4640} (upper panel), we compare the equivalent
widths of \NIII$\lambda\lambda$4640.64, 4641.85\footnote{The blue
component of the triplet shows a similar behavior.}, in the following
abbreviated by \NIII\niiir, as a function of \Teff, for MW series `E'
(black/solid) and SMC series `D' models (red/dashed). All models have
the same gravity, \logg~=~4.0, and the same nitrogen content,
[N]~=~7.78 dex (almost solar, \citealt{asplund05, asplund09}). As usual, we define
equivalent widths (hereafter EWs) to be positive for absorption and to
be negative for emission lines. The result of this comparison is
similar to our findings from Paper~I. Even when we account for
consistently scaled mass-loss rates, the low-$Z$ models result in more
emission. Thus, lower mass-loss rates associated with low-$Z$
environments do not compensate the increase in emission strength due to
a lower degree of line-blocking.

So far, we neglected the fact that a lower $Z$ also implies a
difference in the [N] baseline abundance, which needs to be considered
in a final comparison. In the middle panel of Fig.~\ref{gal_smc_4640},
we compare the same series of models as in the upper one, accounting
now for more consistent abundances (MW: black/solid, SMC:
red/dashed). Here we used [N] values in agreement with the
theoretically expected maximum enrichment as provided by
\citet{brott11a}, for MW and SMC models of 40~\msun\ and initial
rotation velocities of 270~\kms. This roughly corresponds to models
with [N] = 8.2 and 7.6 for the MW and SMC, respectively.
\footnote{Unfortunately, there are only few abundances studies in the O-star
regime to confirm these assumptions, and most of them are biased towards
late O-types (reviewed by, e.g. \citealt{herrero03, herrero04,
morel09}). \citet{heap06}, following previous work by~\citet{bouret03},
found most of their 18 SMC sample stars to be enriched, and more than
half of the sample displayed [N] $>$ 7.5. 
Regarding Galactic objects, 
\citet{martins11a} (see also \citealt{martins11b}) recently found a typical
enrichment by 0.4-0.6 dex above the Galactic baseline abundance,
corresponding to [N] $\approx$ 8.0-8.2.} 

It turns out that with the inclusion of more realistic abundance
conditions, the increase of emission strength due to less blocking
becomes strongly suppressed, and now the MW models display stronger
emission that the SMC ones. This might explain the relatively low
number of "f" objects among the SMC O-stars (see
Sect.~\ref{SMC_Of_stars}). 

For convenience, the lower panel of Fig.~\ref{gal_smc_4640}
displays a direct comparison of SMC models with expected maximum
enrichment ([N] = 7.58, dashed-dotted) and a solar nitrogen content
([N] = 7.78, dashed). Obviously, the emission strength of \NIII\niiir\
increases with increasing nitrogen content (see also Fig.~15 in
Paper~I).

\subsection{\NIV\nivem\ -- dependence on background abundance}
\label{z_dep} 
\begin{figure}
\resizebox{\hsize}{!}
  {\includegraphics{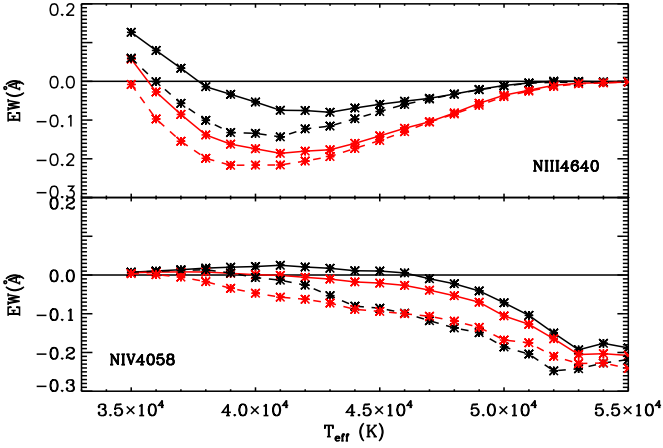}} 
\caption{Equivalent width of \NIII\niiir\ (upper panel) and
\NIV\nivem\ (lower panel) as a function of \Teff, for MW (black) and
SMC (red) models, at \logg=4.0 and [N]=7.78 dex. Solid lines: series `A'
(thin winds), dashed lines: series `E' (prototypical for Galactic
supergiants).}
\label{ew_4640_4058}
\end{figure}

Though we studied the response of the \NIII\ triplet on different
background abundances already in Paper~I, a similar analysis for \NIV\nivem\
is still missing. 
This is now done in Fig.~\ref{ew_4640_4058}, by means of our
model-grid. We display models with thin winds (series `A'; solid) and 
with wind-strengths typical for Galactic supergiants (series `E';
dashed), and compare the impact of MW (black) and SMC (red) background
abundances. All models have the same gravity, \logg = 4.0, and the
same [N] = 7.78. To enable a better comparison, we show the effects
for both \NIII\niiir\ (upper panel) and \NIV\nivem\ (lower panel). We
clearly see that \NIII\niiir\ is much more influenced by $Z$ than
\NIV\nivem, irrespective of wind-strength and \Teff. For LMC
background abundances, $Z$ = 0.5 (not displayed), the effect on
\NIV\nivem\ becomes almost negligible.

With respect to the formation of this line, all `A' and also the
`cooler' `E' models (with \Teff\ $\leq$ 45~kK) behave as discussed in
paper~II. A lower background $Z$ induces a more depopulated ground
state, owing to somewhat higher ionizing fluxes around the \NIV\ edge
at $\lambda$~=~160~\AA.
Consequently, the drain of the lower level of the \nivem\ transition,
3p, via the `two electron' transitions becomes enhanced, and more
emission is produced at lower $Z$.
For `E' models with \Teff\ $\geq$ 45~kK the situation changes, and now
the higher $Z$ (MW) models produce slightly more emission: At higher
\Teff\ and \mdot, the resonance line towards 3p (at $\lambda$~=~247
\AA) leaves detailed balance and 3p becomes pumped, stronger at low
$Z$ because of higher EUV fluxes. Thus, less emission at \NIV\nivem\
is produced, compared to a Galactic environment.

Overall, however, background abundances have a weak effect on \NIV\nivem,
much weaker than wind-strength effects (e.g., red solid vs. red
dashed). This is related to the much higher sensitivity of the \NIV\
continuum on mass-loss rate. Accordingly, also the helium content has
a certain impact on the \NIV\ emission strength, since this parameter
controls the overall flux level of the \HeII\ continuum including the
\NIV\ edge. In particular, an increase of \YHe\ decreases the \NIV\
emission strength.
\footnote{A similar behavior is found for the \NIII\ triplet emission, for
\Teff\ $\leq$ 40~kK.} 

As already discussed above and in Paper~I, \NIII\niiir\ reacts much
stronger on different background abundances (e.g., upper panel of 
Fig.~\ref{ew_4640_4058}, black vs. red ), and in a similar way
for all mass-loss rates. Because of this different behavior, the
(theoretical) \NIV\nivem/\NIII\niiir\ emission line ratio, studied in
Sect.~\ref{em_ratio}, is affected by metallicity.

Interestingly, the reaction of \NIII\niiir\ on \mdot\ becomes negligible
for models with \Teff\ $\geq$ 46~kK. At these temperatures, \NIII\ has
become a real trace ion, and the (weak) line emission is formed in
quite deep layers, hardly affected by mass loss and velocity field
(see also Fig. 10 in Paper~I). In this case, the relative
overpopulation is caused by recombination cascades and depopulation of
the lower level by `two electron' drain (particularly for low $Z$
conditions), whilst dielectronic recombination remains unimportant.

\subsection{\NIV\nivem\ -- dependence on nitrogen abundance}
\label{ab_niv}
\begin{figure}
\resizebox{\hsize}{!}
  {\includegraphics{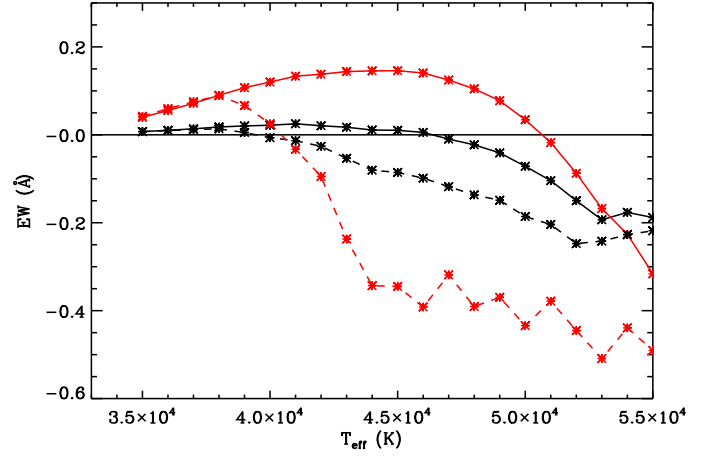}} 
\caption{Equivalent width of \NIV\nivem\ as a function of \Teff\ for
models with [N]~=~7.78 (black) and [N] = 9.0 (red).  All models
correspond to \logg=4.0 and $Z$~=~\Zsun, with mass-loss rates
according to series `A' (solid) and `E' (dashed). See text.}
\label{ew_4058_ab}
\end{figure}
\begin{figure*}
\begin{minipage}{9cm}
\resizebox{\hsize}{!}
  {\includegraphics{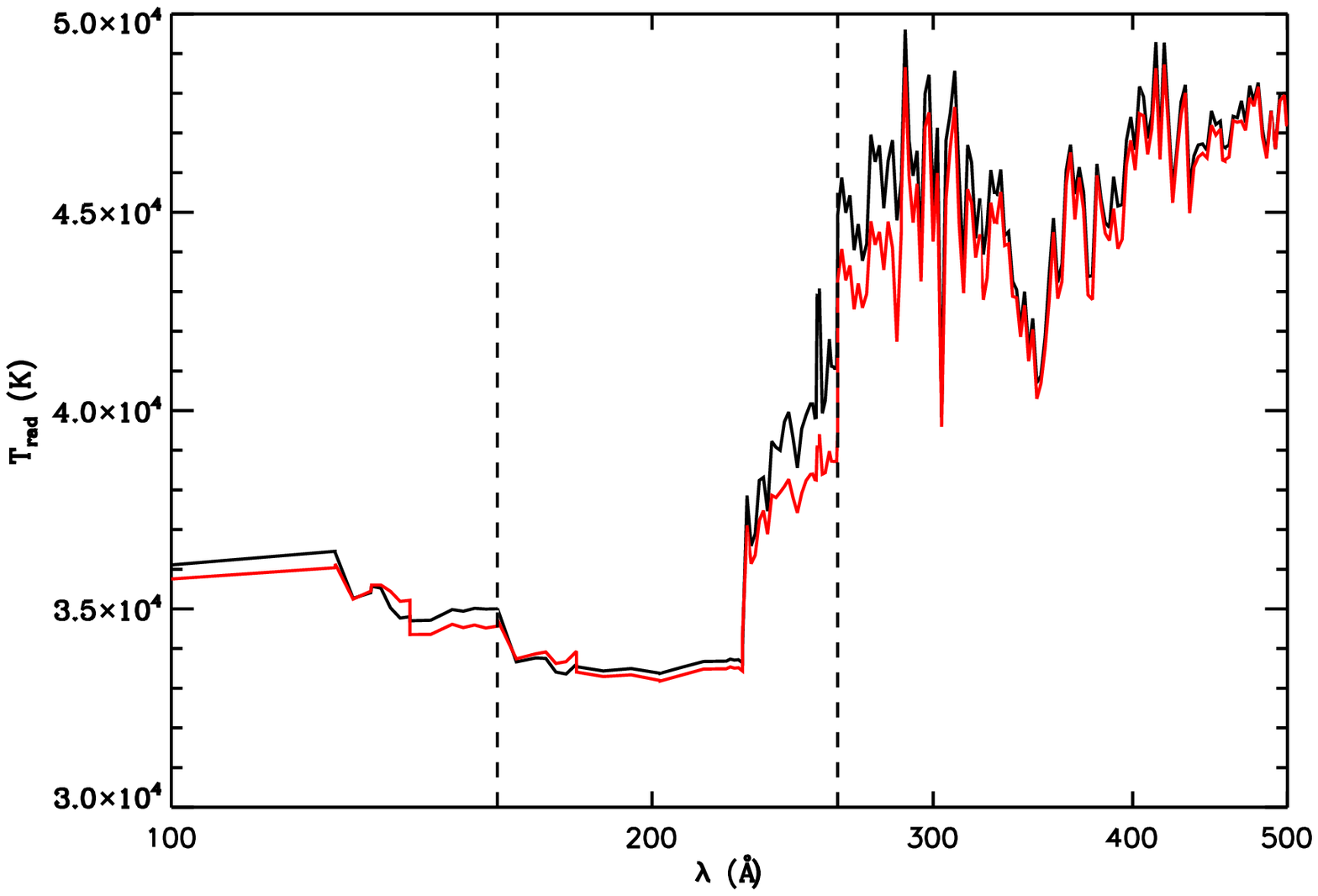}} 
\end{minipage}
\hspace{-.5cm}
\begin{minipage}{9cm}
\resizebox{\hsize}{!}
  {\includegraphics{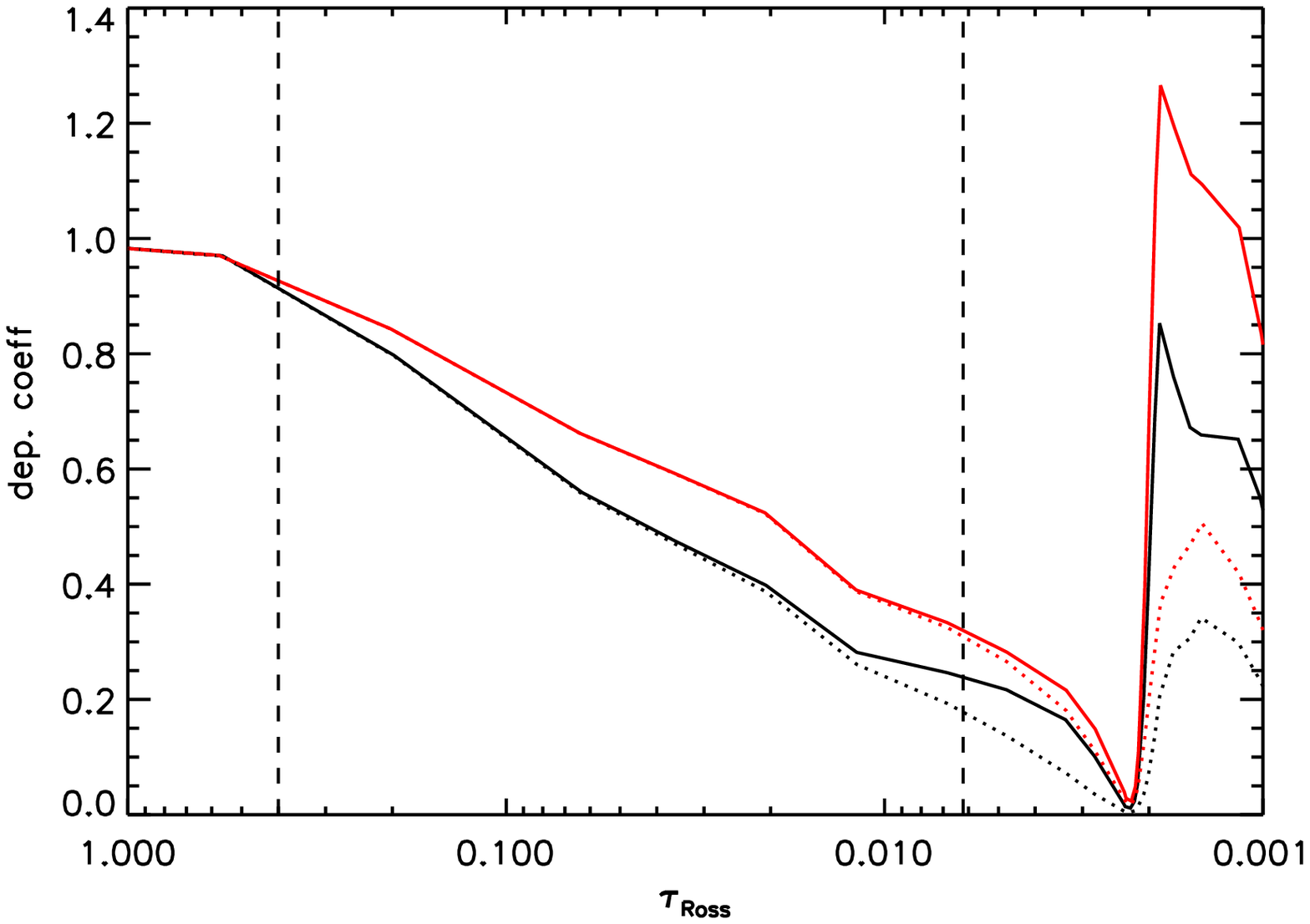}} 
\end{minipage}
\caption{Radiation temperatures and NLTE departure coefficients for
two MW-models at \Teff = 45~kK, \logg = 4.0, {\it and a low density
wind} (series `A'), with solar (black) and strongly enhanced nitrogen ([N] =
9.0, red). Left panel: Enhanced nitrogen leads to lower \NIV\ (and
\NIII) continuum-fluxes, here expressed as radiation temperatures. 
Corresponding edges at 160~\AA\ and 261~\AA\
indicated by vertical lines. Right panel: The lower fluxes give rise to
less depopulated \NIV\ ground- (solid) and 2p$^2$ (dotted) states, where the
latter are the lower levels of the important two-electron
transitions draining \NIV\ 3p. Consequently, there is more
absorption/less emission at \NIV\nivem\ when [N] becomes increased.
Formation region of \NIV\nivem\ indicated by dashed lines.}
\label{traddep_A4540}
\end{figure*}
Figure~\ref{ew_4058_ab} displays the reaction of \NIV\nivem\ on
variations of [N] and $\log Q$ for MW models. As in
Fig.~\ref{ew_4640_4058}, solid and dashed lines correspond to model
series `A' and `E', respectively. We compare models with [N]~=~7.78
(black) and highly enhanced nitrogen, [N] = 9.0 (red), selected to
demonstrate extreme effects.

First, we concentrate on the influence of [N] for model series `E'
(supergiant mass-loss rates, dashed). These models behave as expected.
As for the \NIII\ triplet (e.g., Fig.~\ref{gal_smc_4640}, lower
panel), we obtain more emission when we increase the nitrogen
content. This is mostly because the formation zone, due to the
increased number of absorbers/emitters, is shifted outwards into the
transition region photosphere/wind where the relative overpopulation
becomes very large or even inverted (cf. Fig.~4 in Paper~II).
Small inaccuracies in the population ratio (e.g., due to inappropriate
gridding) can lead to sizeable effects in line-strength when close to
inversion, and this is the reason for the non-monotonicity in EW
encountered for the high [N] `E'-models.

Somewhat unexpectedly, however, we found the opposite reaction for
low-\mdot\ models (solid). For large nitrogen content (red), we either
obtain more absorption or less emission than solar-[N] models
(black), for almost the whole temperature range. Furthermore, the
turning point from absorption to emission occurs at hotter \Teff\ (by
5 kK) than for the solar-[N] models,
whilst for the `E' series this turning point remains rather unaffected
by nitrogen content.  The underlying mechanism is again related to a
diminished drain of the lower level when [N] becomes increased. Here,
a higher [N] is responsible for a lower \NIV\ continuum-flux
(Fig.~\ref{traddep_A4540}, left), leading to less depopulated ground-
and 2p$^2$ states (Fig.~\ref{traddep_A4540}, right), and thus to a
higher population of 3p, i.e., to more absorption/less emission at
\NIV\nivem. Note that this process is similar to the effect produced
by a higher $Z$ (Sect.~\ref{z_dep}). 

These findings imply an important consequence. According to our
predictions, in a certain \Teff\ range \NIV\nivem\ might switch, for
growing [N], from emission to absorption, provided the wind-strength
is not too high! The other way round (and exploiting the results from
the complete grid): If \NIV\nivem\ is observed in absorption at 44~kK
$\leq$ \Teff\ $\leq$ 50~kK (for Galactic stars), this would be an
indication of strong nitrogen enrichment (and mass-loss rates below
`D' corresponding to $\log Q < -12.8$). The lower limit in \Teff\
corresponds to the absorption/emission turning point for `D' models
(at higher \Teff\ and similar or higher \mdot, only emission lines are
predicted, independent of [N]), whilst the upper one refers to the
same point for model series `A'.

\section{Predictions on the \NIV/\NIII\ emission line
ratio}
\label{em_ratio}

\subsection{Overview}

The complete \citet{walborn02b} classification scheme proposed for `normal'
O-stars,\footnote{\citet{crowther11} have updated the classification
scheme for O2-3.5 If*/WN5-7 stars using the morphology of \Hb.} which
also covers the O4 type, is summarized in Table~\ref{wal_scheme}.
Walborn et al. suggested to use the \NIV/\NIII\ emission line ratio as the primary
classification criterion for the earliest spectral types, instead of the
\HeI/\HeII\ absorption line ratio.
This scheme has been used in a variety of studies during the past
years to classify spectra of early O-stars (e.g., \citealt{walborn04},
\citealt{massey04, massey05, massey09}, \citealt{evans06}, \citealt{sota11}),
though there are still some controversial issues. 
(i)~The classification criteria are not quantitative and involve 
secondary statements regarding the strength of \HeI$\lambda$4471, see
Table~\ref{wal_scheme}. (ii)~\citet[hereafter Mas05]{massey05} have
criticized that relying on the strength of the nitrogen emission lines
lacks a theoretical basis, i.e., it is not clear whether \Teff\ is the
only parameter that differentiates the newly defined spectral types.
Indeed, our previous and present work implies that the emission
strength of \NIV\nivem\ (and also that of \NIII\niiir, at least until
\Teff\ $\approx$ 46~kK) crucially depends on \mdot. Based on
their analysis of early LMC and SMC stars, Mas05 pointed out that
for stars with similar \Teff\ and \logg\  the \NIV/\NIII\ ratio could
vary by the full range defined for the scheme (but see Sect.~\ref{comp_mass}).
Furthermore, they suggested that any spectroscopic
classification should be able to constrain \Teff\ without knowledge of other
important parameters such as \logg\ or \mdot. (iii) Already
\citet{walborn02b} pointed out that even though there were no
indications of any correlation between the newly defined spectral
types and the corresponding host galaxy, effects of $Z$ on the emission
line ratio might need to be considered, accounting for the results from
\citet{Crowther00} for WR-stars. Using synthetic WN models, Crowther 
had found that earlier spectral types are predicted at lower
metallicity, following the \citet{Smith96} classification scheme for
WN stars. Since both nitrogen emission lines seem to react differently
on variations of $Z$ (Sect.~\ref{z_dep}), it is clear that such a
dependence cannot be ruled out. 

\begin{table}
\caption{Classification scheme for
spectral types O2-O4 using the emission line ratio
\NIV\nivem/ \NIII\niiir\ and the strength of \HeI$\lambda$4471, 
as defined by \citet{walborn02b}.}
\label{wal_scheme}
\tabcolsep1.6mm
\begin{center}
\begin{tabular}{ll}
\hline
\hline
\multicolumn{1}{c}{Spectral type}
&\multicolumn{1}{c}{Criteria}\\
\hline
\multicolumn{1}{l}{}
&\multicolumn{1}{c}{Supergiants}        \\
\hline
O2 If*      & \NIV\ $\gg$ \NIII, no or very weak \HeI\\
O3 If*      & \NIV\ $>$ \NIII, very weak or no \HeI\\
O3.5 If*    & \NIV\ $\sim$ \NIII, very weak \HeI \\
O4 If+      & \NIV\ $<$ \NIII, weak \HeI \\
\hline
\multicolumn{1}{l}{}
&\multicolumn{1}{c}{Giants}        \\
\hline
O2 III(f*)      & \NIV\ $\gg$ \NIII, no or very weak \HeI \\
O3 III(f*)      & \NIV\ $>$ \NIII, very weak or no \HeI\\
O3.5 III(f*)    & \NIV\ $\sim$ \NIII, very weak \HeI \\
O4 III(f+)      & \NIV\ $<$ \NIII, weak \HeI \\
\hline
\multicolumn{1}{l}{}
&\multicolumn{1}{c}{Dwarfs}        \\
\hline
O2 V((f*))      & \NIV\ $\gg$ \NIII, no \HeI \\
O3 V((f*))      & \NIV\ $>$ or $\sim$ \NIII, very weak \HeI\\
O3.5 V((f+))    & \NIV\ $<$ \NIII, very weak \HeI \\
O4 V((f+))      & no \NIV, weak \HeI \\
\hline
\end{tabular}
\end{center}
\tablefoot{Luminosity classes defined as follows. Supergiants (I):
\HeII$\lambda$4686 in emission; giants (III): \HeII$\lambda$4686 in
weak absorption/P-Cygni profile; dwarfs (V): \HeII$\lambda$4686 in
strong absorption. Note that the f+ designation recently became
obsolete since the \SiIV\ emission at $\lambda\lambda4089-4116$ has 
been established as a common feature in normal O-type spectra
\citep{sota11}.}
\end{table}

In the following, we will address these and other problems, first
by means of theoretical predictions for the \NIV/\NIII\ emission line
ratio, and later on by a comparison with observed spectra. 

\begin{figure*}
\begin{minipage}{9cm}
\resizebox{\hsize}{!}
  {\includegraphics{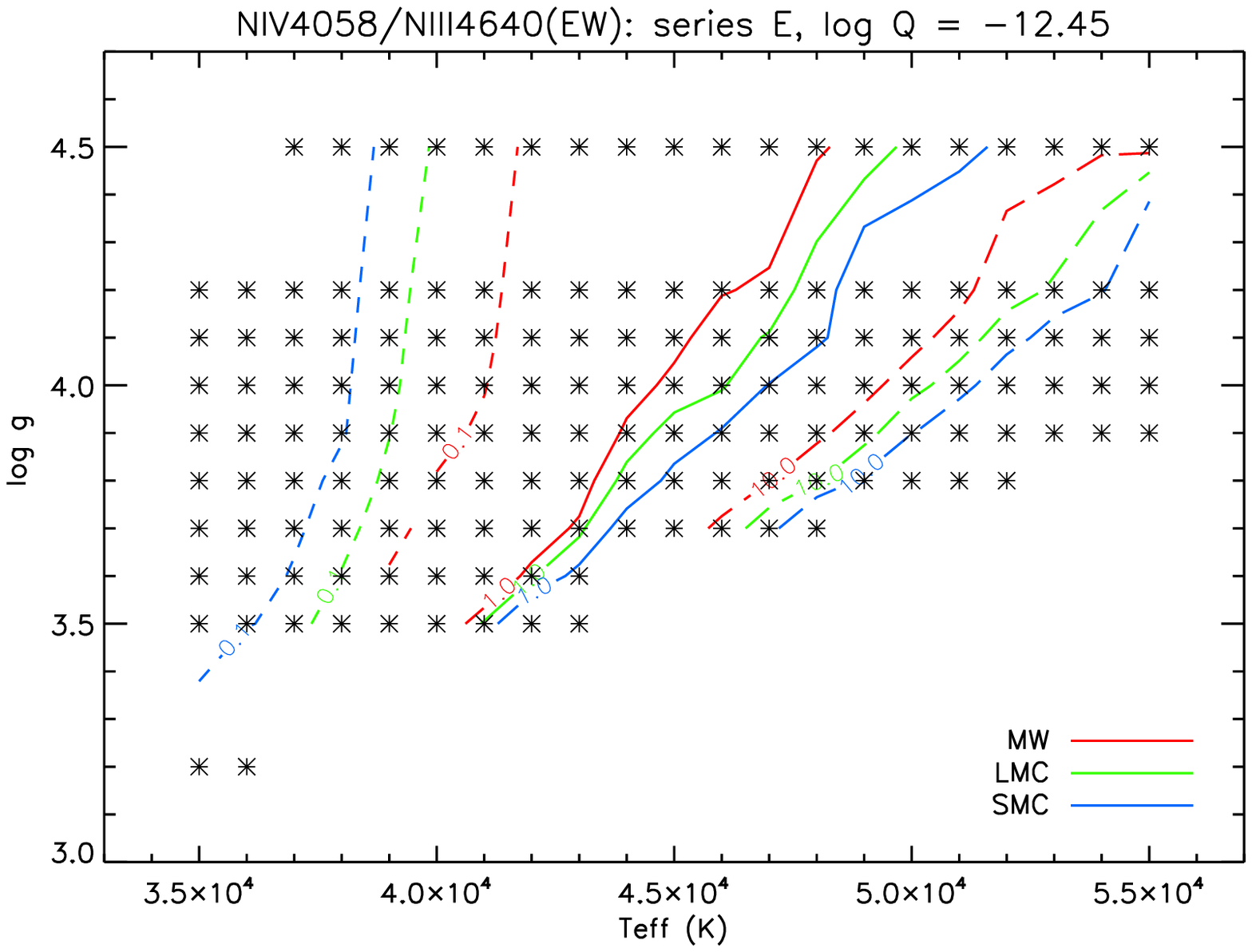}} 
\end{minipage}
\hspace{-.5cm}
\begin{minipage}{9cm}
\resizebox{\hsize}{!}
  {\includegraphics{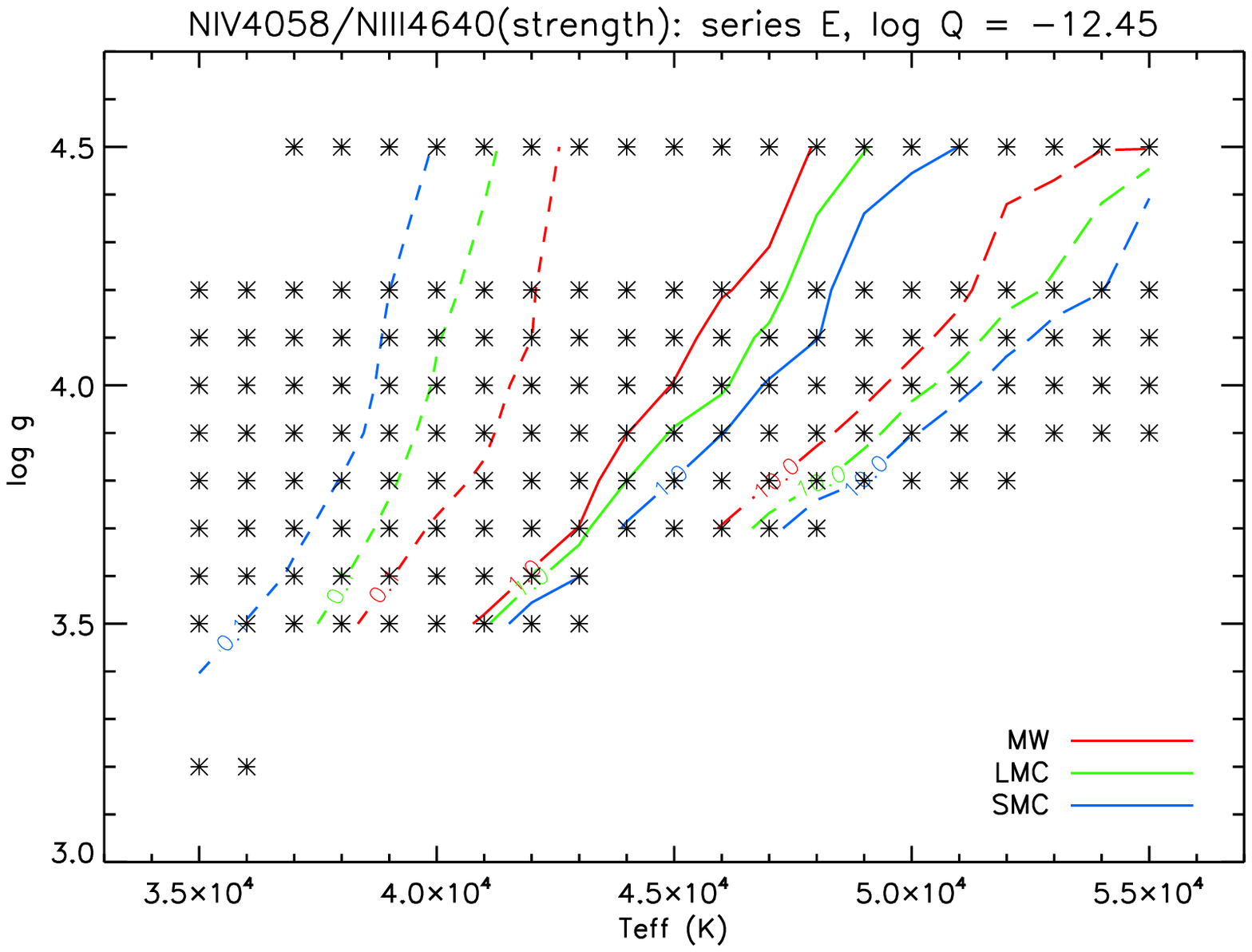}} 
\end{minipage}
\caption{\NIV\nivem/\NIII\niiir\ for Galactic (red), LMC (green) and
SMC (blue) O-stars. Displayed are
iso-contours of the emission line ratio in the \Teff-\logg\ plane
(model series `E', [N] = 7.78), for values of \NIV/\NIII\ = 0.1
(dashed), 1 (solid) and 10 (long-dashed). The
asterisks indicate the position of the grid-models used.
Left panel: emission line ratio calculated using EWs; right panel:
emission line ratio calculated using line-strengths.
}
\label{iso_ratio_z}
\end{figure*}

\subsection{Basic considerations}
We took advantage of the large model-grid described in
Sect.~\ref{model grid}, and analyzed the influence of various 
parameters ($Z$, $\log Q$, [N]) by studying the iso-contours of
specific emission line ratios in the \Teff-\logg\ plane
(Figs.~\ref{iso_ratio_z} to~\ref{iso_ratio_abun}). Here and in the
following, we always used models with \YHe\ = 0.1.

To discriminate the specific spectral types by nitrogen lines alone, 
one has basically to account for five different, qualitatively
defined ranges with respect to the line strengths of \NIV\nivem\ vs.
\NIII\niiir\ (see Table~\ref{wal_scheme}). 
Note that the ranges for luminosity classes I/III are somewhat shifted relative
to class V. To allow for a quantitative description, we investigated
the behavior of 
three extreme values, namely \NIV/\NIII\ = 0.1 (lower limit for O4
I/III and O3.5 V), \NIV/\NIII\ = 1 (O3.5 I/III and O3 V) and
\NIV/\NIII\ = 10 (representative for O2 I/III/V).

Before going into the details of our analysis, we compare in
Fig.~\ref{iso_ratio_z} the \NIV/\NIII\ emission line ratios expressed
in terms of EW (left) and line-strength (right, quantified in terms of
emission peak height), to ensure that different
definitions\footnote{Note that line ratios for \HeI/\HeII\ are partly
defined using EWs.} would not lead to different conclusions. Except
for small subtleties, there is no significant change in the run of
the iso-contours, regardless whether EWs or line-strengths are used to
derive the line ratio. The encountered subtle differences are mostly
caused by the considerable wind-strength (model series `E') used in
Fig.~\ref{iso_ratio_z}: Particularly for \NIV\nivem, enhanced EWs
owing to extended wings are produced, whilst the peak heights remain
unaffected. We have also convinced ourselves that a convolution of the
theoretical spectra with typical rotational speeds, \vsini\ =
100~\kms, and/or a degrading to a resolving power of 6000 have a
rather low impact (but see \citealt{markova11}): Generally, the
iso-contours are shifted to somewhat lower \Teff, by roughly 500 to
1000~K in extreme cases of high wind-strength and high [N]. In the
following we concentrate on the ratio of line-strengths alone, as
originally defined by~\citet{walborn02b}.

The inspection of the different iso-contours displayed in the
\Teff-\logg\ plane, Figs.~\ref{iso_ratio_z} to~\ref{iso_ratio_abun},
allows us to infer an important characteristics for the emission line
ratio. As it is true for the \HeI/\HeII\ line ratio (e.g., Mas05),
also the \NIV/\NIII\ line ratio is a sensitive function of surface
gravity. The hotter the temperature, the larger the \logg-value
necessary to preserve a specific ratio (ionization vs. recombination).
This trend seems to have vanished for the cooler models (\Teff\ $\leq$
40~kK) at high \logg\ in Fig.~\ref{iso_ratio_z}, but here the
influence of gravity is counteracted by the rather strong wind (see
Sect.~\ref{ratio_wind}). 

\subsubsection{Impact of background metallicity}
\label{ratio_z}
%
%
%
%

Figure~\ref{iso_ratio_z} shows the dependence of the line
ratio on \Teff\ and \logg\ for a fixed wind-strength, $\log Q =
-12.45$, and [N]~=~7.78, for iso-contours corresponding to \NIV/\NIII\
= 0.1, 1 and 10. To study the influence of $Z$, we display the
predicted behavior for MW (red), LMC (green), and SMC (blue) O-stars. 

Even a first inspection indicates the potential of this line ratio as
a temperature diagnostics. For a fixed surface gravity, e.g.,
\logg~=~4.0, we find difference on the order of 3~kK (for MW models)
to 8~kK (for SMC models) between iso-contours at \NIV/\NIII\ = 0.1 and
1, roughly corresponding to O4 (lower limit) vs. O3.5~I/III or O3~V. For
low-\mdot\ models this spread becomes somewhat smaller, 6 kK for SMC
stars, whilst for Galactic objects the \NIV/\NIII\ = 0.1 iso-contour
is no longer present since \NIV\ turns into weak absorption 
(Fig.~\ref{ew_4640_4058}).
The difference between the \NIV/\NIII\ = 1 and 10 iso-contours (O3.5
I/III or O3 V vs. O2) is on the order of 4-5~kK for all metallicities,
decreasing to 2.5-3~kK for low \mdot.



The specific differences of the emission line ratios as a function of
$Z$ can be easily explained in terms of our conclusions from
Sect.~\ref{z_dep} and by remembering the basic features displayed in
Fig.~\ref{ew_4640_4058}. First, \NIII\niiir\ is strongly affected by
$Z$ (more emission at lower $Z$, irrespective of temperature), in
contrast to \NIV\nivem. The latter line is increasing in strength over
almost the complete range in \Teff, whilst the former is increasing
until a certain maximum located at $\approx$ 40~kK, and decreasing
thereafter. Taken together, this implies that {\it lower} \Teff\ are
required to produce similar line ratios for low-metallicity objects at
cooler temperatures, compared to Galactic objects (lower \Teff\ means less
\NIII\ emission, to compensate for
the basically larger emission due to lower $Z$). Beyond 40~kK, on the other
hand, {\it higher} \Teff\ are required instead to reduce the increased
\NIII\ emission. This explains why the SMC/LMC iso-contours at
\NIV/\NIII\ = 0.1 are located at cooler \Teff\ than their Galactic 
counterpart, whilst the corresponding \NIV/\NIII\ = 1 and 10
iso-contours are on the hotter side. Because the $Z$ effect is largest
for SMC objects, also the \Teff\ differences between the \NIV/\NIII\ = 0.1 and
1 levels are largest in this situation. 

Finally, let us mention that the influence associated with $Z$ (around
2-3~kK for a given line ratio) is comparable with the typical spread
present in early spectral types, and thus in agreement with the 
non-detection of any correlation between the new spectral types and
the host galaxy (see above).

\subsubsection{Impact of wind-strength}
\label{ratio_wind}

\begin{figure}
\resizebox{\hsize}{!}
  {\includegraphics{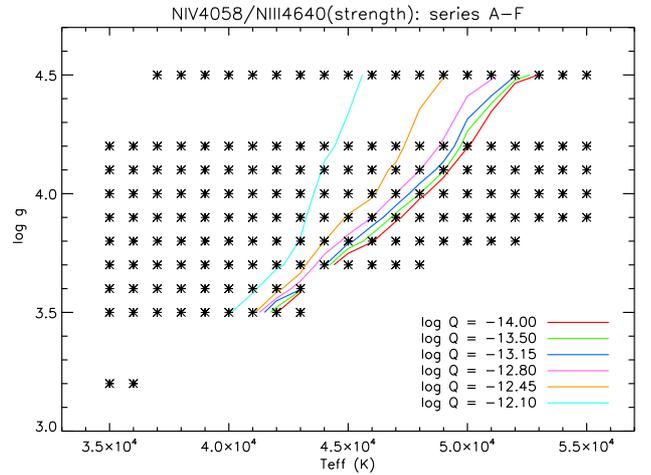}}
\caption{Dependence of \NIV\nivem/\NIII\niiir\ on wind-strength
(series `A'-'F'), for LMC models with [N]~=~7.78. The iso-contours 
correspond to a nitrogen emission line ratio of unity.}
\label{iso_ratio_wind}
\end{figure}

%
To test the influence of $\log Q$ on the emission line ratio, in
Fig.~\ref{iso_ratio_wind} we compare the iso-contours corresponding to
\NIV/\NIII~=~1 (O3.5 I/III or O3 V) for different wind-strengths, from
thin (series `A') to dense winds (series `F'). All models have LMC 
background abundances and [N]~=~7.78. 
Since lower metallicities counteract the impact of wind-strength,
corresponding iso-contours for SMC conditions vary to a lesser degree,
whilst the variations are larger for Galactic conditions, in both
cases by few hundreds of Kelvin.

Generally, the impact of wind-density on the \NIV/\NIII\ emission line
ratio is quite strong, particularly for larger surface gravities. 
From Fig.~\ref{iso_ratio_wind} it is evident that for a fixed \logg\
cooler \Teff\ are required to produce similar line ratios at higher
wind densities. Basically, we can combine the reaction into three 
groups, namely `A' to `C' models, `D' models, and models with winds
stronger than `D'. The first group consists of low-\mdot\ models with 
line ratios which are almost insensitive to wind-strength.  Models
belonging to the `D' series display a slight reaction, since the wind
begins to affect the line emission (either from both lines or, for
hotter stars, only from \NIV\nivem, see Fig.~\ref{ew_4640_4058}), and
in a different way, so that the line ratio becomes modified. From the
`E' series on, we found larger differences, e.g., at \logg\ = 4.0, of
roughly 2 and 4.5 kK for `E' and `F' models compared to low-\mdot\
ones. 

In conclusion, objects with the same \Teff\ and \logg\ but substantially
different wind-strengths are predicted to display significantly
different line ratios. Consequently, the assigned spectral types would
be no longer monotonic in \Teff\ if there would be a large scatter in
wind-strength. One has to note, however, that a certain \Teff-\logg\
pair also implies a certain luminosity class, and that the wind
densities per luminosity class are only mildly varying. Thus, the
monotonicity of a spectral type-\Teff\ relation might still be
warranted for a {\it specified} luminosity class, as long as the
corresponding luminosity class indicators (for the earliest O-stars,
\HeII$\lambda$4686) allow for a reliable classification (but see
Sect.~\ref{summary}).

\subsubsection{Impact of nitrogen abundance}
\label{nitro_ratio}

\begin{figure}
\begin{minipage}{8cm}
\resizebox{\hsize}{!}
  {\includegraphics{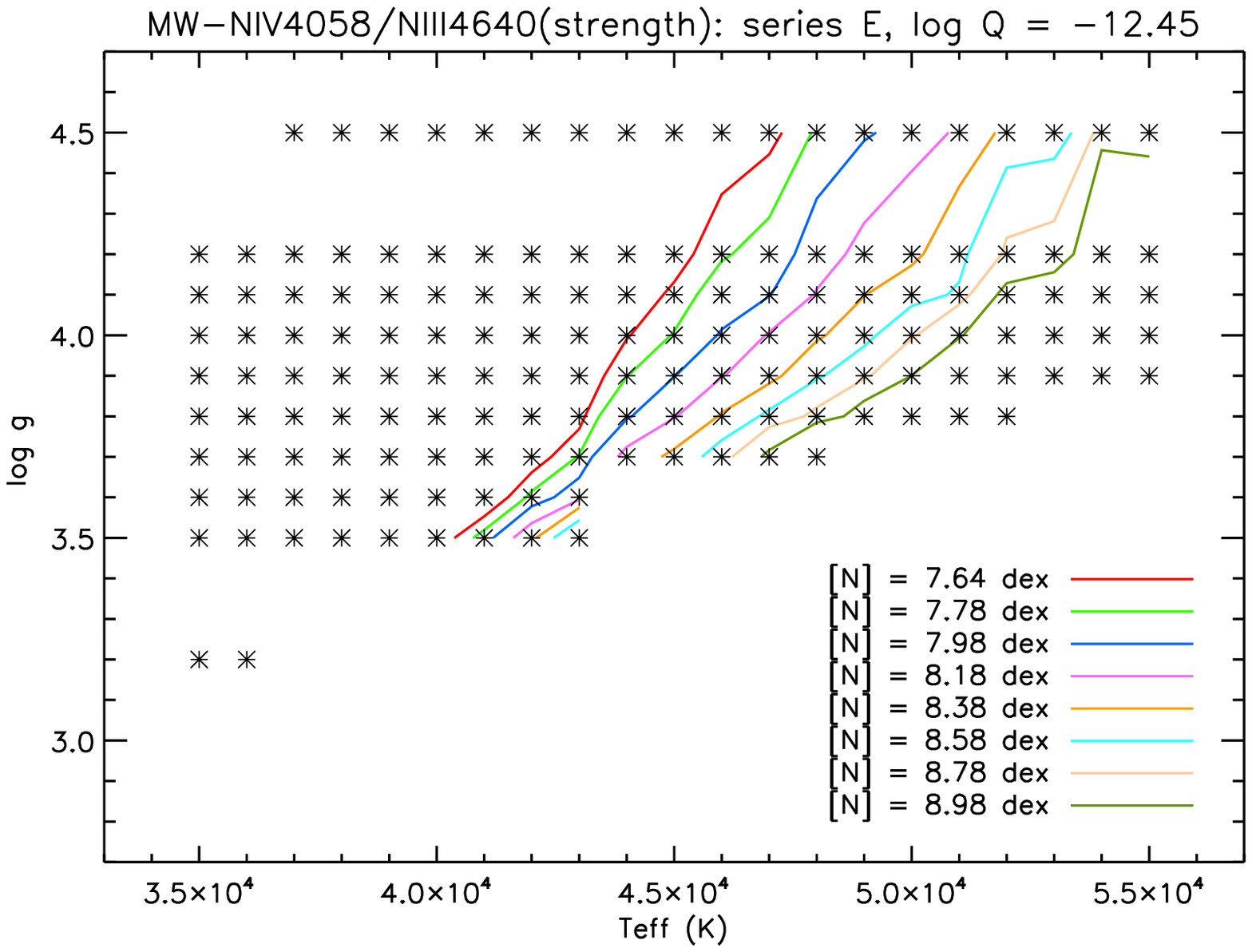}} 
\end{minipage}
\begin{minipage}{8cm}
\resizebox{\hsize}{!}
  {\includegraphics{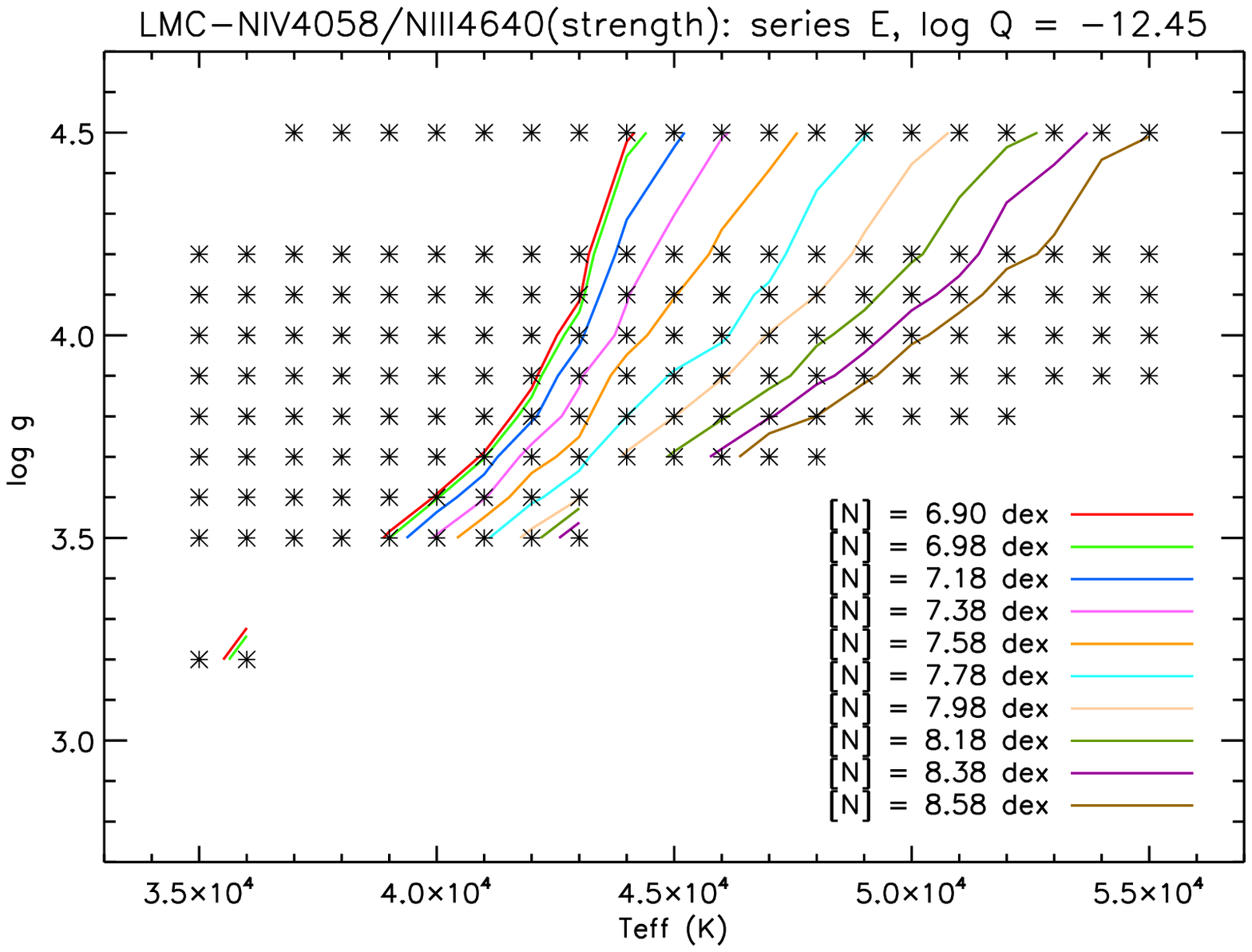}} 
\end{minipage}
\begin{minipage}{8cm}
\resizebox{\hsize}{!}
  {\includegraphics{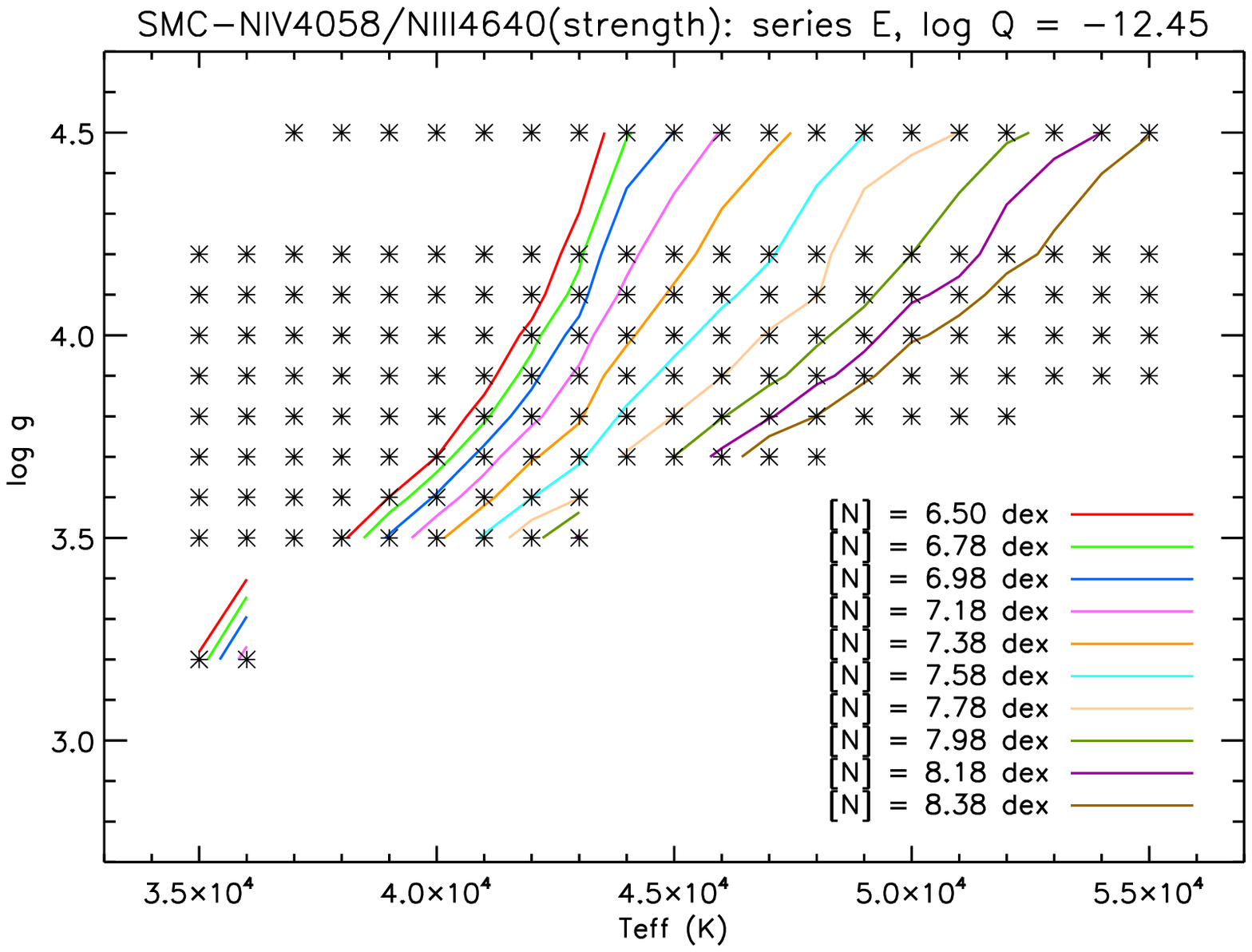}} 
\end{minipage}
 \caption{Dependence of \NIV\nivem/\NIII\niiir\ on nitrogen abundance,
for three different background metallicities: Galactic (upper panel), LMC
(middle panel), and SMC (lower panel). All models have $\log Q = -12.45$ 
(series `E'). The iso-contours correspond to a nitrogen emission line ratio of
unity.}
\label{iso_ratio_abun}
\end{figure}

After investigating the impact of background metallicity and
wind-strength, we now concentrate on the influence of nitrogen
content. At first glance, this is somewhat surprising. Abundance
should have a marginal or only small effect on line {\it ratios},
since it should cancel out as long as the lines are not too strong and
the ionization equilibrium is not disturbed. This statement, however,
is only true if the lines form in a `typical' way, i.e., are simple
absorption lines (e.g., \HeI$\lambda$4471/\HeII$\lambda$4541 used for
the classification of not too early O-stars). In the case considered
here, however, the lines are formed by complex and {\it different}
NLTE processes, and different abundances might have a different impact
on the formation of both lines (see Sect.~\ref{ab_niv}), so that the
ratio might become affected. Moreover, the variation of [N] in early
type stars can be much larger than, e.g., the variation of \YHe, which
amplifies the effect. Thus, it is mandatory to test for the impact of
nitrogen content on the \NIV/\NIII\ emission line ratio.

For this purpose, we display in Fig.\ref{iso_ratio_abun} the response
of the line ratio on different nitrogen abundances as present in our
model-grid, for each $Z$ (Galactic, LMC, and SMC background
abundances). To allow for a fair comparison, all models belong to
series `E'. Again, we display only contours with \NIV/\NIII~=~1, as a
representative value. Note that for each $Z$, the range of nitrogen
content is different, from the corresponding baseline
abundance to enrichments of roughly 1.3, 1.6, and 1.9 dex for MW,
LMC, and SMC objects, respectively (see legend and Table~\ref{tab_grid}).

\begin{table}
\caption{Stellar and wind parameters of models used to compare
synthetic nitrogen line profiles from {\sc fastwind} and {\sc cmfgen}.
The grid is a subset of the grid presented by \citet{lenorzer04}. For
details, see Paper~I. Even entries provide those parameters from
our {\sc fastwind} grid models which reproduce best the H/He spectra
from {\sc cmfgen}.}
\label{grid-cmfgen}
\tabcolsep1.8mm
\begin{center}
\begin{tabular}{llcccc}
\hline 
\hline
\multicolumn{1}{c}{Model}
&\multicolumn{1}{c}{code}
&\multicolumn{1}{c}{\Teff}
&\multicolumn{1}{c}{\Rstar}
&\multicolumn{1}{c}{\logg}
&\multicolumn{1}{c}{$\log Q$}\\
\multicolumn{1}{c}{}
&\multicolumn{1}{c}{}
&\multicolumn{1}{c}{(K)}
&\multicolumn{1}{c}{(\rsun)}
&\multicolumn{1}{c}{(cgs)}
\\
\hline
d2v    & {\sc cmfgen} & 46100 & 11.4 & 4.01 & -12.43 \\
E4740  & {\sc fastwind} & 47000 &      & 4.00 & -12.45 \\
d4v    & {\sc cmfgen}   & 41010 & 10.0 & 4.01 & -12.75 \\
D4140  & {\sc fastwind} & 41000 &      & 4.00 & -12.80 \\
\hline
s2a    & {\sc cmfgen}   & 44700 & 19.6 & 3.79 & -11.99 \\
F4540  & {\sc fastwind} & 45000 &      & 3.80 & -12.15 \\
s4a    & {\sc cmfgen}   & 38700 & 21.8 & 3.57 & -12.15 \\
F3935  & {\sc fastwind} & 39000 &      & 3.50 & -12.15 \\
\hline
\end{tabular}
\end{center}
\tablefoot{All models were calculated with \vmic = 10~\kms, and
with four different nitrogen abundances, [N] = 7.78, 7.98, 8.38, 8.78.
Wind strength parameter, $Q$,
calculated in units of $M_{\odot}\, yr^{-1}$, \rsun\ and \kms.}
\end{table}

Generally, an enhancement of [N] at fixed \logg\ shifts the
iso-contours towards higher temperatures. We checked that this shift
indeed is related to the different [N]-dependencies of the specific 
formation mechanisms of \NIV\nivem\ and \NIII\niiir\ rather than to a
modified ionization equilibrium, which remains quite unaffected from
variations in [N]. Moreover, the shift is quite similar for all $Z$,
producing an increase of roughly 1~kK for a change of +0.2 dex in
nitrogen abundance (at \logg = 4.0; lower $\Delta$\Teff\ are found at
lower \logg). This is an interesting result because objects with
considerable differences in [N] would display, for the same line
ratio, i.e., spectral type, large differences in \Teff. If we, e.g.,
consider LMC models at \logg~=~4.0, a line ratio of unity
(corresponding to an O3 dwarf) would be obtained at temperatures
differing by 6~kK if either [N] is at the baseline abundance (red,
\Teff\ $\approx$ 42.5 kK) or if [N] = 8.18 (dark green, \Teff\
$\approx$ 48.5 kK), corresponding to a typical enrichment of LMC stars
(paper~II)! Such a difference is much larger than the typical spread
of temperature per spectral type, and we conclude that the present
classification scheme for the earliest O-stars might be strongly
biased by nitrogen abundance. Only if the nitrogen content/enrichment
would be rather similar for a given position in the \Teff-\logg\
plane\footnote{If, e.g., nitrogen would be enriched by rotational
mixing alone and the initial rotational speeds were similar.}, this
bias would not contaminate a spectral-type-\Teff\ relation, similar to
our argumentation regarding the bias by wind-strength from above.

\section{Comparison with results from {\sc cmfgen}}
\label{comp_cmfgen}
In Paper~I we compared our results for \NIII\ lines with those from
the alternative code {\sc cmfgen},
for a grid of models comprising dwarfs and supergiants in the \Teff\ range
between 30 and 45 kK, and with `old' solar abundances according to
\citet{grevesse98}. In the following, we examine the consistency also
with respect to \NIV\ and \NV, concentrating on the hotter models
({\tt d4v, d2v, s4a, s2a}, see Table~\ref{grid-cmfgen}), but allowing
for different nitrogen abundances to check the predictions from the
previous sections.

{\sc cmfgen} models were computed with a modified
photospheric structure following the approach from \citet{santo97},
smoothly connected to a beta velocity law. In our approach, the
Rosseland mean from the original formulation was replaced by
the more appropriate flux-weighted mean. Several comparisons using
`exact' photospheric structures from {\sc tlusty} \citep{hubeny95}
showed excellent agreement with our method (see also
\citealt{Najarro11}). A turbulent velocity of 10~\kms\ was assumed
when computing both the level populations and line profiles. Our
models account for the presence of H, He, C, N, O, Si, P, S, Fe, and
Ni, totaling 3965 full levels (1319 super-levels) and $\approx$ 70,000
lines.

\begin{figure}
\resizebox{\hsize}{!}
  {\includegraphics{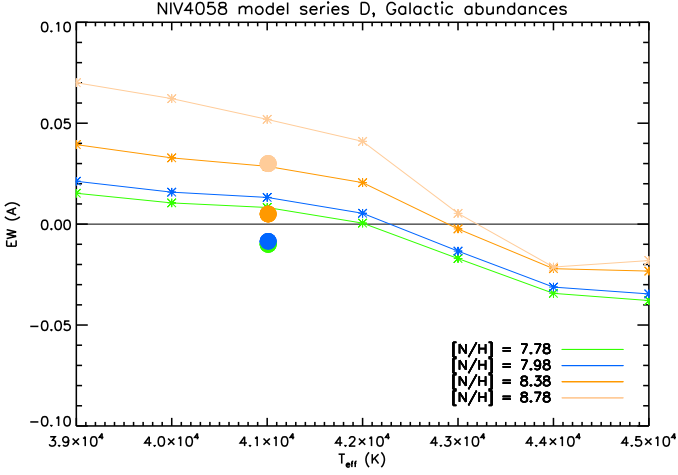}} 
\caption{\NIV\nivem, switching from emission to absorption with
increasing abundance. Model series `D' ($\log Q = -12.8$), $\logg =
4.0$, compared to results from {\sc cmfgen}, model {\tt d4v} (at
\Teff\ $\approx $ 41~kK, filled circles).} 
\label{nivemtoabs}
\end{figure}

First, we examine the potential switch of \NIV\nivem\ from emission
to absorption when increasing the abundance. From
Fig.~\ref{ew_4058_ab}, this effect should occur at lower
wind-strengths and not too high \Teff, i.e., potentially for model
{\tt d4v}. Indeed, the predicted effect is clearly present in the
corresponding {\sc cmfgen} models, where $\lambda$4058 is in emission
for [N] = 7.78 and 7.98, and in absorption for [N] = 8.38 and 8.78
(filled circles in Fig.~\ref{nivemtoabs}). The total variation in
equivalent width as a function of [N] compares very well to results
from {\sc fastwind}, which appear at somewhat higher \Teff = 42.5 kK,
as shown by the solid lines (based on our {\sc fastwind} grid for
\logg = 4.0 and model series `D'). Thus we conclude that the predicted
effect is more or less code independent.

\begin{figure}
\resizebox{\hsize}{!}
  {\includegraphics{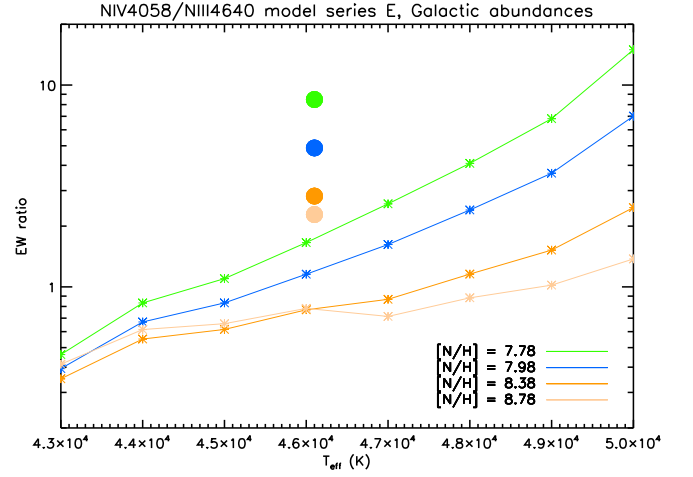}} 
\caption{Equivalent width ratio \NIV\nivem/\NIII$\lambda$4640, as a
function of nitrogen abundance. Model series `E' ($\log Q = -12.45$),
$\logg = 4.0$, compared to results from {\sc cmfgen}, model {\tt d2v}
(at \Teff\ $\approx$ 46~kK, filled circles).} 
\label{ew_ratio_d2v}
\end{figure}

In Fig.~\ref{ew_ratio_d2v} we compare the influence of the nitrogen
abundance on the emission line ratio \NIV/\NIII\ (calculated from
EWs), by means of model {\tt d2v}. Again, {\sc cmfgen} (filled
circles) predicts a similar effect as {\sc fastwind}, i.e., the EW ratio
increases with [N]. The actual values of these ratios, however, are
quite different from those calculated by {\sc fastwind}. Similar
ratios are only reached at higher \Teff, because, as discussed below,
{\sc fastwind} predicts stronger \NIII\ and weaker \NIV\ emission
lines for hot objects. Nevertheless, the `[N]-effect' is clearly
visible in the {\sc cmfgen} models, not only for model {\tt d2v} but
also for the others. Thus, the position of a certain emission line
ratio in the \Teff-\logg\ plane
depends on the actual nitrogen abundance. As already discussed, this
can lead to an ambiguity of the spectral-type-\Teff\ relation. We come
back to this problem in Sect.~\ref{discussion}.

In Appendix~\ref{app_cmfgen} we provide a detailed comparison of
strategic H/He/N lines predicted by {\sc cmfgen} and {\sc fastwind},
for all models from Table~\ref{grid-cmfgen}, and for different [N].
{\sc fastwind} spectra have been taken from our grid, to show that the
provided resolution in parameter space is sufficient in most cases.

Let us first concentrate on the H/He spectra (upper panels of
Figs.~\ref{d4v_8.78_hhen}, \ref{d2v_8.78_hhen}, \ref{s4a_8.78_hhen}
and \ref{s2a_8.78_hhen}), which turn out to remain unaffected by
typical variations in [N]. In most cases, there is an excellent
agreement between the H/He spectra from {\sc cmfgen} and the closest
or almost closest {\sc fastwind} grid-model, even for the \HeI\
singlet lines (but see \citealt{najarro06b}). The Stark-wings of the
Balmer lines are well reproduced at a similar gravity. The largest
discrepancies occur for model {\tt d2v} where
\HeI$\lambda\lambda$4471,4713 result in a better fit at \Teff\ = 47~kK
instead at the nominal value of 46~kK, and for model {\tt s4a} where
\logg\ = 3.5 produces better consistency than a model at \logg\ = 3.6
which would lie closer to the nominal value of \logg\ = 3.57. Finally,
\Ha\ ({\sc fastwind}) shows less emission for model {\tt s2a}, mostly
because the closest grid model has a lower wind-strength of $\log Q$~=~-12.15
than the nominal value, $\log Q$~=~-11.99. 
Figure~\ref{d4v_8.78_hhen} illustrates how the H/He spectrum changes
when \Teff\ is modified by $\pm$1000~K. In the considered
temperature range, the major reaction occurs in \HeI, predominantly in
the singlet lines ($\lambda\lambda$4387, 6678). 

The lower panels of Figs.~\ref{d4v_8.78_hhen}, \ref{d2v_8.78_hhen},
\ref{s4a_8.78_hhen} and \ref{s2a_8.78_hhen} display important nitrogen
lines for the four investigated models, at [N] = 8.78 (ten
times solar), whilst Figs.~\ref{d4v_7.78_n}, \ref{d2v_7.78_n},
\ref{s4a_7.78_n} and \ref{s2a_7.78_n} display these lines for [N]~=~7.78
(solar).

\NIII\ lines have been already compared in Paper~I, for [N]~=~7.92
(and \vmic = 15 \kms), and these findings remain valid also at higher
nitrogen abundance.

For the dwarf models {\tt d4v} and {\tt d2v}, the emission lines at
\trip\ are stronger than predicted by {\sc cmfgen}, and indeed
stronger than displayed in Paper~I, where the disagreement was less
obvious. This discrepancy could be tracked down to a different value
of \vmic\ used in the present models. A lower value (10 vs. 15 \kms)
results in narrower resonance zones, and can lead to higher emission
peaks.\footnote{A similar influence of \vturb\ has been seen, e.g., in
synthetic spectra of \Bra\ when in emission due to {\it photospheric}
processes \citep{Najarro11}.} Moreover, differences in the occupation
numbers - when close to inversion in certain regions - become more
pronounced, due to less averaging. For models {\tt s4a} and {\tt s2a}
we note similar discrepancies, though at a tolerable level.

As discussed in paper~I, the \NIII\ absorption line at $\lambda4097$ is
slightly stronger in {\sc fastwind} models at hot temperatures. Note
that for {\tt d2v} the profiles become identical when comparing with a
model that reproduces the {\sc cmfgen} \HeI\ lines, i.e., for \Teff~=~47~kK
instead of 46~kK.

The quartet lines\footnote{Preferentially used by~\citet{martins11b}
to derive nitrogen abundances in magnetic O-stars of late and
intermediate spectral type.} at \qua\ behave quite interestingly.
Among the low [N] models, these lines are in absorption only at {\tt
d4v}, whilst for the other models (higher \Teff\ and/or higher \mdot)
they appear in emission. In a certain parameter range (models {\tt
d2v} and {\tt s4a}) and in analogy to \NIV\nivem, these lines switch,
for increasing [N], from emission to absorption. Regarding these major
trends, {\sc cmfgen} and {\sc fastwind} behave similarly, and the
lines agree quite well for the dwarf models. For the supergiant
models, on the other hand, {\sc fastwind} predicts more emission/less
absorption than {\sc cmfgen}, in particular for {\tt s2a}. Thus, and
due to their complex behavior, the quartet lines need to be treated
with care when analyzing {\it early} O-stars.

\NIII$\lambda4003$ and $\lambda4195$ (in the blue wing of
\HeII$\lambda$4200 and additionally blended by \SiIII\ at hotter
temperatures) show a mostly reasonable agreement. \NIII$\lambda$4195 
matches almost perfectly when predicted to be in emission, and [N] is
low. The biggest discrepancy for $\lambda$4003 is found in
the low [N] model of {\tt s4a}, where {\sc cmfgen} predicts much more
absorption.

\NIV\ and \NV\ lines have not been compared so far. \NIV\nivem\ from
{\sc fastwind} shows always less emission than produced by {\sc
cmfgen}, though for models {\tt s2a} and for the low [N] model {\tt
s4a} the agreement becomes satisfactory.

In combination with more emission from the \NIII\ triplet at hotter
temperatures, this leads to a lower \NIV/\NIII\ EW/line-strength
ratio
in {\sc fastwind} (see Fig.~\ref{ew_ratio_d2v}), i.e., effective
temperatures deduced from this line-ratio {\it alone} would be higher
compared to {\sc cmfgen} results. This general trend is independent of
abundance. 

\NIV$\lambda$3480 is in good agreement for most models, except for
{\tt d4v} where {\sc fastwind} predicts more absorption. In contrast, 
\NIV$\lambda$6380 shows considerable differences, and for almost all
models (except for the high [N] model of {\tt d4v}) our grid predicts
much more absorption. Since this line provided no difficulties when
comparing with observations (neither in Paper~II nor in the present
investigation), we suggest that it might suffer from certain problems
in {\sc cmfgen}. We are currently working on this problem, which we
think is connected to the desaturation of the \NIV\ resonance line
around 247 \AA\ at the base of the wind.

Finally, the \NV\ doublet $\lambda\lambda$4603-4619 compares very well
in most cases, and only for the low [N] models of {\tt s2a} and {\tt
s4a} we find more absorption by {\sc fastwind}.

Summarizing the major discrepancies, we note that in the early O-type
domain {\sc fastwind} produces more emission in the \NIII\ triplet,
less emission at \NIV\nivem, and mostly much more absorption at
\NIV\nivab, compared to {\sc cmfgen}. We now ask how such
discrepancies might affect an abundance/\Teff\ analysis, and how one
might proceed to diminish the impact of corresponding uncertainties.

Since the H/He lines are in very good agreement, the stellar
parameters should be constrained quite well, as long as \HeI\ remains
visible. For the hottest stars with no or negligible \HeI, the
situation might become more difficult, since \Teff\ needs to be
constrained mostly from the nitrogen lines, and we would deduce lower
\Teff\ from {\sc cmfgen} if concentrating on the \NIII\ triplet and
\NIV\nivem\ alone (see above). In turn, this would lead to rather
different abundances. Fortunately, however, there is also \NV\ which
is very sensitive on \Teff\ (see below and Figs.~\ref{d4v_8.78_hhen},
\ref{d2v_7.78_n} and \ref{d2v_8.78_hhen}), and seems to be quite
code-independent\footnote{And not affected by X-rays from wind
embedded shocks under typical
conditions, see Paper~II.}. In conclusion, \Teff\ for `cooler' objects
should be mostly determined from the \HeI/\HeII\ ionization balance. 
Potential discrepancies regarding \NIII\trip\ and \NIV\nivem\ because
of `erroneous' predictions can be easily identified then. For the
hottest objects, on the other hand, \NV\ should obtain a strong weight
when determining \Teff. As long as either \NIV\nivab\ or
\NIV$\lambda$3480 are observed, these lines might be used as abundance
indicators, whilst any inconsistency of \NIV\nivem\ will tell about
the quality of that line.  

\begin{figure*}
\begin{center}
\resizebox{\hsize}{!}
  {\includegraphics[angle=90]{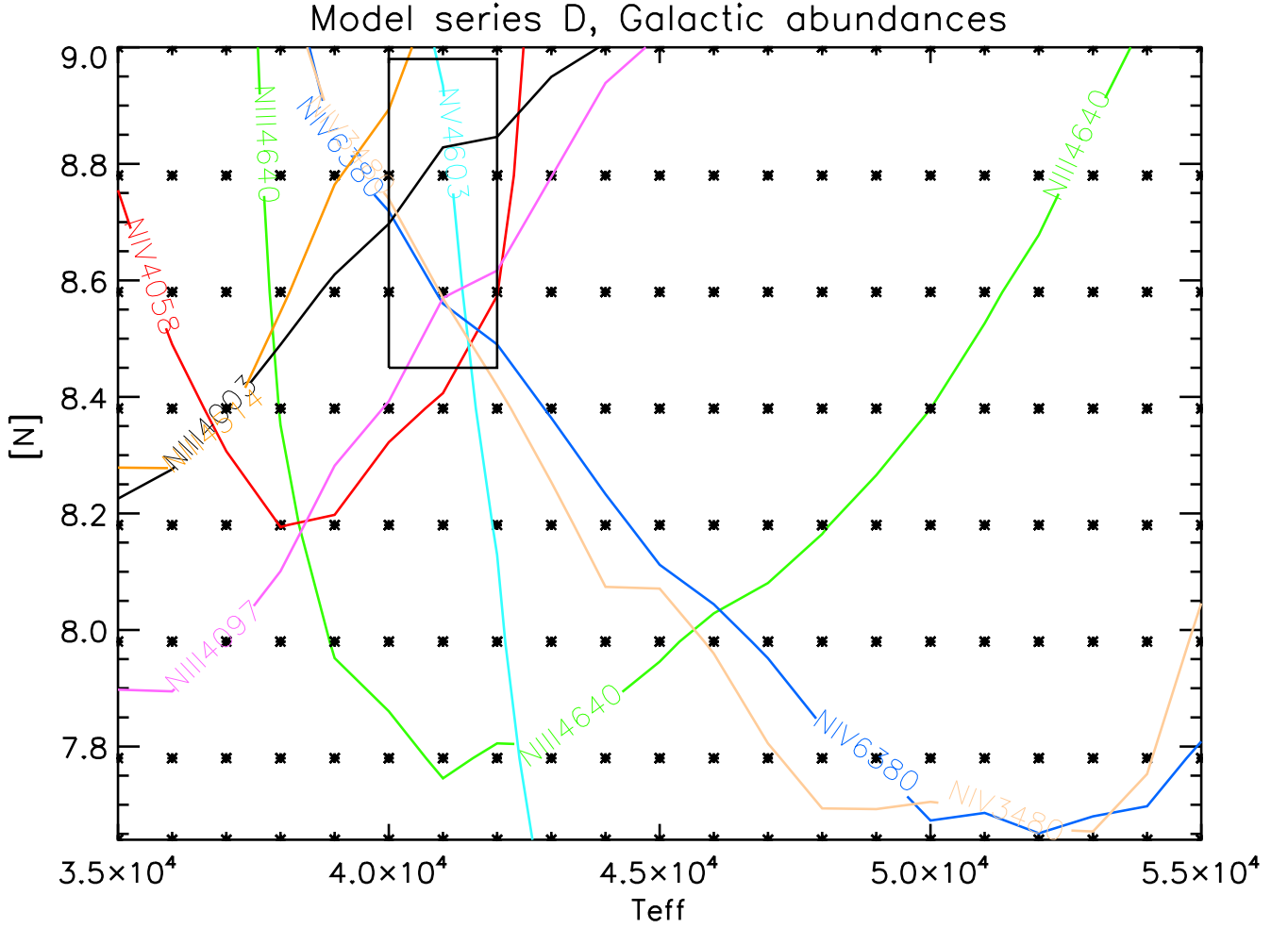}} 
\end{center}
\caption{Nitrogen abundance `analysis' of {\sc cmfgen}
model {\tt d4v} with [N] = 8.78. EW iso-contours for important nitrogen
lines are displayed in the \Teff-[N] plane of our {\sc fastwind} model
grid (series `D', $\log Q = -12.8$, \logg = 4.0). The displayed
iso-contours refer to EWs measured from the {\sc cmfgen} spectra of
model {\tt d4v}. In the ideal case, i.e., if {\sc cmfgen} and {\sc
fastwind} produce identical results, all iso-contours would cross at a
single point located at \Teff = 41kK and [N] = 8.78. The black box
marks the smallest region in parameter space where almost all lines
(except for \NIII\niiir) are predicted at the measured value. In
particular, five out of eight lines cross around the point \Teff\ =
41.5~kK / [N] = 8.55, i.e., at a lower abundance (by roughly 0.2 dex) than 
present in the original {\sc cmfgen} model. 
}
%
%
\label{d4v_isoew}
\end{figure*}

In Fig.~\ref{d4v_isoew} we now `analyze' synthetic spectra from
{\sc cmfgen} by means of our {\sc fastwind} grid, to check the overall
consistency when accounting for {\it all} important lines. This is
done for the high [N] model of {\tt d4v}, which is a prototypical
case, and by means of EW iso-contours in the \Teff-[N] plane (for
the appropriate model series `D' with $\log Q$ = -12.8, and a gravity of
\logg = 4.0). Equivalent width for all analyzed lines have been
measured from the {\sc cmfgen} spectra. These EWs are displayed as
iso-contours with respect to our {\sc fastwind} grid and should cross,
in the ideal case of a perfect consistency between both codes, at one
point corresponding to the model parameters, i.e., \Teff\ = 41~kK and
[N] = 8.78. Because of the various discrepancies discussed above, this
is not the case, but at least almost all iso-contours are close to
each other in quite a confined region, marked by a black box. 
In particular, five out of eight lines
cross around the point \Teff\ = 41.5~kK / [N] = 8.55. At the
original ({\sc cmfgen}) values, \Teff\ = 41~kK and [N] = 8.78, on the
other hand, there is only \NV$\lambda$4603 and \NIII$\lambda$4003. Thus, 
a {\sc fastwind} analysis of this model spectrum would yield lower nitrogen 
abundances, by roughly 0.2 dex (more on this below). 

Figure~\ref{d4v_isoew} displays a number of interesting aspects. Most
iso-contours have a parabola-like shape with a minimum at low [N],
where the `left' branch with negative slope relates to increasing
ionization fractions (the same EW at higher \Teff\ implies a lower
abundance), and the `right' branch with positive slope to decreasing
fractions. Because the right branch applies for lines from lower
ionization stages (\NIII), and the left branch for lines from higher
ones (\NIV, \NV), these different slopes allow to constrain the actual
parameter region quite well, even though both codes produce different
line-strengths when compared at the same location in parameter space. 

The only feature which is quite outside the enclosed region is the
\NIII\ triplet, which is {\it significantly} different in both codes
(and stronger in {\sc fastwind}, thus formed at lower \Teff\ when
compared to the {\sc cmfgen} EWs). 

At the comparatively `cool' temperature of model {\tt d4v},
\NV$\lambda\lambda$4603-4619 becomes almost
independent of abundance (almost vertical left branch), and thus a
very sensitive temperature indicator. Indeed, it perfectly fits at the
actual \Teff\ (which is also true for the \HeI/\HeII\ lines, see
Fig.~\ref{d4v_8.78_hhen}). Interestingly as well, the iso-contours for
\NIV\nivab\ and \NIV$\lambda$3480 occupy an almost identical region,
i.e., give very similar results, though being members of different
spin systems (singlet and triplet system, respectively). Insofar, the
information provided by both lines is similar, and the
observation/analysis of either of these lines should be sufficient,
which is fortunate since the 3500~\AA\ region is scarcely observed. 

Overall, when `analyzing' [N] from the {\sc cmfgen} spectrum and
accounting for the rather fixed \Teff\ and \logg\ values from H/He
and \NV,  we would derive a lower or roughly similar nitrogen
abundance from most lines, compared to the actual value used by {\sc cmfgen},
except for the \NIII\ quartet lines which would imply a somewhat
higher abundance. These results are also valid for the other models
from Table~\ref{grid-cmfgen}. We thus conclude that in those cases
where \Teff\ can be additionally constrained, {\sc fastwind} analyses
of early O-type stars will yield mostly lower nitrogen abundances than
analyses performed by means of {\sc cmfgen}.\footnote{At the
present stage of knowledge, we do not know which code does a `better'
job in reproducing reality.}


\section{Analysis of a sample of LMC/SMC early-type O-stars} 
\label{ana_obs}

So far, we mainly provided theoretical predictions on the \NIV/ \NIII\
emission line ratio as used by~\citet{walborn02b} to classify the
earliest O-stars, and concentrated on the impact of various 
parameters. For testing our predictions, we need to confront these results with
the analysis of a significant stellar sample. The LMC sample analyzed
in Paper~II contained only a few early-type objects (BI237, BI253,
N11-026, N11-031, and N11-060). These stars, biased towards O2, did
not cover all spectral types and luminosity classes within the
classification scheme. To allow for a larger sample, we added the
earliest stars from the analyses of LMC/SMC objects by \citet[hereafter 
Mas04, Mas05, and Mas09] {massey04, massey05, massey09}.  

\subsection{The stellar sample}
\label{sample} 
\begin{table}
\begin{center}
\caption{Sample stars used within this study, along with galaxy
membership and spectral type.}
\vspace{0.3cm}
\label{tab_sample}
\begin{tabular}{llll}
\hline 
\hline
Star & Cross-IDs & Galaxy & Spectral Type \\
\hline
AV 177        & -                & SMC   & O4 V((f))\tablefootmark{a}  \\
AV 435        & -                & SMC   & O3 V((f$^*$))\tablefootmark{a} \\
NGC 346-355   & NGC 346 W3       & SMC   & ON2 III(f$^*$)\tablefootmark{b} \\
LH 81:W28-5   & -                & LMC   & O4 V((f+))\tablefootmark{a}\\
LH 81:W28-23  & -                & LMC   & O3.5 V((f+))\tablefootmark{a} \\
LH 90:ST 2-22 & -                & LMC   & O3.5 III(f+)\tablefootmark{a} \\
LH 101:W3-24  & ST 5-27          & LMC   & O3.5 V((f+))\tablefootmark{a} \\
LH 101:W3-19  & ST 5-31          & LMC   & O2 If$^*$\tablefootmark{a} \\
R136-018      &   -              & LMC   &  O3 III(f$^*$)\tablefootmark{a} \\
R136-040      & R136a-535        & LMC   &  O2-3.5 V\tablefootmark{a} \\
Sk--65$^{\circ}$ 47  & LH 43-18  & LMC   & O4 If\tablefootmark{a} \\
Sk--67$^{\circ}$ 22  & BAT 99-12 & LMC   & O2 If$^*$/WN5\tablefootmark{c}\\
\hline
BI237        &      -            &LMC    & O2 V((f$^*$))\tablefootmark{d} \\
BI253        &      -            &LMC    & O2 V((f$^*$))\tablefootmark{d}\\
N11-026      & -                 & LMC   &O2 III(f$^*$)\tablefootmark{d} \\
N11-031      & P3061/LH10-3061   & LMC   &ON2 III(f$^*$)\tablefootmark{d} \\
N11-060      & P3058/LH10-3058   & LMC   & O3 V((f$^*$))\tablefootmark{d} \\
\hline
\end{tabular}
\end{center}
\tablefoottext{a}{Mas05};
\tablefoottext{b}{\citet{walborn04}};
\tablefoottext{c}{\citet{crowther11}};
\tablefoottext{d}{\citet{mokiem07a} and references therein}
\tablefoot{The horizontal line separates stars analyzed within this work 
(drawn from Mas04/05/09) from stars previously analyzed
in Paper~II. Identifications are as follows: "AV"
from~\citet{azzopardi82}; "BAT"
from~\citet{breysacher99}; "BI"from~\citet{Brunet75}; "LH" from~\citet{Lucke72}
except "LH10-3058" that is from~\citet{walborn02b, walborn04}; "N11"
from~\citet{evans06}; "NGC" from~\citet{Massey89}; "P" from~\citet{parker92};
"R136" from~\citet{massey98}; "R136a" from~\citet{malumuth94}; "Sk" from
~\citet{Sanduleak70}.}
\end{table} 

In particular, we selected objects with spectral types earlier than
O5. Table~\ref{tab_sample} gives a complete list of all early O-type stars
considered in the following, along with their spectral type and 
galaxy membership. The final sample consists of seventeen stars,
fourteen from the LMC and three from the SMC. Thirteen objects have been
drawn from Mas04 and Mas05, plus one object from Mas09 (the only
early-type star of that sample). Two of the objects analyzed in
Paper~II (BI237 and BI253) were also studied by Mas05. For these
objects, as well as for the remaining sample members (N11-026,
N11-031, and N11-060), we refer to our analysis from Paper~II.

From the original subsample of early-type stars studied by Mas04 and
Mas05, we selected representative objects. Owing to various reasons,
eight stars have been discarded. (i) Three dwarfs (R136-033,
R136-040 and R136-055) did not show any trace of either \NIII\trip\ or
\NIV\nivem\ in their spectra, and Massey and co-workers could
not classify them according to the~\citet{walborn02b} scheme. Therefore, we 
analyzed only one prototypical example for these stars, R136-040.
(ii) From the four early giants compiled in Mas05, we discarded two of
them, R136-047 (O2 III(f$^*$)) and LH~64-16 (O2N III(f$^*$)). The FOS data used for
the R136 stars, see Sect.~\ref{obs}, suffered from 
an intermittent behavior of some of the FOS diodes \citep{massey98},
which could result in the appearance of spurious features
contaminating the spectra.  Owing to a rather bad quality of the
spectra, we were not able to obtain a satisfactory fit for R136-047,
and we decided to discard it from the analysis. This problem was not
met for the remaining R136 stars used within this work (R136-018 and
R136-040). The other giant discarded, LH~64-16, shows similar severe
discrepancies as we had found in Paper~II for another star of this
class, N11-031. Owing to extreme difficulties to fit either
\NIII/\NIV\ or \NIV/\NV\ lines, 
we excluded it from the
present analysis. (iii) Regarding supergiants, we selected three out
of seven stars. For typical conditions (\mdot\ $\geq 10^{-5}$\msunyr,
\vinf$\sim$ 3000~\kms, and \Rstar $\sim$~15 \rsun), the corresponding
wind-strength, $\log Q \geq -12$, is well outside of our model-grid
(Table~\ref{tab_grid}), and we concentrated on representative objects
since analyzing all of them would have been extremely time-consuming,
even with a fast code such as {\sc fastwind}.

In summary, our sample comprises 9 dwarfs, 5 giants, and 3
supergiants.
The source of the corresponding spectral types for each object is listed in
Table~\ref{tab_sample}.

\subsection{Observational data}
\label{obs} 

Spectra for the 12 early O-stars from Massey et al. were taken in three
ranges, the UV, the blue and the red
visual band. Only optical data were used within this work.
 
(i) For the bulk of the stars, these data were obtained at the Cerro
Tololo Inter-American Observatory (CTIO) 4m Blanco telescope with the
RC spectrograph, during January 1999 (P.I. P. Massey). The blue and
red spectra cover the ranges 3750-4900~\AA\ and 5400-7800~\AA,
respectively. The spectral resolution was 1.4~\AA\ for the blue and
2.8~\AA\ for the red band, with a S/N between 200 to 560 per
resolution element, corresponding to about 4 pixels, in the blue 
(measured at 4500~\AA), and a typical S/N of about 150 per resolution
element in the red. 

For LH 101:W3-24,
additional spectra for the \Ha\ region were taken by means of the Hubble
Space Telescope (HST) using STIS/CCD (as
part of programme GO-9412, P.I. P. Massey), to amend the ground-based
observations which were heavily contaminated by strong nebulosity.
These data cover 6300-6850~\AA, with a spectral resolution of
0.84~\AA\ and S/N of 50.

(ii) For the R136 stars, we used two different data sets obtained with
the HST. One was taken with the G400 FOS/RD configuration under
GO-6417 (P.I. P.~Massey), with a resulting wavelength coverage from
3250 to 4820 \AA, 3 \AA\ resolution and a S/N of 60 at 4400 \AA.
Additional observations were taken with STIS/CCD under GO-7739 (P.I.
P. Massey). Blue observations covered the wavelength range
4310-4590~\AA, at a S/N of 100 for a spectral resolution of 0.4~\AA.
Observations centered around \Ha\ were made with the same setup as for
LH 101:W3-24 described above. Since the STIS spectra have better S/N
and resolution, we used them preferentially for our analysis. Because
of their limited wavelength coverage, excluding the spectral region
that comprises the most important nitrogen lines, we nevertheless had
to consider the FOS data, see Figs.~\ref{R136-040}
and~\ref{R136018}.

(iii) Spectra for NGC 346-355 were collected with the Clay 6.5m
(Magellan II) telescope at the Las Campanas Observatory, using the
Boller \& Chivens Spectrograph (P.I. P. Massey). The coverage was
3410-5040~\AA\ with a S/N of 600 at 4500~\AA\ in the blue, and from
5315 to 6950~\AA\ in the red, achieving a S/N of 400 at 6500~\AA.
For both ranges, the spectral resolution was 2.4~\AA.

All spectra were reduced with IRAF,\footnote{IRAF is distributed by
the National Optical Astronomy Obser\-vatories, which are operated by
the Association of Universities for Research in Astronomy, Inc., under
cooperative agreement with the National Science Foundation.} and we
performed additional re-normalizations for different wavelength ranges
within this work. 

As outlined in Sect.~\ref{sample}, our present sample includes two
objects (BI237 and BI253) which have been studied both in Paper~II and
by Mas05. The UVES spectra used in the former analysis have a
considerably higher resolving power, 40,000, compared to a resolving
power of 3,000 as provided by the RC spectrograph used by Mas05. After
comparing both data sets amongst each other and with our fits from
Paper~II, it turned out that the nitrogen lines are quite similar,
whilst the cores of the hydrogen (except for \Ha) and \HeII\ lines are
somewhat deeper in the UVES data, more than expected from the
higher resolution.
Because of the high quality of the UVES data, we kept the parameters
as derived in Paper~II, though a re-analysis based on the Mas05 data
would have provided rather similar values.

\subsection{Analysis}
\label{analys}
\begin{table*}
\caption{Fundamental parameters for the early O-star sample, assuming unclumped
mass-loss.}
\label{tab_param}
\tabcolsep0.7mm
\begin{center}
\begin{tabular}{lllrccrccccccccc}
\hline
\hline
\multicolumn{1}{l}{Star}
&\multicolumn{1}{l}{Spectral type}
&\multicolumn{1}{l}{Nitrogen}
&\multicolumn{1}{c}{\Teff}
&\multicolumn{1}{c}{\logg}
&\multicolumn{1}{c}{$\logg_{\rm true}$}
&\multicolumn{1}{r}{\Rstar}
&\multicolumn{1}{c}{log \Lstar}
&\multicolumn{1}{c}{\mdot\tablefootmark{c}}
&\multicolumn{1}{c}{log $Q$}
&\multicolumn{1}{c}{$\beta$}
&\multicolumn{1}{c}{\vinf}
&\multicolumn{1}{c}{\vsini}
&\multicolumn{1}{c}{\vmac}
&\multicolumn{1}{c}{$Y_{\rm He}$}
&\multicolumn{1}{c}{[N]}\\
\multicolumn{1}{l}{}
&\multicolumn{1}{l}{}
&\multicolumn{1}{l}{\Teff\ diag.}
&\multicolumn{1}{c}{(kK)}
&\multicolumn{1}{c}{(cgs)}
&\multicolumn{1}{c}{(cgs)}
&\multicolumn{1}{r}{(\rsun)}
&\multicolumn{1}{c}{(\lsun)}
&\multicolumn{1}{c}{}
&\multicolumn{1}{c}{}
&\multicolumn{1}{c}{}
&\multicolumn{1}{c}{(\kms)}
&\multicolumn{1}{c}{(\kms)}
&\multicolumn{1}{c}{(\kms)}
&\multicolumn{1}{c}{}
&\multicolumn{1}{c}{}\\
\hline
                                      &                 &                  &     &     &     &     & Dwarfs    &   & & & & & & \\
\hline
BI253\tablefootmark{a}                &O2 V((f$^*$))    & \NIV/\NV         &54.8 &4.18 &4.20 &10.7 &5.97 &1.53 &-12.61 &1.21 &3180 &230 &- &0.08 & 7.90 \\
BI237\tablefootmark{a}                &O2 V((f$^*$))    & \NIV/\NV         &53.2 &4.11 &4.12 & 9.7 &5.83 &0.62 &-12.98 &1.26 &3400 &140 &- &0.09 & 7.38 \\
R136-040                              &O2-3.5 V         &        -      &$>$51.0 &4.01 &4.02 &10.3 &5.81 &1.93 &-12.53 &0.80 &3400 &120 &- &0.08 & 6.90 \\
N11-060\tablefootmark{a}              &O3 V((f$^*$))    & \NIII/\NIV/\NV   &48.0 &3.97 &3.97 & 9.5 &5.63 &0.51 &-12.92 &1.26 &2740 &68  &40&0.12 & 8.20 \\
                                      &                 & \NIV/\NV         &51.0 &4.10 &4.10 & 9.2 &5.71 &0.48 &-12.92 &     &     &    &  &     & 8.15 \\
AV 435\tablefootmark{b}               & O3 V((f$^*$))   & \NIII/\NIV       &46.0 &3.90 &3.91 &13.8 &5.88 &0.21 &-13.15 &0.80 &1500 &110 &- &0.10 & 7.58 \\
LH 81:W28-23                          & O3.5 V((f+))    & \NIII/\NIV/\NV   &47.0 &3.80 &3.82 &10.0 &5.65 &1.57 &-12.53 &1.00 &3050 &146 &- &0.25 & 8.40 \\
LH 101:W3-24                          & O3.5 V((f+))    &     -            &47.0 &4.00 &4.01 &8.1  &5.46 &0.27 &-13.00 &0.80 &2400 &120 &- &0.10 & 7.78 \\
LH 81:W28-5                           & O4 V((f+))      &  \NIII/\NIV/\NV  &44.0 &3.80 &3.81 &9.8  &5.51 &1.08 &-12.60 &0.80 &2700 &120 &80 &0.15 & 8.38 \\
AV 177\tablefootmark{b}               & O4 V((f))       &     -            &44.0 &3.80 &3.85 &8.8  &5.42 &0.23 &-13.19 &0.80 &2650 &220 &- &0.15 & 7.78 \\
\hline
                                      &                 &                  &     &     &     &     & Giants    &     &     &    &  &     &  \\
\hline
N11-031\tablefootmark{a}              &ON2 III(f$^*$)   & \NIII/\NIV       &47.8 &3.95 &3.95 &13.4 &5.92 &2.02 &-12.64 &1.08 &3200 & 71 &60&0.11 & 7.83 \\
                                      &                 & \NIV/\NV         &56.0 &4.00 &4.00 &12.2 &6.12 &2.20 &-12.54 &     &     &    &  &     & 8.30 \\
NGC 346-355\tablefootmark{b}          & ON2 III(f$^*$)  & \NIV/\NV         &55.0 &4.00 &4.01 &12.0 &6.08 &2.50 &-12.39 &0.80 &2800 &140 &- &0.10 & 8.10 \\
                                      &                 & \NIII/\NIV       &51.0 &4.00 &     &12.5 &5.98 &2.00 &-12.51 &     &     &    &- &     &  7.98    \\
N11-026\tablefootmark{a}              &O2 III(f$^*$)    & \NIII/\NIV/\NV   &49.0 &4.00 &4.00 &11.3 &5.82 &1.56 &-12.63 &1.08 &3120 &72  &60&0.10 & 7.80 \\
                                      &                 & \NIV/\NV         &52.0 &4.10 &4.10 &11.0 &5.89 &1.49 &-12.63 &     &     &    &  &     & 7.75 \\
R136-018                              & O3 III(f$^*$)   & \NIII/\NIV/\NV\  &47.0 &3.90 &3.92 &14.2 &5.95 &1.68 &-12.76 &0.80 &3200 &180 &- &0.10 & 8.18 \\
LH 90:ST 2-22                         &O3.5 III(f+)     & \NIII/\NIV/\NV   &44.0 &3.70 &3.71 &18.8 &6.08 &4.56 &-12.36 &0.80 &2560 &120 &80&0.15 & 8.58 \\
\hline
                                      &                 &                  &     &     &     &     & Supergiants & & &     &    &  &     &  \\
\hline
Sk--67$^{\circ}$ 22                   & O2 If$^*$/WN5   & \NIII/\NIV/\NV   &46.0 &3.70 &3.74 &12.4 &5.80 &15.00&-11.60&0.80 &2650 &200 &- &0.30 & 8.78 \\
LH 101:W3-19                          & O2 If$^*$       & \NIII/\NIV/\NV  &44.0 &3.90 &3.91 &25.3 &6.34 &20.86&-11.97&0.80 &2850 &180 &- &0.10 & 8.18 \\
Sk--65$^{\circ}$ 47                   & O4 If           & \NIII/\NIV       &40.5 &3.60 &3.62 &19.8 &5.98 &11.00&-11.89&0.80 &2100 &160 &- &0.12 & 7.78 \\
\hline
\end{tabular}
\end{center}
\tablefoottext{a}{Stellar/wind parameters derived in Paper~II}; 
\tablefoottext{b}{SMC star}; \tablefoottext{c}{in units of \mdu}
\tablefoot{$\logg_{\rm true}$ is the surface gravity corrected for
centrifugal effects. For each star, the additional (or even primary) 
\Teff\ diagnostics used in parallel with the \HeI/\HeII\ ionization
equilibrium is indicated.  For four stars, two parameter sets are
provided, because we had problems to reach a simultaneous fit for all
considered lines from three nitrogen ionization stages. For N11-026
and N11-060 the first entry is our preferred one, and the second entry
is a (disfavored) alternative, whereas for N11-031 (see Paper~II) and
NGC 346-355 we consider both solutions as possible. For errors, see text.}
\end{table*}

Stellar and wind parameters were derived following the methodology
outlined in Paper~II, and are listed in Table~\ref{tab_param}.
Comments on individual objects and corresponding H/He/N line fits are
provided in Appendix~\ref{comments}.

In brief, we proceeded as follows. In accordance with Paper~II and the
investigations by Massey and collaborators, we assumed unclumped mass-loss
throughout the analysis.\footnote{Regarding the impact of clumping on
the synthetic nitrogen spectra, see Paper~II.} First, we determined
\vsini\ for each object using the Fourier
method~\citep{gray76}.\footnote{As implemented and tested in the
OB-star range by \citet{simon-diaz06} and \citet{simon-diaz07}.}
Subsequently, we roughly constrained the stellar/wind parameters as
well as the helium and nitrogen abundances by means of our model-grids
for LMC and SMC background metallicities. For deriving \Teff, we used
both the helium and nitrogen ionization balance when possible, i.e.,
when nitrogen lines from at least two ionization stages were visible
For the cooler objects of our sample (\Teff\ $\leq$ 44~kK), we mainly
relied on the helium ionization balance, and used nitrogen as a
consistency check. For hotter objects, where \HeII$\lambda$4541 begins
to loose its sensitivity to \Teff, we gave larger weight to the
nitrogen balance, as long as a reasonable fit to the \HeI/\HeII\ lines
could be maintained. Note that any \Teff/[N] degeneracy (as
present, e.g., for the line-strength {\it ratio} of \NIV/\NIII) is
broken by the absolute strengths of the lines, and the fact that we have
usually more than two lines at our disposal, which react quite
differently on changes in \Teff\ and [N] (cf. Fig.~\ref{d4v_isoew}).
Thus, as long as lines from at least two nitrogen ionization stages
are available, \Teff\ and [N] can be determined in parallel from
fitting these lines (as long as the other stellar/wind parameters can
be inferred from independent diagnostics). 
If nitrogen lines were not available (too weak), we either
relied on \HeI/\HeII\ alone (LH 101:W3-24 and AV 177), or, when even
\HeI\ was no longer visible, we settled for a lower limit on \Teff\
(R136-040). We refer to Appendix~\ref{comments} for a detailed discussion
on the specific diagnostics used (see also Table~\ref{tab_param}),
and on particular problems for individual stars.

Gravities were inferred from the wings of the Balmer lines, and $\log
Q$ from fitting \Ha\ and \HeII$\lambda$4686, with nitrogen lines
serving as a consistency check for both quantities. Because of the
rather low resolution and the mostly high \vsini\ values, we were able
to constrain the `macro-turbulence', \vmac, for only two objects (LH
81:W28-5 and LH 90:ST 2-22).  For the remaining ones, our fits were
acceptable without any need for extra-broadening.

At this point, we adopted terminal velocities as derived by Massey et
al. from UV lines, updating \mdot\ to preserve $Q$ if necessary. For the bulk
of the stars, we kept the velocity field exponent used within our model-grid,
$\beta$~=~0.8, because of reasons of consistency, since also Massey et al. 
used this value. Only for LH~81:W28-23 we adopted $\beta$~=~1.0, to
better reproduce \HeII$\lambda$4686 (see Fig.~\ref{LH81W23}).
For those objects with \Ha\ and \HeII$\lambda$4686 in emission (LH
101:W3-19, Sk--67$^{\circ}$ 22, and Sk--65$^{\circ}$ 47), it would
have been also possible to derive $\beta$ from the profile shape.
Because of the quite good fit-quality for both lines already with
$\beta$ = 0.8 (see Figs.~\ref{Sk67-22}, \ref{LH101W19}, and
\ref{Sk65-47}), we had no real reason to change this value though.

Final parameters were derived from a grid of much higher resolution
around the initial constraints. Stellar radii were calculated from
$M_V$ as provided by Mas04/05/09 and synthetic fluxes, with a
corresponding final update of \mdot. Following our experience for the
earliest O-stars from Paper~II, we used a micro-turbulent velocity,
\vmic~=~10~\kms, for all objects considered here. 

Errors on stellar/wind parameters are adopted following Paper~II. In
brief, we estimate an uncertainty of $\pm$1~kK in \Teff\ (for those
objects with a unique solution, see Table~\ref{tab_param}), $\pm$0.1~dex 
in \logg, and $\pm$0.05 dex in $\log \Rstar$, adopting a typical
error of $\pm$0.25~mag in $M_V$ (see Eq.~8 in \citealt{repolust04}).
Together with the errors for \Teff, this adds up to an uncertainty of
$\pm$0.11~dex for $\log L/\lsun$. Larger errors, plus/minus a factor of two, 
are present for \mdot, when \Ha\ is in absorption.\footnote{In this case, one is
unable to derive $\beta$ from line-fitting, and the uncertainty
in $\beta$ reflects in quite a large error for \mdot\
(\citealt{markova04}).} For objects with mass-loss indicators in
emission, the error is somewhat lower. 
Errors on \vinf\ are on the order of $\pm$100~\kms, taken from Massey
et al., and we estimate the errors on \vsini\ and \vmac\ as $\pm$10 to
20~\kms.
A rough estimate on the error of \YHe\ is $\pm$0.01 to 0.02. Regarding
[N], we adopt a conservative value of $\pm$0.15 to 0.20 dex to account
for the dependence on stellar and wind parameters.

\subsection{Comparison with results from Massey et al.}
\label{comp_mass}

In the following, we briefly discuss overall differences between the
stellar/wind parameters derived within this work and by Massey et al.
for overlapping objects, also regarding BI237 and BI253 which were
already analyzed in Paper~II. For details, see Appendix~\ref{comments}.

For the majority of objects, where we still could use the \HeI/\HeII\ 
balance as a primary temperature indicator (but also accounting
for nitrogen, see Sect.~\ref{analys}), we derived slightly cooler
\Teff, on the order of 0.5 to 1~kK. This is not too disturbing,
however, as one often finds systematic differences of this order when
different elements are used to derive \Teff\ for hot stars, as, e.g.,
in the case of using Si vs. He for early B-/late O-type stars
(e.g., \citealt{hunter07}). The maximum difference amounts to $\Delta\Teff$ =
$-2$~kK for LH~81:W28-23, on the margin of the adopted errors. 

For a few objects (BI237, BI253, NGC 346-355, R136-018, and
Sk--67$^{\circ}$ 22) where we needed to exploit the nitrogen
ionization balance to break the helium degeneracy, we inferred
considerably hotter temperatures ($\Delta$\Teff~$\sim$~2-7~kK).
Admittedly, and owing to the restricted quality of the data for some
of the earliest objects with weak \NIII\ triplet emission and very
weak \HeI$\lambda$4471, the effective temperatures derived might by
affected by uncertainties due to noise and continuum placement.
Nevertheless, the fact that also here the differences are systematic
suggests that the solution to this problem may not be purely
observational.

We also derived quite similar values for \logg, except for the five
objects explicitly mentioned above, which indicated higher gravities
(mostly because of the higher \Teff), with a maximum difference of
roughly 0.25 dex for BI253. Our \mdot\ values are systematically
lower, mostly because of the lower \Teff. For those stars with
increased \Teff, the reduction is caused by lower \Rstar\ and/or
higher $\beta$, where the latter effect is particularly strong for
BI253 which displays the largest difference, $\Delta \log \mdot =
-0.51$~dex.
Helium abundances agree quite well except for those few stars with a
larger change in \Teff, where we had to adapt \YHe\ to preserve the
fit quality of the helium lines. The largest difference found amounts
to about 0.05 in \YHe\ for LH 90:ST 2-22. Finally, most of our \vsini\
are in good agreement with the Massey et al. values, and only three
objects displayed substantial differences, with the largest one of
about 30~\kms for BI253.

One of the major implications of our re-analysis of the earliest
O-stars as considered by Massey et al. can be stated already now: We
do not find the pronounced degeneracy of the \NIV/\NIII\ emission line
ratio as claimed by Mas05 (for details, see Sect.~\ref{line-ratios}).
For instance, these authors questioned the monotonicity of the
classification scheme with respect to \Teff, because, among other 
problems, they inferred similar \Teff\ for two dwarfs, LH~101:W3-24
(O3.5) and BI237 (O2), which on the other hand displayed considerable
differences in their \NIV/\NIII\ emission line ratios.
These \Teff\ were derived by a pure H/He analysis. By exploiting the
nitrogen ionization balance, we are now able to break this
degeneracy, since, e.g., we find a considerably hotter temperature for
BI237 (see table~\ref{tab_param}), similar to the temperature
derived for BI253, the other O2 dwarf from the sample.

\section{Discussion}
\label{discussion}

\subsection{Nitrogen abundances}
\label{nitro} 

Though not the major topic of our present work, let us briefly comment
on the nitrogen abundances derived within this analysis. Globally,
these are consistent with our results from Paper~II. The bulk of the LMC stars
displays a considerable enrichment, stronger than expected from
tailored evolutionary calculations ~\citep{brott11b}, and supporting
the idea of an efficient mixing during very early phases of O-star
evolution. We confirm the tight correlation between the derived
helium and nitrogen content.

Only three dwarfs (note that only one prototypical case of
nitrogen-weak objects, R136-040, has been
analyzed) seem to be unenriched, visible already in the absence of
nitrogen lines in their spectra. The star with the highest enrichment
within the sample is Sk--67$^{\circ}$ 22 (O2 If$^*$/WN5), and the
derived [N] lies roughly 1.9 dex above the LMC baseline abundance, 
larger than any of the values found in Paper~II.
In analogy, also its helium content is extreme, and both findings
support the evolved nature indicated by its `slash' designation.

Although the nitrogen content (in absolute numbers) of the few SMC
sample members lies below that of most of our LMC stars, all of them
are strongly enriched, by more than one dex above the SMC nitrogen
baseline abundance\footnote{According to~\citet{hunter07},
[N]$_{\rm{baseline}}$~=~6.5}. The largest enrichment (by $\sim$1.6
dex, [N] = 7.98{\ldots}8.10)
was found for NGC 346-355, in agreement with its `ON' designation. 
A very similar abundance, [N] = 7.92, together with similar
stellar parameters, has been previously derived by both \citet{bouret03}
and \citet{heap06} for this star, see Appendix~\ref{comments}.

\subsection{Effective temperatures vs. spectral types}
\label{teff-spt}

\begin{figure}
\resizebox{\hsize}{!}
   {\includegraphics{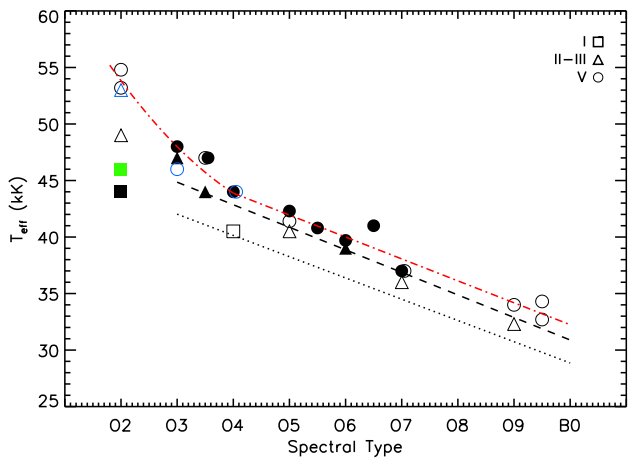}} 
\caption{Effective temperatures as a function of spectral type, for
LMC and SMC O-stars analyzed within this work and Paper~II (N11-031
and R136-040 discarded, see text). For NGC~346-355 (ON2 III(f$^*$),
SMC), \Teff\ according to our `average' solution, see text. Black
symbols: LMC objects; blue symbols: SMC objects; green symbol:
Sk--67$^{\circ}$ 22 (O2 If$^*$/WN5, LMC). Squares refer to
supergiants, triangles to (bright) giants, and circles to dwarfs.
Filled symbols correspond to objects with [N]~$\geq$ 8.0, and open
ones to objects with [N]~$<$ 8.0. The dashed-dotted (red) line
displays a least square linear and quadratic fit to all dwarfs,
according to Eq.~\ref{fit_dwarfs}. The \Teff\ calibration from
\citet{martins05a} for Galactic O-dwarfs (dashed) and supergiants
(dotted) is displayed for comparison. For clarity, some objects have
been slightly shifted horizontally.} 
\label{teffspt}
\end{figure}

In Fig.~\ref{teffspt}, we display the derived effective temperatures as
a function of spectral type for our present stellar sample
(Table~\ref{tab_sample}), augmented by all (LMC-) stars later than O4
from Paper~II to extend the sampling toward cooler types analyzed in a
uniform way.

Because it was not possible to assign a specific spectral type to
R136-040 (see above), this star is not contained in
Fig.~\ref{teffspt}. We also discarded N11-031 (ON2 III(f$^*$)) from
this diagram, owing to severe problems in the determination of its
\Teff. In Paper~II we were not able to derive a unique effective
temperature from \HeI\ and all three nitrogen ionization stages in
parallel, whilst using either \HeI/\NIII/\NIV\ or \NIV/\NV\ resulted
in a difference of $\Delta$\Teff = 8~kK, which is too large to allow
for further conclusions. Nevertheless, our sample contains another ON2
giant, NGC~346-355 from the SMC. Also for this object, we were only
able to obtain a hotter (based on \NIV/\NV) and a cooler (based on
\NIII/\NIV) solution, but the difference is much smaller,
$\Delta$\Teff = 4~kK. In Fig.~\ref{teffspt} we assigned a mean value,
\Teff=~53~kK (accounting for larger errors than typical, on the order
of 3~kK), to remain unbiased from a somewhat subjective view, but
note that both the hot and the mean temperature are considerably
higher than what we would have derived using just the \HeII/\HeI\ lines alone.
Further comments on the ON2 stars are given at the end of this
section.

To check the impact of [N] on the \Teff\ estimates for a
given spectral type, in Fig.~\ref{teffspt} we denote
objects with nitrogen abundances above and below [N] = 8.0 by
filled and open symbols, respectively. This threshold has been chosen
according to our findings from Paper~II, roughly separating
LMC objects with mild or typical enrichment from those with
unexpectedly strong enrichment.
We note that this value would be too high for SMC objects if one would
be interested in displaying the actual enrichment,\footnote{Indeed,
all SMC objects in Fig.~\ref{teffspt} appear with open symbols, 
despite their strong enrichment. Because of the lower baseline
abundance, a threshold of [N] = 8.0 corresponds to an enrichment of 1.5
dex, which is very unlikely to occur except for extreme objects such
as NGC~346-355 located close to this value.} but for our purpose of
testing and comparing the impact of [N], only the absolute value and
not the enrichment is decisive (see Sect.~\ref{nitro_ratio}).

The complete LMC sample considered in Fig.~\ref{teffspt} covers 26
stars spread over (almost) the full range of spectral subtypes,
comprising 16 dwarfs, 7 (bright) giants, and 3 supergiants.
On the other hand, only three early SMC objects
(two dwarfs and one giant) have been analyzed, preventing firm
conclusions.

Before we concentrate on the earliest types, we highlight some general
trends. From a first glance, and even though the number of supergiants
and (bright) giants within our sample is lower than the number of
dwarfs, it is obvious that \Teff\ increases from supergiants to dwarfs for
all spectral types, similar to Galactic conditions (e.g.,
\citealt{repolust04, martins05a}). At least for types later than O3.5,
giants and supergiants are about 1~kK and 4~kK cooler than dwarfs,
respectively. The latter difference is similar to what has been found
in previous studies, \eg\ by Mas05, \citet{mokiem07a}, and Mas09. For
the earliest types, on the other hand, this difference becomes much
larger, about 10~kK at O2.

We can also clearly distinguish different behaviors of the
spectral-type-\Teff\ relation. For luminosity classes V and II-III,
later types follow a linear trend, whilst the increase in \Teff\ is
much steeper for the earliest ones. On the other hand, the few early
supergiants of our sample seem to follow a linear trend with a slope
similar to cooler objects. 

At least for the dwarfs, we are able to provide a {\it typical} relation,
when ignoring any differences in $Z$ and [N].
Using a linear and a quadratic least-square fit for the
objects later/including and earlier than O4, respectively, we find
\beq
\label{fit_dwarfs}
\Teff = \left\{ \begin{array}{ll}
  51.64 -1.94 \times ST &\mbox{{if ST $\ge$ 4}} \\
  70.87 -10.29 \times ST + 0.88 \times ST^2
 &\mbox{{if 2 $\leq$ ST $<$ 4,}}
       \end{array} \right.
\label{dwarfs_guide}
\eeq 
where ST is the spectral type for O-dwarfs, and \Teff\ is expressed in
kK. At the present state of knowledge, this relation might be applied
to LMC stars only, since the low number of analyzed SMC-dwarfs as well as 
other arguments (see Sect.~\ref{caveats}) prohibit an application for stars
from the SMC.

For comparison, we show in Fig.\ref{teffspt} the observed
spectral-type-\Teff\ relation for Galactic dwarfs (dashed) and
supergiants (dotted) from~\citet{martins05a}. For the dwarfs, there is
a typical offset of roughly 1~kK, whereas the (cooler) O2 and
O4-supergiant seem to follow the Galactic calibration.

A similar comparison between their LMC sample and Galactic dwarfs was
provided by \citet{mokiem07a}, who found a somewhat larger offset by
$\sim$ 2~kK. This difference is caused by lower \Teff\ as resulting 
from the updated version of {\sc fastwind} for later spectral types 
(see Paper~II).

Reassuringly, the scatter in the spectral-type-\Teff\ relation for
objects later than O4 is small ($\sim$1~kK), since the spectral types
(as well as our primary \Teff\ indicator) 
rely on the \HeI/\HeII\ line strength (or E.W.) ratio, which is a
rather monotonic function of \Teff\ for a given luminosity class (controlling
gravity and wind-strength) and background abundance.\footnote{All our
cool objects are LMC stars.} The only outlier is N11-065, an O6.5
dwarf. The analysis of this star was quite difficult, and for the
determination of \Teff\ (and other parameters) we used also the
\NIII/\NIV\ ionization balance, in parallel with \HeI/\HeII.
Nevertheless, we found considerable problems in fitting the emission
at \NIII\trip\ together with the pronounced absorption at \NIV\nivem\
and the remaining nitrogen lines. Even though this problem became
somewhat relaxed by allowing for wind-clumping, it seems that there
was indeed a problem in the analysis of this star, and that it might
be cooler. 

Our sample includes two O4 dwarfs, one from the LMC (LH~81:W28-5) and
the other from the SMC (AV 177). For these stars we derive similar
\Teff\ and \logg, see Table~\ref{tab_param}, though the SMC star
(because of lower $Z$ and weaker winds) should be hotter (e.g.,
\citealt{bouret03, heap06}, \citealt{mokiem06}). Mas04/05, who came to
the same general conclusion, found from a pure H/He analysis the same
\Teff\ for AV177 (SMC), whilst they derived an even higher \Teff\ for
LH~81:W28-5 (LMC). We note that the present spectrum of the latter
shows weak emission in \NIV\nivem\ (Fig.~\ref{LH81W5}), so that the
star might be alternatively classified as O3.5. Since also the 
gravities are quite low for lc~V stars, the situation for both objects
remains somewhat unclear, even though the derived parameters are fully
consistent with the observed \NIV\ and \NIII\ lines.

We concentrate now on the impact of different parameters such as [N],
$Z$, and \mdot, on the spectral-type-\Teff\ relation for the earliest
O-types (O2-O3.5), for which the spectral classification depends
crucially on the \NIV/\NIII\ emission lines, whilst our \Teff\
diagnostics includes additional lines from those ions as well as from
\NV\ when visible. To allow for an easy understanding of the
following, we summarize the results from our parameter study in
Sect.~\ref{em_ratio}: For a given emission line ratio (i.e., spectral
type), the derived \Teff\ should increase with [N] and \logg, and
decrease with $Z$ (at least for higher \Teff, see Fig.~\ref{iso_ratio_z})
and \mdot\ (more precisely, $\log Q$).

Thus, the general \Teff-difference between dwarfs, (bright) giants,
and supergiants can be easily attributed to differences in \logg\ and
$\log Q$, which increase and decrease for increasing luminosity
class, respectively. In this way, both effects add up,
leading to \Teff(lc V) $>$ \Teff(lc III) $>$ \Teff(lc I), at least for
comparable [N] and $Z$. Regarding this general trend, a
classification in terms of nitrogen (early O-types) and helium (late
O-types) provides similar effects.

\begin{itemize}
\item[$\bullet$]
For the two LMC O3.5 dwarfs (LH~81:W28-23 and LH~101:W3-24), we find
the same \Teff, though \logg\ differs by 0.2 dex. The effect produced
by the larger [N] of LH~81:W28-23 ($\Delta$[N] $\approx$ 0.6~dex compared to
LH~101:W3-24, implying a shift towards higher \Teff) is counteracted by both
its denser wind and a lower \logg. Anyway, it is not clear
whether LH~81:W28-23 is correctly designated as a dwarf. Indications of a
giant nature are its low surface gravity (\logg~=~3.8), the
wind-strength, and the trace of a P-Cygni profile at
\HeII$\lambda$4686 (Fig~\ref{LH81W23}). If this would be the case, the
inferred \Teff\ might be too high for this luminosity class when
compared to the `other' O3.5 giant, LH 90:ST~2-22.

\item[$\bullet$]
The two O3 dwarfs, N11-060 from the LMC and AV~435 from the SMC, show
different \Teff. Astonishingly, the LMC star is hotter than the SMC
one, by 3~kK, contrary to what might be expected. Here, the [N]
effect outweighs the corresponding one associated to $Z$\footnote{In
cooler stars, the \HeI/\HeII\ ionization balance used for
classification is almost only affected by background abundance 
(less $Z$, less EUV line-blocking), see \citet{repolust04}.}, whilst
differences in \logg\ and wind-strength
compensate each other. We will come back to this finding and probable
consequences in Sect.~\ref{caveats}.

\item[$\bullet$]
The largest \Teff\ spread seen in our analysis occurs for the O2
stars, with \Teff\ ranging from 44 to 55~kK when accounting for all
luminosity classes. Again, the more enriched of the two dwarfs (BI253
vs. BI237) is the hotter one, since the [N] effect 
dominates over the larger wind-strength. 
For the two supergiants (Sk--67$^{\circ}$ 22 and LH 101:W3-19), we
find a similar \Teff\ difference, consistent with our predictions for
a combination of [N] and $\log Q$. Interestingly, the effect from a
lower \logg\ (3.70 vs. 3.90), as seen for the O3 dwarfs, becomes
inhibited by the extreme mass-loss. In Fig.~\ref{iso_ratio_wind}, we
noted that the \NIV/\NIII\ emission line ratio begins to loose its
sensitivity on \logg\ at the `F'-series of our model-grid. The
wind-densities of the two supergiants are even higher (roughly
corresponding to `G'), and in this situation the surface gravity does
not play any role for determining the line ratio. Finally, the
difference between two O2 giants (N11-026 from the LMC and NGC 346-355
from the SMC), $\Delta$\Teff~=~3~kK with respect to the `average' 
solution for NGC 346-355, is larger than expected, since all other
parameters except for $Z$ are quite similar, and the $Z$ effect alone
should amount to 1~kK (see Fig.~\ref{iso_ratio_z}). Note,
however, that the errors in \Teff\ for both objects are larger than
typical ($\ge$ 3~kK), related to the problems we encountered for 
ON2 (III) stars. 
\end{itemize}

\paragraph{The ON2 (III) stars.} 
Let us point out already here that there seems to be a general, severe
problem with the ON2 III class, as indicated from our results for
N11-031 and NGC~346-355, and the inspection of LH~64-16 (see
Sect.~\ref{sample}). This problem must relate to some unknown physical
process allowing for the presence of strong \NIII, \NIV\ and \NV\
lines in combination with weak, but still visible \HeI. Having
explored hundreds of models (both from our grid and additional,
fine-tuned ones), varying the N abundances as well as other parameters,
it turned out that these lines could not be synthesized in parallel. 
Moreover, we explored a variety of potential sources also from the {\sc
cmfgen} side, manipulating the photosphere-wind transition zone,
including wind-clumping with a variety of clumping laws, and
near-photospheric X-rays,\footnote{which do not help because of
the destruction of \NIII\ and \HeI, see Paper~II.} etc., and found
that none of these would solve the problem. Thus, we can be also sure
that the problem is neither related to {\sc fastwind}
nor to our present nitrogen model atom.

Also the mass-discrepancy found for these objects
underpins their problematic nature. Using the hotter solutions for
N11-031 and NGC~346-355, we derived spectroscopic masses of 60 and
53~\msun, respectively. Evolutionary masses from non-rotating tracks
by \citet{Charbonnel93} and \citet{Schaerer93}\footnote{For the
hottest stars, the inclusion of rotation has only a modest effect,
resulting in $\sim$10~\% lower masses, e.g., Mas05, as long as the
initial rotation is not high enough to induce quasi-homogeneous
evolution.} yield
90-95~\msun\ for both stars. Note however that also for a cooler
solution for NGC 346-355, at \Teff\ = 49.5~kK, already Mas05 stated
quite a mass discrepancy, in this case 50 (spectroscopic) vs.
75~\msun\ (evolutionary). For another LMC ON2~III star, LH~64-16, not
analyzed during the present work, Mas05 found an even larger discrepancy, 26
vs. 76~\msun. The authors suggested that the latter star might be the
result of binary evolution, since it shows highly processed material
at the surface, and since this star appeared to be located to the left
of the ZAMS. This seems also to be the case for N11-031 (at least for
the hot solution), but not for NGC~346-355. A similar problem was also
found for some of the close binaries in the R136 cluster
\citep{Massey02b}, supporting the idea that these stars suffered from
binary interactions. An alternative to binarity might be
quasi-homogeneous evolution, but the low \vsini\ values measured for
our objects (on the order of 100~\kms) render this possibility as
unlikely.

\subsection{\NIV/\NIII\ emission line ratio}
\label{line-ratios}

To further test the significance of the~\citet{walborn02b}
classification scheme, we investigated the relation between the
\NIV/\NIII\ emission line ratio and \Teff\ for our O2-O4 sample stars
in a quantitative way, similar to previous work by Mas05, but using
our updated parameters (see Sect.~\ref{comp_mass}). Unlike Mas05 who
used EW ratios, we employed the line-strength ratios to be consistent
with the classification scheme. Later on, we also consider an
alternative line ratio, \NV$\lambda$4603/\NIV$\lambda$6380 (hereafter
\NV/\NIV), to check its potential for classification purposes, and to
test whether this ratio might break potential degeneracies inherent to
\NIV/\NIII. 

Table~\ref{tab_ratios} lists the line ratios for our targets, as
derived both from the observations and from the synthetic models
associated to our best-fitting solutions. Again, we discarded N11-031
and R136-040 (see Sect.~\ref{teff-spt}). Note that at the lower and
upper end of the scheme the errors on the observed line ratios can
become significant, due to noise and/or absence of \NIV\nivem\ or
\NIII$\lambda$4640. In those cases, we provide corresponding lower or
upper limits and their uncertainties. Typical errors 
are on the order of 0.1 to 0.2 dex.

\begin{figure}
\begin{minipage}{9cm}
\resizebox{\hsize}{!}
  {\includegraphics{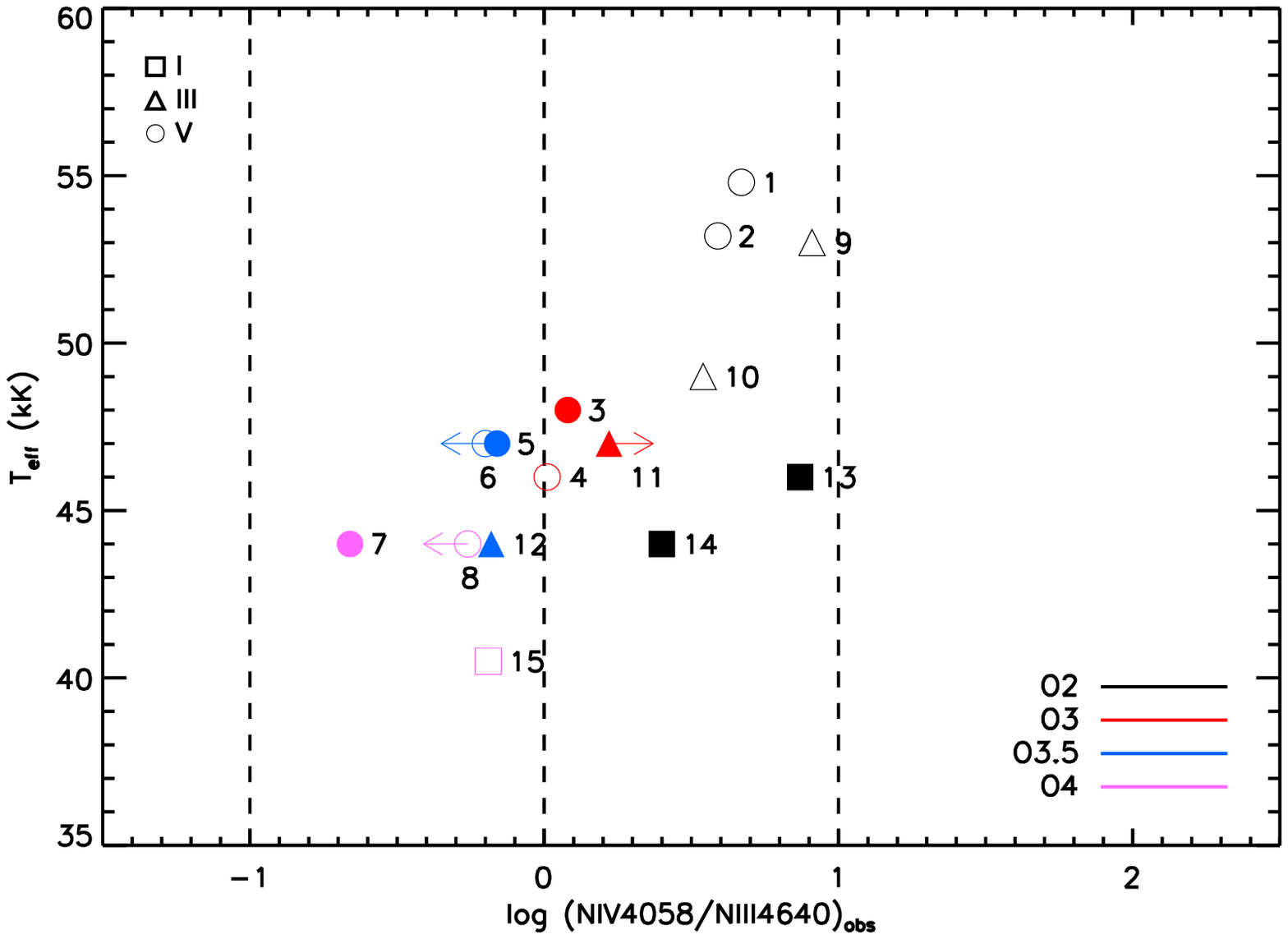}} 
\end{minipage}
\hspace{-.5cm}
\begin{minipage}{9cm}
\resizebox{\hsize}{!}
   {\includegraphics{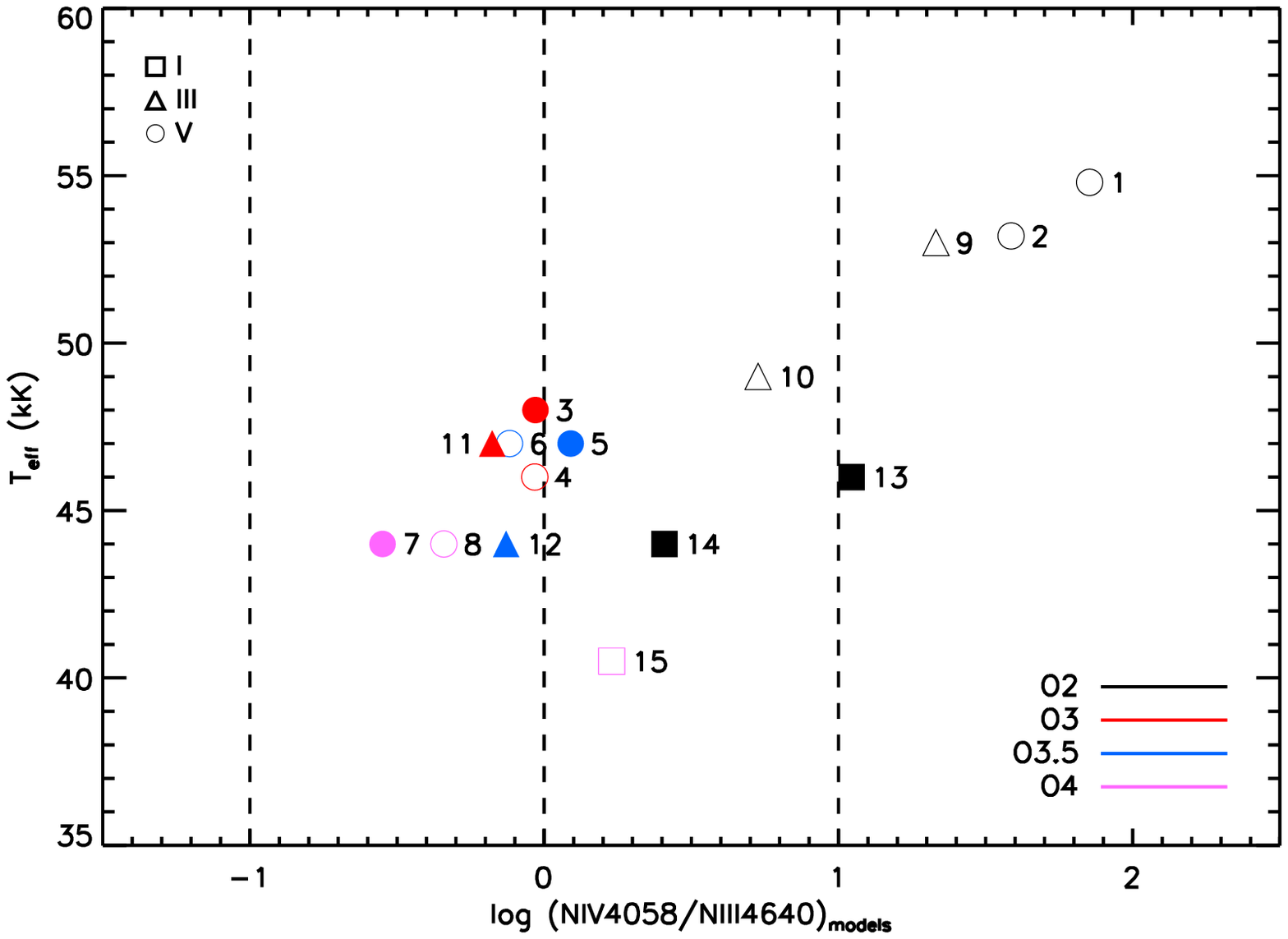}} 
\end{minipage}
\caption{Effective temperatures as a function of $\log$ \NIV/\NIII,
for O2-O4 sample stars. Spectral types distinguished according to
legend. Upper panel: observed emission line ratios, limits indicated
by arrows; lower panel: emission line ratios predicted from the
(globally) best-fitting models. Objects numbered according to
Table~\ref{tab_ratios}. Filling and symbols as in Fig.~\ref{teffspt}.
Dashed vertical lines (at $\log$ \NIV/\NIII\ = -1, 0, 1) correspond to
the limits used in Fig.~\ref{iso_ratio_z}. For object \#9
(NGC~346-355), \Teff\ and theoretical line ratio according to our
`average' solution. Stars \#4, \#8, and \#9 are SMC stars.}
\label{ratio_4058_4640}
\end{figure}

Figure~\ref{ratio_4058_4640} displays the derived effective
temperatures as a function of the observed (upper panel) and `model'
(lower panel) \NIV/\NIII\ line ratios, expressed logarithmically. 
Number designations for each object refer to Table~\ref{tab_ratios}.
Different colors are used for each spectral type: O2 (black), O3
(red), O3.5 (blue), and O4 (purple). Symbols referring to luminosity
classes and filling referring to [N] are as in Fig.~\ref{teffspt}.

\begin{table*}
\tabcolsep0.8mm
\begin{center}
\caption{Observed and predicted line-strength ratios for \NIV/\NIII\ and
\NV/\NIV, for the O2-O4 stars analyzed within this work. Observed
ratios (or limits) inclusive errors. Numbers refer
to Figs.~\ref{ratio_4058_4640} and \ref{ratio_4603_6380}.}
\label{tab_ratios}
\begin{tabular}{lllcrcrrcrl}
\hline 
\hline
\multicolumn{1}{l}{Star}
&\multicolumn{1}{l}{\#}
&\multicolumn{1}{l}{SpT}
&\multicolumn{1}{c}{\Teff}
&\multicolumn{2}{c}{$\log (\frac{\rm{NIV}}{\rm{NIII}})$}
&\multicolumn{1}{c}{$\log (\frac{\rm{NIV}}{\rm{NIII}})$}
&\multicolumn{2}{c}{$\log (\frac{\rm{NV}}{\rm{NIV}})$}
&\multicolumn{1}{c}{$\log (\frac{\rm{NV}}{\rm{NIV}})$}
&\multicolumn{1}{c}{Comments}\\
\multicolumn{1}{l}{}
&\multicolumn{1}{l}{}
&\multicolumn{1}{l}{}
&\multicolumn{1}{c}{(kK)}
&\multicolumn{2}{c}{obs}
&\multicolumn{1}{c}{model}
&\multicolumn{2}{c}{obs}
&\multicolumn{1}{c}{model}
&\multicolumn{1}{c}{}\\
\hline
BI253               &    1  & O2 V((f$^*$))   &  54.8  &   0.67   &$\pm$0.15   &  1.85    & 0.53   & $\pm$0.23     & 0.43  &\NIII$\lambda$4640 and \NIV\nivem\ underpredicted\tablefootmark{a} \\
BI237               &    2  & O2 V((f$^*$))   &  53.2  &   0.59   &$\pm$0.14   &  1.58    & 0.36   & $\pm$0.14     & 0.21  &\NIII$\lambda$4640 underpredicted\tablefootmark{a}\\
N11-060             &    3  & O3 V((f$^*$))   &  48.0  &   0.08   &$\pm$0.02  &$-$0.03  &$-$0.07    &$\pm$0.03   &$-$0.30  &\NIV\nivem\ and \NV$\lambda$4603 underpredicted\tablefootmark{a,b}\\
AV 435              &    4  & O3 V((f$^*$))   &  46.0  &   0.01   &$\pm$0.18  &$-$0.03 &$<-$0.30     &($+$0.10)  &$-$0.46  &\NV$\lambda$4603 diluted in noise\\
LH 81:W28-23        &    5  & O3.5 V((f+))    &  47.0  &$-$0.16   &$\pm$0.04   &  0.09  &$-$0.01    &$\pm$0.05   &$-$0.18  &\NIII$\lambda$4640 underpredicted\\
LH 101:W3-24        &    6  & O3.5 V((f+))    &  47.0 &$<-$0.20   & ($+$0.12) &$-$0.12 &$<-$0.28     &($+$0.14)  &$-$0.38  &\NIII$\lambda$4640 underpredicted \\
LH 81:W28-5         &    7  & O4 V((f+))      &  44.0  &$-$0.66   &$\pm$0.10  &$-$0.55  &$-$0.53    &$\pm$0.13   &$-$0.60  &Satisfactory fits\\
AV 177              &    8  & O4 V((f))       &  44.0 &$<-$0.26   &($+$0.13)  &$-$0.34 &$<-$0.36     &($+$0.09)  &$-$0.57  &Only \NIII$\lambda$4640 and \NIV\nivab\ visible \\
NGC 346-355         &    9  & ON2 III(f$^*$)  &  53.0  &   0.91   &$\pm$0.09   &  1.33    & 0.19    &$\pm$0.04     & 0.21  &\NIII$\lambda$4640, \NIV\nivem, \NV$\lambda$4603 underpred.\tablefootmark{c} \\
N11-026             &   10  & O2 III(f$^*$)   &  49.0  &   0.54   &$\pm$0.06   &  0.73    & 0.14    &$\pm$0.02   &$-$0.08  &\NIII$\lambda$4640, \NIV\nivem, \NV$\lambda$4603 underpred.\tablefootmark{a,b}\\   
R136-018            &   11  & O3 III(f$^*$)   &  47.0 &$>+$0.22   &( $-$0.21) &$-$0.18 &$>-$0.20     &($-$0.13)  &$-$0.32  &\NIII$\lambda$4640 and \NIV\nivem\ underpredicted          \\
LH 90:ST 2-22       &   12  & O3.5 III(f+)    &  44.0  &$-$0.18   &$\pm$0.06  &$-$0.13  &$-$0.20    &$\pm$0.04   &$-$0.39  &Satisfactory fits\\
Sk--67$^{\circ}$ 22 &   13  & O2 If$^*$/WN5    &  46.0  &   0.87   &$\pm$0.14   &  1.04    & 0.35    &$\pm$0.03     & 0.31  &\NV$\lambda$4603 under, \NIV\nivem\ overpredicted\\
LH 101:W3-19        &   14  & O2 If$^*$       &  44.0  &   0.40   &$\pm$0.07   &  0.41    & 0.05    &$\pm$0.14     & 0.00  &\NV$\lambda$4603 overpredicted\\
Sk--65$^{\circ}$ 47 &   15  & O4 If            &  40.5  &$-$0.19   &$\pm$0.12   &  0.23 &$<-$0.38     &($+$0.08)  &$-$0.52  &\NIV\nivem\ overpredicted\\
\hline
\end{tabular}
\end{center}
\tablefoottext{a}{For fits, see Paper~II, Appendix~C}; 
\tablefoottext{b}{Compromise solution, see Paper~II}; 
\tablefoottext{c}{'Average' solution}.
\tablefoot{Errors in brackets provide uncertainties of lower or upper
limits. Predicted line ratios (`model') drawn from best-fitting
synthetic spectra. N11-031 and R136-040 discarded, see text.} 
\end{table*}

If we examine the {\it observed} emission line ratio, we find quite a 
monotonic behavior, confirming its \Teff\ sensitivity, and similar
trends as in the previous section. On the one hand, dwarfs and giants
behave rather similar (though the giants seem to be a little cooler),
except for the O2 types where the spread is larger. 
On the other, the observed relation for supergiants lies in parallel
to the relation for dwarfs, but at considerably lower \Teff. As
already pointed out, this is caused by the different gravities and
wind-strengths. E.g., from Table~\ref{tab_param} we find $\log Q$
values in the ranges $[-13.00,-12.53]$ for the LMC dwarfs, around $-13.2$
for the SMC dwarfs, $[-12.76,-12.36]$ for the giants, partly
overlapping with the dwarfs, and $[-11.97,-11.60]$ for the supergiants.

The various spectral subtypes are located in quite different ranges
(by definition, consistent with the \citealt{walborn02b} scheme): All
O2 stars are located at $\log$ \NIV/\NIII\ $\geq$ 0.4, and O3 dwarfs
and O3.5 giants around $\log$ \NIV/\NIII\ $\sim$ 0 (\NIV\ $\sim$
\NIII). Finally, one of the O4 dwarfs is located at $\log$~\NIV/\NIII\
$\approx -0.7$, for the other one (\#8) we are only able to provide an
upper limit, $\log$~\NIV/\NIII\ $\la -0.3$, whilst the O4 supergiant
(\#15) lies close to the O3.5 border.

We see small [N] effects\footnote{slightly contaminated by
mass-loss effects, see below.} on the line ratio, where a larger
abundance tends (but not necessarily) to increase the emission
line ratio within a given spectral subtype, in particular for the
following pairs of stars: O3.5 dwarfs -- LH~101:W3-24 (\#6, upper
limit) and LH~81:W28-23 (\#5); O3 dwarfs -- AV435 (\#4, SMC) and
N11-060 (\#3); O2 dwarfs -- BI237 (\#2) and BI253 (\#1); O2
supergiants -- LH~101:W3-19 (\#14) and Sk--67$^{\circ}$ 22 (\#13).
Since, on the other hand, the emission line ratio should decrease for
enhanced [N] when keeping all other parameters fixed, a consistent
solution requires these objects to have a {\it higher} \Teff. For
the pair of O4 dwarfs -- AV177 (\#8, SMC) and LH 81:W28-5 (\#7), we
note that the star with the higher abundance (\#7) seems to have a
lower emission line ratio,\footnote{but note also the quite large
uncertainty for object \#8.} but this is consistent with the {\it
similar} \Teff\ of these stars. 

Indeed, there are not only [N] effects, but also mass-loss effects on
the emission line ratio. For all pairs considered above, the object
with higher [N] has also a higher wind strength, $\log Q$ (cf.
Table~\ref{tab_param}), which potentially counteracts the [N] effect.
Comparing the differences in [N] with the differences in $\log Q$, it
turns out that the [N] effect should dominate in all cases though.

Taken together, our findings explain the almost monotonic increase of
\Teff\ with \NIV/\NIII\ also within the individual spectral subtypes, 
consistent with the derived scatter of \Teff\ per subtype.
Differences in background metallicity and wind-strength seem to play a
secondary role, compared with the larger impact of [N].

If we now inspect the line ratios {\it predicted} by our best-fitting
models (lower panel), we mostly find quite similar values and trends.
For the bulk of the stars there are only small shifts due to minor
problems in properly fitting the lines (see Table~\ref{tab_ratios} for
particular comments on each star). However, more severe differences
are present for BI237 (\#1), BI253 (\#2), and NGC 346-355 (\#9,  
SMC).  For
the former O2 dwarfs, we predict too weak \NIII\ emission, thus
overestimating the line ratio, but note that the {\it observed}
emission is also rather weak for these stars. For NGC 346-355, the
problem is different since we are not able to reproduce both lines
using the `average' solution (see Fig.\ref{NGC346-355}) displayed
here. 

We conclude that, for a given luminosity class, the effective
temperatures are a rather monotonic function of $\log$ \NIV/\NIII,
that the scatter within a spectral subtype is mostly due to abundance
effects, and that our models are in fair agreement with the observed
line-ratios, except for the hottest objects where we underestimate
the observed (low) \NIII\ emission strength.

\begin{figure}
\begin{minipage}{9cm}
\resizebox{\hsize}{!}
  {\includegraphics{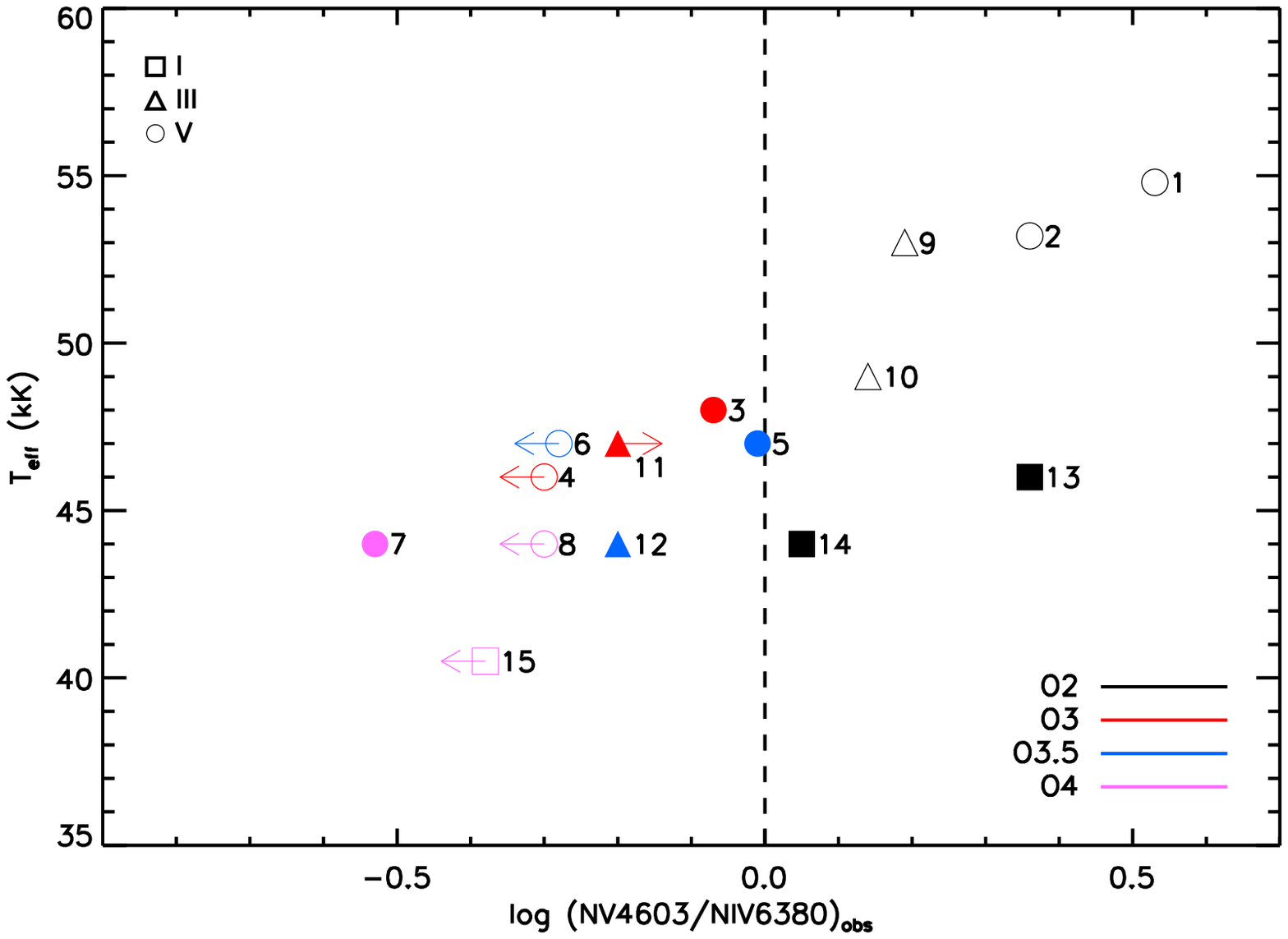}} 
\end{minipage}
\hspace{-.5cm}
\begin{minipage}{9cm}
\resizebox{\hsize}{!}
   {\includegraphics{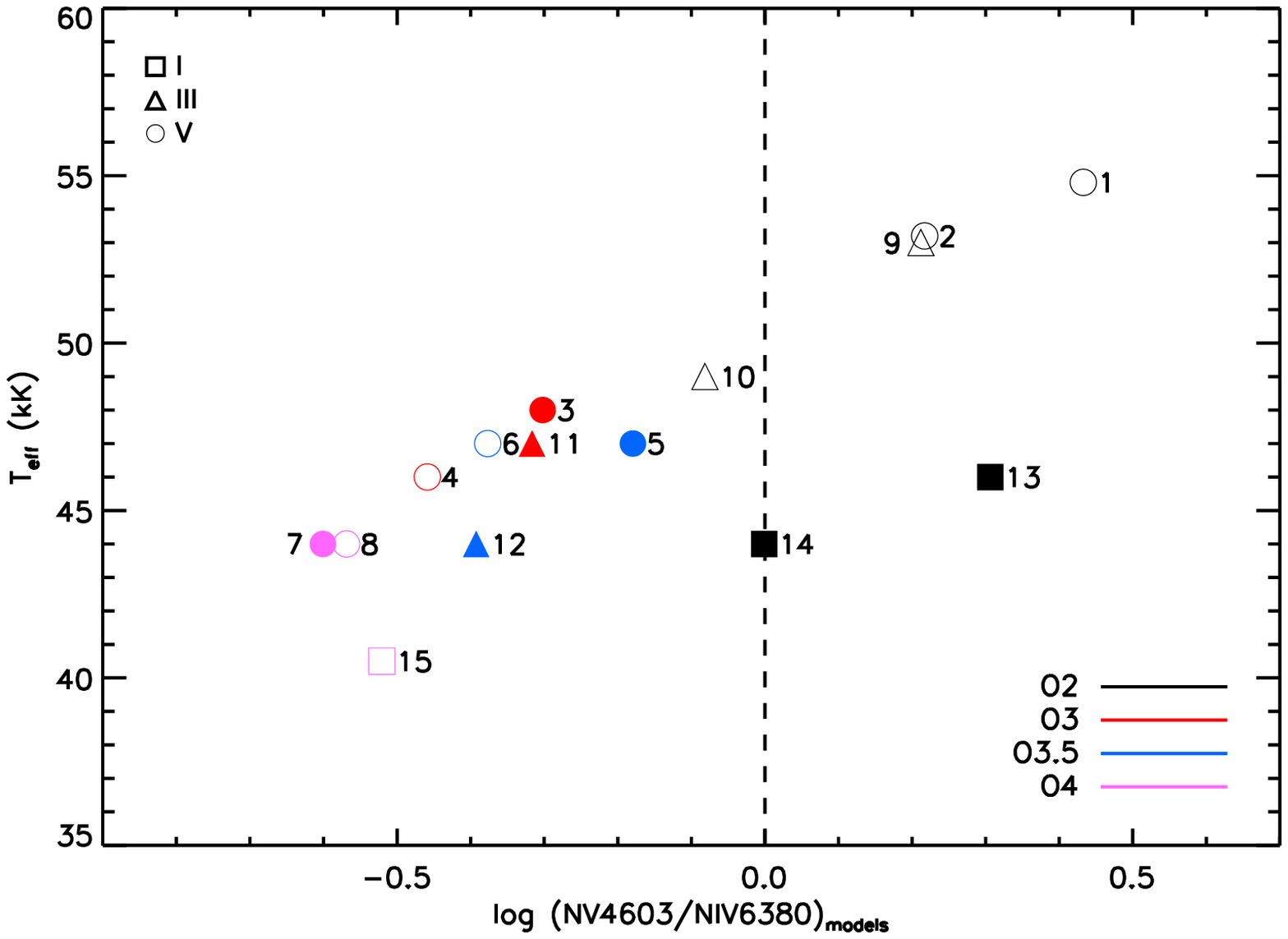}} 
\end{minipage}
\caption{As Fig.~\ref{ratio_4058_4640}, but for the
\NV$\lambda$4603/\NIV$\lambda$6380 line ratio.} 
\label{ratio_4603_6380}
\end{figure}

\subsection{\NV/\NIV\ line ratio}
\label{line-ratios_nv_niv}
Since both \NIV\nivab\ and \NV$\lambda$4603 are absorption lines
(i.e, less affected by complex formation processes), since they turned
out to be quite reliable during our analyses, and since
\NV$\lambda$4603 is very \Teff\ sensitive (Sect.~\ref{comp_cmfgen}), we
checked the corresponding line ratio as a potential diagnostic tool,
which might be even used for future classification purposes.

From Fig.~\ref{ratio_4603_6380}, we see that the relation \Teff\ vs.
log \NV/\NIV\ is remarkably monotonic. By inspection of the {\it
observed} line ratios (upper panel), we find again two different
trends, one for dwarfs and (bright) giants and another one for
supergiants. The objects are basically grouped together within three
regions: O2 stars with $\log$~(\NV/\NIV)$_{\rm{obs}} > 0$, O3/O3.5 dwarfs
and giants around $\log$~(\NV/\NIV)$_{\rm{obs}} \approx -0.3 {\ldots}
0$, and the O4 stars with
$\log$~(\NV/\NIV)$_{\rm{obs}} \la -0.4$. The only discrepant object seems
to be the SMC O3-dwarf AV 435 (\#4), which appears at the edge of the
O4-region. However, this `erroneous' position could be
tracked down to a considerable error in the measured line ratio,
because of very weak \NV\ lines diluted in the continuum
(Fig.~\ref{AV435}). 

Compared to the \NIV/\NIII\ emission line ratio, there seems to be a 
clearer separation between the different subtypes, e.g.,
Sk--65$^{\circ}$ 47 (\#15) is located closer to the remaining O4
objects, whilst regarding \NIV/\NIII\ it is closer to the O3 V/O3.5
III group. Moreover, there is a clear separation between the two O2
dwarfs (\#1,2) and the ON2 giant (\#9, SMC) because of weaker
\NIV\nivab.

Comparing now with the {\it predicted} line ratios (lower panel), we
see that predicted and observed ones agree quite well {\it over the
complete range.} Still, there are certain shifts because of a
non-perfect representation by our synthetic lines
(Table~\ref{tab_ratios}), but interestingly we do no longer find the
extreme differences for BI237, BI253, and NGC 346-355 as present in
the \NIV/\NIII\ diagram, which indicates that we are able to obtain a
satisfactory representation of the \NIV/\NV\ ionization balance. So
far, we cannot comment on any [N] related bias, since we did not
perform corresponding theoretical studies. In case the \NV/\NIV\ line
ratio might be used in future classification schemes, this will become
certainly necessary.

The monotonic behavior of the \NV/\NIV\ line ratio illuminates its
promising potential, particularly for the hottest objects (O2), simply
because there is more \NV\ than \NIII\ present at these temperatures.
We suggest to investigate this possibility when larger samples become
available.

\subsection{Caveats for low-metallicity stars}
\label{caveats}
Summarizing our findings, we conclude that already the present 
\citet{walborn02b} classification scheme allows for a reasonable
relation between spectral type and effective temperature, {\it as long
as it possible to discriminate the luminosity class}.
The only significant bias in the scheme might be produced by nitrogen 
abundance effects (or {\it extreme} variations in wind
strength). E.g., if the nitrogen abundance is not the same in an
O3~III and an O3.5~III star, then the O3~III star is not necessarily
hotter than the O3.5 III.

However, there are also important caveats regarding
low-metallicity (e.g., SMC) stars. As it is well-known, effective
temperatures increase with decreasing $Z$ for a given spectral type if
the classification is based on the helium ionization balance, i.e.,
the \HeI/\HeII\ line-strength ratio (O4-O9.7 stars, e.g., 
\citealt{bouret03}, Mas04/05, \citealt{heap06}, \citealt{mokiem06,
mokiem07b}). 
According to our predictions, this no longer needs to be true for the
earliest O-stars classified by means of nitrogen. 
\begin{itemize}
\item[(i)] {\it O2/O3 stars.}
Though for similar [N] a lower metallicity implies a higher \Teff\ (at
least for spectral types O3(dwarfs)/O3.5(giants and supergiants) and
earlier, see Fig.~\ref{iso_ratio_z}), this effect might be
counteracted by a different nitrogen content if we assume typical
(maximum) enrichments, of roughly +0.6, +0.9, and +1.0~dex above the
corresponding MW, LMC, and SMC baseline abundance, as predicted by
\citet{brott11a} for a 40~\msun\ star at an initial rotation of
270~\kms. As visible from Fig.~\ref{iso_ratio_abun}, for an O3 dwarf
with \logg~=~4.0~dex, our predictions indicate
\Teff~$\approx$~47-48~kK for the Galaxy and \Teff~$\approx$~46~kK for
the LMC/SMC.\footnote{See also Fig.~\ref{teffspt}, where the SMC O3
dwarf AV\,435 indeed is cooler than the corresponding LMC objects.}
Such cooler or at least similar \Teff\ for stars in a lower $Z$
environment should be present only for typical nitrogen abundances;
objects with a considerably different enrichment will contribute to
enlarging the spread.
\item[(ii)] {\it O3.5/O4 stars.}
In view of the results from Sect.~\ref{ratio_z}, there might be an
additional problem around O3.5/O4. From Fig.~\ref{iso_ratio_z}, SMC
stars should be {\it cooler} than corresponding LMC ones, even at a
similar [N], for a line ratio around \NIV/\NIII\ = 0.1 (which
represents a lower limit for O3.5~V and O4~I/III). Since the 
SpT-\Teff\ relation is rather monotonic for LMC stars at {\it all}
spectral types, derived either by helium or nitrogen (see
Fig.~\ref{teffspt}), the corresponding relation for SMC stars might be
not monotonically connected between these two regime: Even though
there seems to be a monotonic relation between \NIV/\NIII\ and \Teff\
on the hot side, the potential `jump' would be caused by the fact that
the SMC SpT-\Teff\ relation on the cooler side (based on
\HeI/\HeII) lies above the LMC relation.
\end{itemize}
To confirm or disprove these predictions and caveats, a thorough
analysis of a large sample of SMC stars is certainly required, 
given the few SMC objects investigated so far.

\subsection{SMC Of-stars}
\label{SMC_Of_stars} 

%
The luminosity criteria for O-type stars are based primarily on their
``Of'' characteristics; i.e., the strength of the \NIII\ triplet and
the \HeII$\lambda 4686$ emission.\footnote{These emission features
were linked with luminosity for Galactic stars (see, for example,
\citealt{walborn72}), a refinement over 
previous luminosity criteria based primarily on the Si{\sc iv}$\lambda
4089$ to \HeI$\lambda 4143$ ratio.} As already emphasized by
Mas04/05/09, this can lead to significant problems with luminosity
classification of O-stars found in low metallicity environments, such
as the SMC. 

The problem is that the emission strengths of the Of features are tied
to mass-loss rates (both for \HeII\ and \NIII), which in turn scale with
metallicity, and to (average) nitrogen abundances
(for \NIII).\footnote{Counteracting the increase of \NIII\ emission
because of less blocking in a low $Z$ environment.} One thus expects a
star in the SMC to show weaker Of-characteristics for the same
physical properties than would a similar star in the Milky Way. As
demonstrated in Sect.~\ref{niii_block_wind}, in particular the \NIII\
emission strength of the SMC star should be lower, unless the nitrogen
abundances of the two objects were similar, which would mean the rare
case of an {\it extreme} enrichment of the SMC star.

Throughout their studies of Magellanic Cloud O-type stars, and
particularly those in the SMC, Mas04/05/09 found numerous
examples where the Of-type properties indeed were weaker than would
be expected given the absolute magnitude of those stars\footnote{The
alternative explanation that all such discrepant stars were too bright
owing to their being binaries was contradicted by the excellent fits
obtained to the spectral features; this would require both components
of a binary to be of identical spectral subtype and brightness.}. We
can demonstrate the weakness of Of-features statistically as follows.
If we restrict ourselves to the SMC O-type stars with spectral
subtypes determined from slit spectroscopy (Table 6 of
\citealt{Massey02}) we find 74 stars, only 5 (7\%) have ``f'' type
designations.  Similarly, of the 83 O-type stars listed by
\citet{EvansHowarth08}, only 7 (8\%) have any ``f'' designation, and
all of these are either ``(f)'' or ``((f))'' indicating that \HeII$\lambda 4686$ 
is weakly in emission or in absorption.  In contrast,
Of-characteristics abound among O-type stars in the Milky Way. Of the
378 Galactic O-type stars catalogued by \citet{Maiz04}, 160 (42\%)
display Of-characteristics. Since Of-type stars are brighter than
non-Of stars (at least in the Milky Way), this may overestimate the
true percentage.  If we instead restrict ourselves just to the sample
of 24 O-type stars with well-established distances by {\sc Hipparcos}
(\citealt{Maiz04} Table 7) we find that 7 (29\%) have
Of-characteristics. 

Thus we conclude that there are strong indications for a significantly
lower percentage of SMC Of-stars, compared to Galactic ones, which we
attribute to the effects outlined in Sect.~\ref{niii_block_wind}.
Remember, however, that a lower wind-strength alone is not sufficient
to explain this finding (Fig.~\ref{gal_smc_4640}, upper panel), but
that a lower average nitrogen content needs to be present as well.

\section{Summary and conclusions}
\label{summary}
We investigated open questions raised by our
previous studies on the formation of \NIII\niiir\ and \NIV\nivem. We 
provided first theoretical predictions for the \NIV/\NIII\
emission line ratio, and confronted these predictions with 
observational findings, concentrating on a sample of
early-type O-stars. The results of this work can be summarized as
follows.
\begin{enumerate}
\item The emission strength of the \NIII\ triplet from objects with similar
\Teff\ and \logg\ depends on their metallicity, associated mass-loss, and
nitrogen content. Whilst even
under SMC conditions lower mass-loss rates alone are not able to
compensate the increase in emission for decreasing $Z$, a lower [N],
coupled to a significantly lower base-line abundance, can easily
outweigh the $Z$ effect and lead to overall lower emission strengths.
This might explain the relatively low number of SMC 'Of' stars.
\item Our models predict an only weak $Z$-dependence of \NIV\nivem\
(contrasted to the \NIII\ triplet). SMC-abundance models with
\Teff~$\leq$~45~kK display slightly more emission than their Galactic
counterparts, and vice versa for hotter temperatures. Much stronger 
is the impact of wind-strength though.
%
\item It turned out that \NIV\nivem\ behaves quite unexpectedly when
[N] is increased in low-\mdot\ models. For almost the whole
temperature range, we either obtain more absorption or less emission,
compared to models with a lower nitrogen content. For a specific
temperature range, \NIV\nivem\ can even switch from emission to
absorption when increasing [N]. For Galactic stars at
44~kK $\leq$ \Teff\ $\leq$ 50~kK and comparatively low \mdot,
our models imply that if \NIV\nivem\ is observed in absorption, 
this would indicate a strong nitrogen enrichment.
\item We provided first theoretical predictions on the \NIV/\NIII\ 
emission line ratio, as a function of $Z$, $\log Q$, and [N], by
studying line-ratio iso-contours in the \Teff-\logg\ plane. For an
emission line ratio of unity (i.e., a spectral type of O3.5~I/III or
O3~V), the corresponding \Teff\ increases with [N] ($\sim$1~kK per
increment of 0.2 dex in [N]) and \logg\ ($\sim$1~kK per increment of
0.1 dex in \logg). In addition, it should decrease with $Z$, at least
for higher \Teff\ ($\sim$2-3~kK difference between SMC and MW
objects), and $\log Q$ ($\sim$2-4.5~kK between low- and high-\mdot\
models). 
%
\item We performed a comparison with results from the alternative 
model atmosphere code {\sc cmfgen}, for a small grid of early O-type dwarfs
and supergiants.
Our basic predictions regarding the impact of [N] on (i) 
\NIV\nivem\ for low-\mdot\ models (see item 3), and (ii) on the \NIV/\NIII\
emission line ratio (see item~4) were confirmed by
corresponding {\sc cmfgen} results. 

Regarding specific line predictions, we found a mostly satisfactory
agreement, except for some systematic deviations: For early O-stars,
{\sc fastwind} produces more emission at the \NIII\ triplet, less
emission at \NIV\nivem, and mostly much more absorption at \NIV\nivab.
This would lead to lower \Teff\ and quite different [N] in analyses
performed by means of {\sc cmfgen}, if concentrating on the \NIII\ 
triplet and \NIV\nivem\ alone. Fortunately, the remarkably good
agreement of the H/He lines in both codes enables an identification of
potential problems regarding the nitrogen lines, as long as the helium
ionization balance is used to constrain \Teff. In the hot O-star
domain, the latter approach is no longer feasible because of
vanishing \HeI. Nevertheless, potential problems should
become obvious also here when relying on the \NV$\lambda\lambda$4603-4619
doublet, which turned out to be very sensitive on \Teff\ as well as
code-independent.

\item We confronted our theoretical predictions with results from an
analysis of a medium-size sample of LMC/SMC O-stars, drawn from
studies by Massey et al. (early types), and from Paper~II. The basic
difference to the Massey et al. analyses is found in the procedure for
deriving \Teff, where we used the nitrogen ionization balance in the
hotter \Teff\ regime instead of the helium one, to avoid any
degeneracy. For the cooler objects of our sample (\Teff $\leq$ 44~kK)
we mainly relied, when possible, on helium, using nitrogen as a
consistency check.
For these stars we found similar or slightly cooler \Teff\ compared to
Massey et al.. Considerably hotter \Teff, on the other hand, were
inferred for the earliest O-stars, by means of the nitrogen
diagnostics. Nevertheless, in most cases the corresponding synthetic 
\HeII/\HeI\ lines were still consistent with the observations, or
indicated only slightly lower temperatures. Notable exceptions are
the ON2~III stars (see below). 
The nitrogen abundances derived within our analysis are
consistent with our results from Paper~II: 
again, the bulk of the stars displays a considerable enrichment.

\item By inspecting the inferred effective temperatures, we saw 
that \Teff\ increases from supergiants to dwarfs for all spectral types, 
consistent with earlier results. For spectral types later than O3.5 
(down to O9.5), LMC giants are cooler by $\sim$1~kK, and
supergiants are cooler by $\sim$4~kK, compared to dwarfs. For types
earlier than O3.5, this difference (and also the scatter) becomes
larger, amounting to $\sim$10~kK at O2 when comparing supergiants and
dwarfs. For LMC dwarfs and giants later than O3.5, and for all
LMC supergiants, we found linear relations between \Teff\ and
spectral type, again consistent with previous work.  The earliest
dwarfs and giants, on the other hand, display a much steeper increase
in \Teff. 
The dominating effect responsible for the scatter in the SpT-\Teff\ relation
at earliest types was attributed to difference in [N], where for a given
spectral type more enriched objects are typically hotter. 

\item The relation between the observed \NIV/\NIII\ emission line ratio and
\Teff\ turned out to be quite monotonic, if discriminating for
luminosity class. Because of the high \Teff\ derived for the earliest
stars by means of the nitrogen ionization balance, we did not find the
pronounced degeneracy of the \NIV/\NIII\ emission line ratio as
claimed by Mas05. The scatter found within a spectral subtype is, again,
primarily produced by abundance effects. Our model predictions are in
fair agreement with the observed line-ratios, except for the hottest
objects where we underestimate the observed (low) \NIII\ emission-strength. 

\item We provided first insights into the relation between \Teff\ and the
\NV$\lambda\lambda$4603-4619/\NIV$\lambda$6380 absorption line ratio, which is
remarkably monotonic, particularly for the hottest objects in our
sample, and we highlighted the promising potential of this line ratio for future
classification schemes.
\end{enumerate}
Both our theoretical predictions and our observational analysis
suggest that the \citet{walborn02b} classification scheme is able to
provide a meaningful relation between spectral type and effective
temperature.
In particular, and as one might have expected, the \NIV/\NIII\
emission line ratio changes with effective temperature, all other
factors being equal. 
However, this ratio is also sensitive to surface gravity, mass-loss
rate, and to nitrogen abundance, which are expected to vary among a
sample of stars.
Thus, the significance of the classification scheme (within the
uncertainties caused by nitrogen abundance) might be only warranted as
long as it is possible to fairly discriminate the luminosity class (as
a proxy to gravity), and as long as there is a strong correlation
between spectral type/luminosity class and wind-strength. If, e.g.,
there would be weak-winded stars (\citealt{Marcolino09, Najarro11} and
references therein) also in the early O-type regime and not only at
later spectral types, the monotonicity with respect to \Teff\ might
become severely disturbed.

A clear identification of early O-type luminosity classes from
spectral morphology alone becomes difficult in low-$Z$ environments
such as the SMC. We emphasize the same point as made by Mas04/05/09,
that the standard luminosity classification criterion, primarily based
on the morphology of \HeII$\lambda$4686, is significantly biased on
mass-loss rates. Owing to lower wind-strengths for SMC conditions,
\HeII$\lambda$4686 is typically in absorption not only for lc~V, but
also for lower gravity objects which under Galactic conditions would
correspond to lc~I/III. Thus, there are O-type stars in the SMC whose
physical properties (visual luminosities and surface gravities) might
be in accord with them being giants or supergiants whereas their
spectroscopically determined luminosity classes may be dwarfs or
giants, respectively. To circumvent this caveat, other information,
such as the absolute visual magnitude of the system, may have to be
appealed to. Without such additional information, an appropriate
classification may rest on a more precise determination of the surface
gravity, e.g., using a visual inspection of the wings of \Hg. Another
possibility is to extend the classification scheme by including the
strength of \Ha\ (which needs to be carefully calibrated, since also
this wind-line remains in absorption).

\smallskip
\noindent
Our study also implies important consequences which need to be
validated or investigated in future work:
\begin{itemize}
\item[(i)] The SpT-\Teff\ scale constrained from our
LMC sample turned out to be, for a given luminosity class, more or
less monotonic. However, the majority of the analyzed objects displayed a
considerable nitrogen enrichment. To quantify the actual impact of
nitrogen abundance would require the analysis of a significant number
of un- or mildly enriched earliest MW/LMC objects\footnote{For SMC
objects, the nitrogen baseline abundance is too low to allow for a
clear-cut classification and spectroscopic analysis by nitrogen lines,
at least in the optical. This was already a problem for unenriched LMC
stars, e.g., R136-040, and the similar objects R136-033 and R136-055.}
that according to our predictions should be cooler than the enriched
ones. 

\item[(ii)] Typical SMC stars of earliest
spectral types might have effective temperatures below corresponding
LMC objects, and the overall SpT-\Teff\ relation for SMC stars might
be non-monotonic around O3.5/O4.

\item[(iii)] Our predictions suggest that there should be
(enriched) Galactic O4 dwarfs with emission at \NIII\niiir\ and
absorption at \NIV\nivem\ (item 3 from above). So far, the
\citet{walborn02b} classification scheme only states that \NIV\nivem\
should be absent for such objects.  A quick inspection of the IACOB
database of Galactic OB stars \citep{simon-diaz11b} allowed us to find
a first indication of this effect, for HD\,46223 (O4 V((f))), and we
were able to reproduce its strong emission at \NIII\niiir\ together
with a pronounced absorption at \NIV\nivem, for a considerable
nitrogen enrichment.

\item[(iv)] As became apparent throughout this work (first indications were 
already found in Paper~II), we encountered severe problems for ON2~III
stars. For these objects, we were not able to find a simultaneous fit
of the pronounced \NIII/\NIV/\NV\ lines and the weak, but still
clearly present \HeI$\lambda$ 4471, independent whether we used {\sc
fastwind} or {\sc cmfgen}. Though we were able to derive a cooler
(fitting \NIII/\NIV\ and \HeI) and a hotter solution (fitting
\NIV/\NV), we were not able to favor one of those. In some cases, the
\Teff\ difference between the hotter and the cooler solution is
extreme, e.g., for N11-031 analyzed in Paper~II. Because of the
restricted quality of our present dataset, new observations with
higher S/N and higher resolution would be extremely valuable to verify
our results using the nitrogen ionization balance and/or the pure
\HeII/\HeI\ results from Massey et al. when \HeI\ (and \NIII) become
extremely weak.

\end{itemize}
Future work is certainly needed to address the aforementioned issues.
Upcoming analyses of extensive O-star samples, e.g., from the
VLT-FLAMES Tarantula survey and the IACOB data base, will be
fundamental for trying to explain these open questions.

\begin{acknowledgements}
{We would like to thank our anonymous referee for very useful comments
and suggestions. Many thanks to Nevy Markova for providing us with her
spectrum of HD\,64568, and to John Hillier for providing the {\sc
cmfgen} code. J.G.R.G. gratefully acknowledges financial support from
the German DFG, under grant 418 SPA 112/1/08 (agreement between the
DFG and the Instituto de Astrof\'isica de Canarias), and from the LMU
Graduate Center (Completion grant). J.P and F.N. acknowledge financial
support from the Spanish Ministerio de Ciencia e Innovaci\'on under
projects AYA2008-06166-C03-02 and AYA2010-21697-C05-01. Partial
support for P.M. was provided by NASA through grant AR-11270 from the
Space Telescope Science Institute, which is operated by the
Association of Universities for Research in Astronomy, Inc., under
NASA contract NAS 5-26555.}
\end{acknowledgements}

\bibliographystyle{aa}
\bibliography{bib_niii}

\Online
\appendix
\onecolumn
\section{Comparison with {\sc cmfgen}} 
\label{app_cmfgen}
\begin{figure}[b]
\begin{center}
{\includegraphics[width=14.5cm]{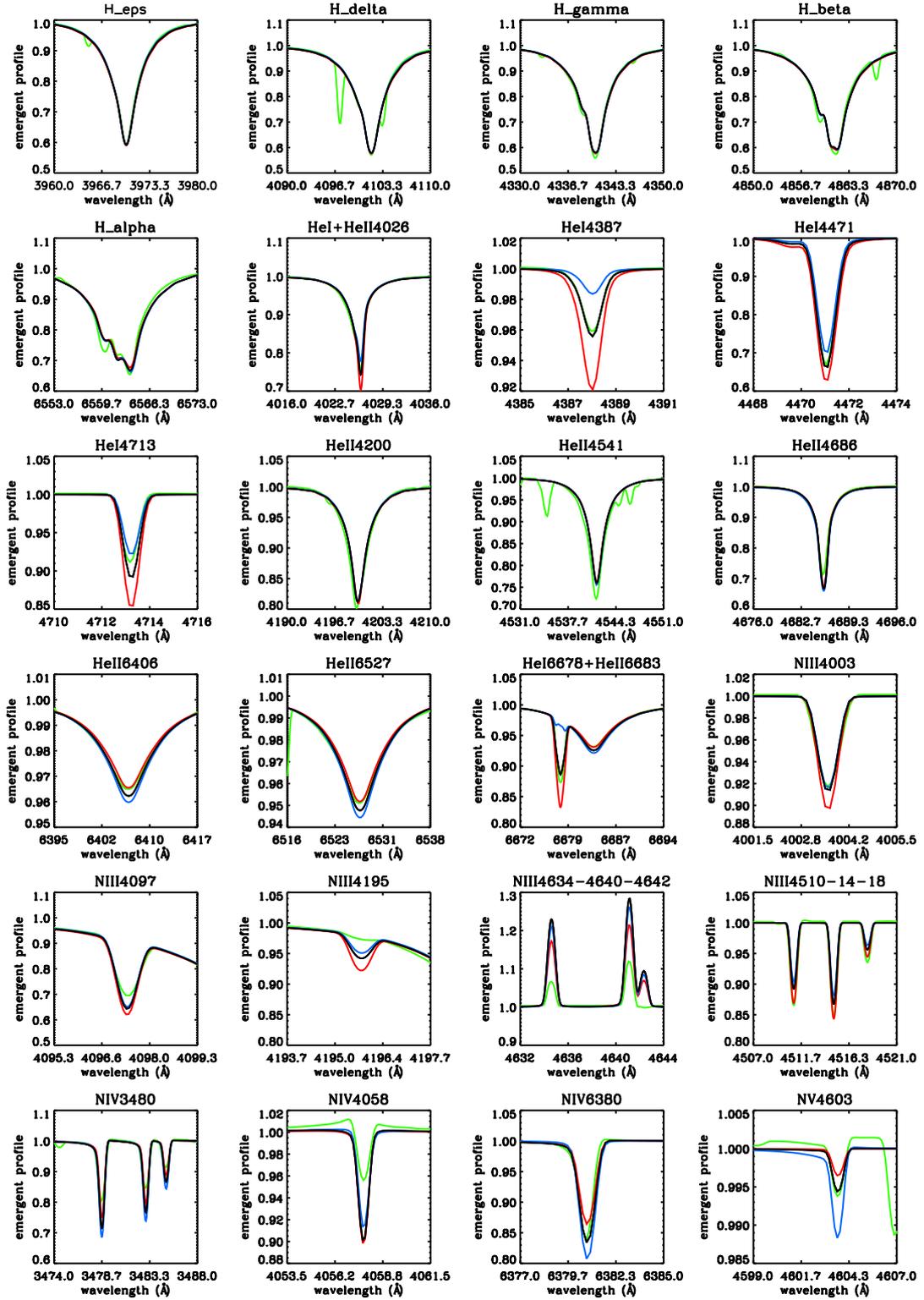}}
\end{center}
\caption{Model {\tt d4v} at [N] =8.78. Comparison of H/He/N spectra
from {\sc cmfgen} (green) and {\sc fastwind}, at the closest
grid-model (black: \Teff~=~41~kK, \logg\ = 4.0, $\log Q = -12.8$, [N]~=~8.78)
and at neighboring grid models with \Teff\ = 40~kK (red) and
\Teff\ = 42~kK (blue). To allow for an easier comparison, all profiles
have been convolved with \vsini\ = 30~\kms.} 
\label{d4v_8.78_hhen}
\end{figure}

\begin{figure*}
\begin{center}
  {\includegraphics[width=15cm]{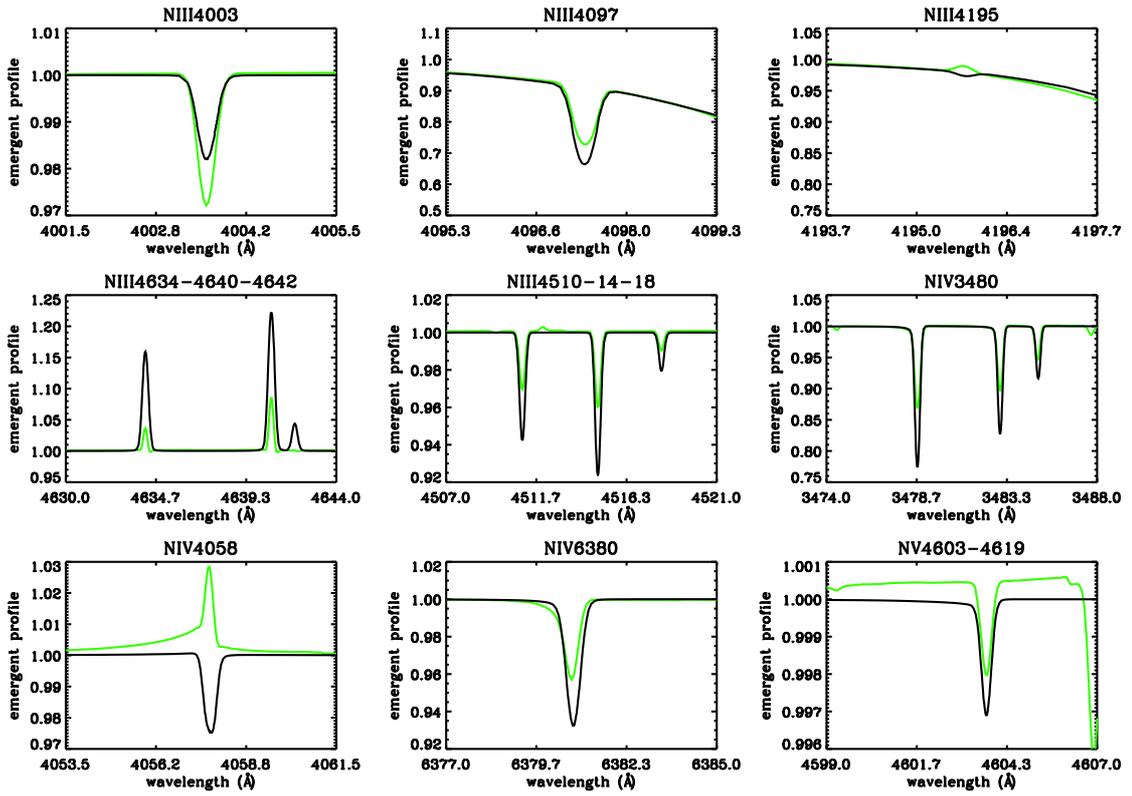}}
\end{center}
\caption{Model {\tt d4v} at [N] =7.78 (solar). Comparison of N spectra
from {\sc cmfgen} (green) and {\sc fastwind}, at the closest
grid-model. (black: \Teff~=~41~kK, \logg\ = 4.0, $\log Q = -12.8$, [N] =
7.78). The H/He spectra remain as in Fig.~\ref{d4v_8.78_hhen}. No
convolution applied.} 
\label{d4v_7.78_n}
\end{figure*}

\begin{figure*}
\begin{center}
  {\includegraphics[width=15cm]{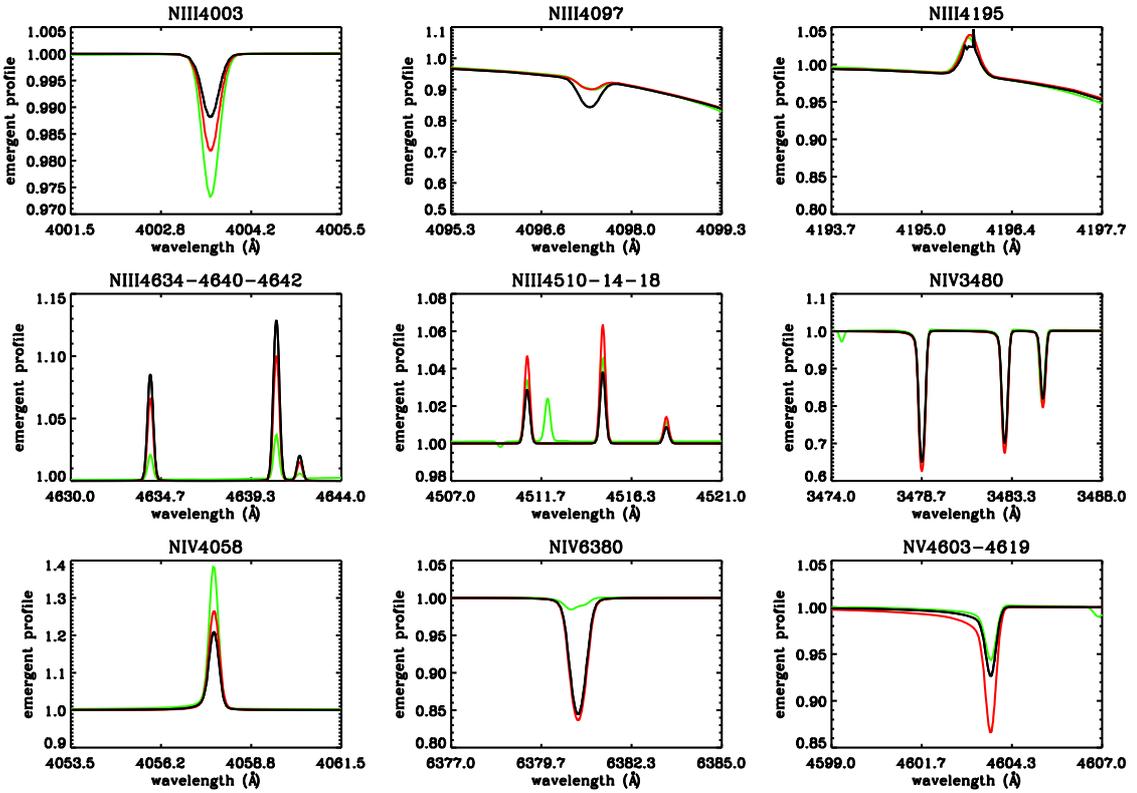}}
\end{center}
\caption{Model {\tt d2v} at [N] =7.78 (solar). Comparison of N spectra
from {\sc cmfgen} (green) and {\sc fastwind}, at the closest
grid-model (black: \Teff~=~46~kK, \logg\ = 4.0, $\log Q = -12.45$, [N] =
7.78), and at the neighboring grid model with \Teff\ = 47~kK (red). 
The H/He spectra remain as in Fig.~\ref{d2v_8.78_hhen}.} 
\label{d2v_7.78_n}
\end{figure*}

\begin{figure*}
\begin{center}
 
{\includegraphics[width=16cm]{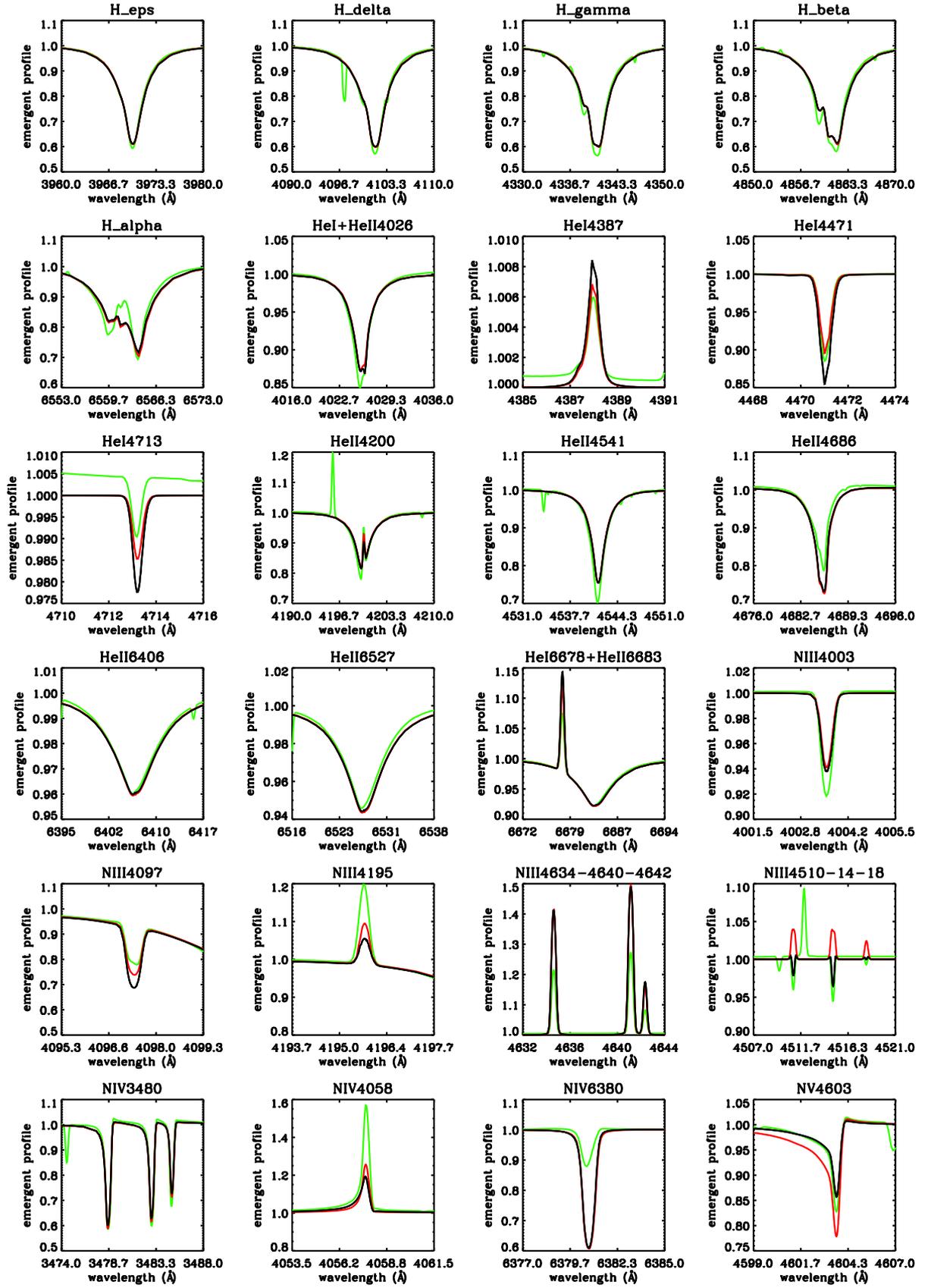}}
\end{center}
\caption{Model {\tt d2v} at [N] =8.78. Comparison of H/He/N spectra
from {\sc cmfgen} (green) and {\sc fastwind}, at the closest
grid-model (black: \Teff~=~46~kK, \logg\ = 4.0, $\log Q = -12.45$, [N] =
8.78) and at neighboring grid model with \Teff\ = 47~kK (red). No
convolution has been applied.} 
\label{d2v_8.78_hhen}
\end{figure*}

\begin{figure*}
\begin{center}
  {\includegraphics[width=16cm]{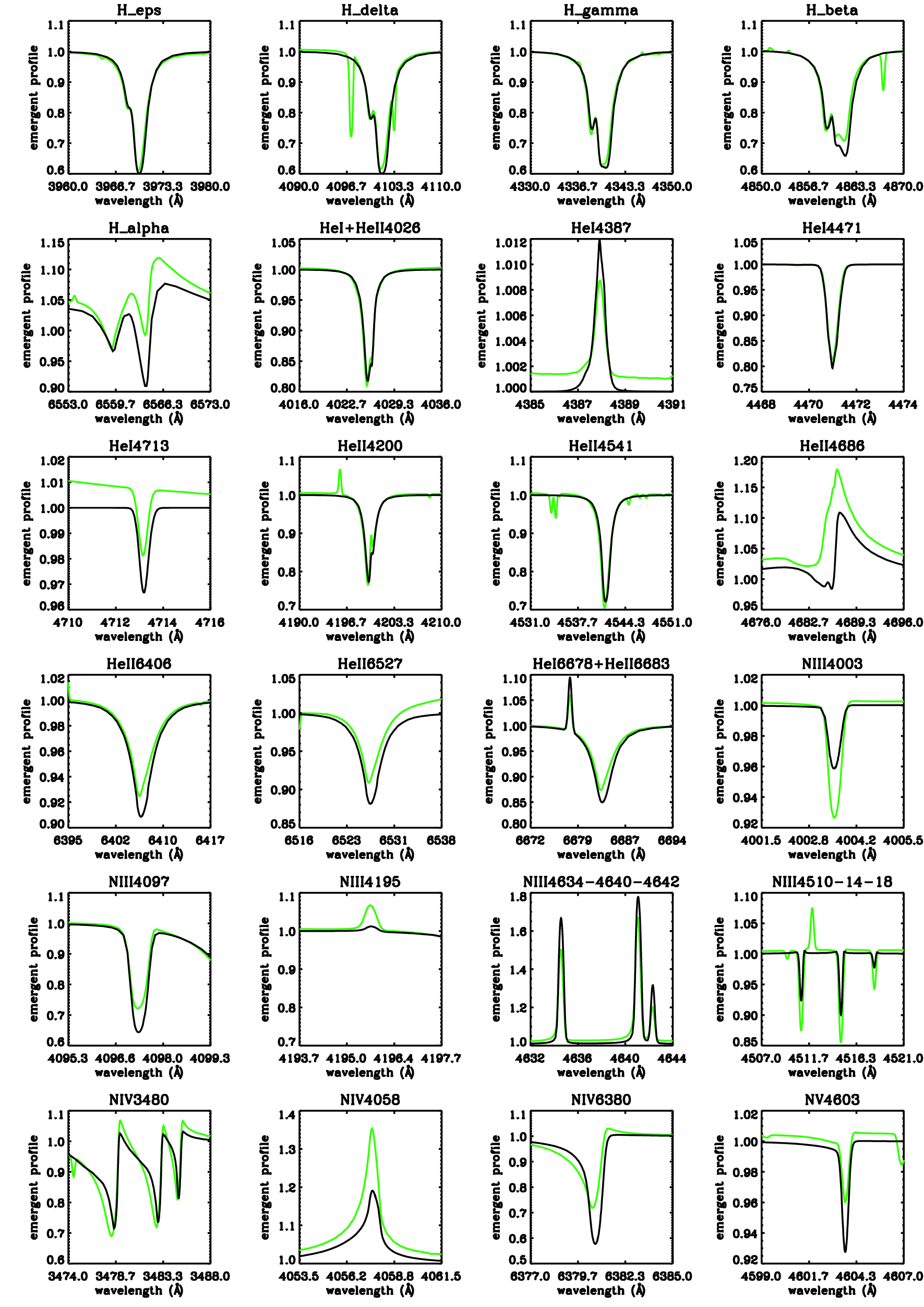}}
\end{center}
\caption{Model {\tt s4a} at [N] =8.78. Comparison of H/He/N spectra
from {\sc cmfgen} (green) and {\sc fastwind}, at the closest
grid-model (black: \Teff~=~39~kK, \logg\ = 3.5, $\log Q = -12.10$,
[N] = 8.78).} 
\label{s4a_8.78_hhen}
\end{figure*}

\begin{figure*}
\begin{center}
  {\includegraphics[width=15cm]{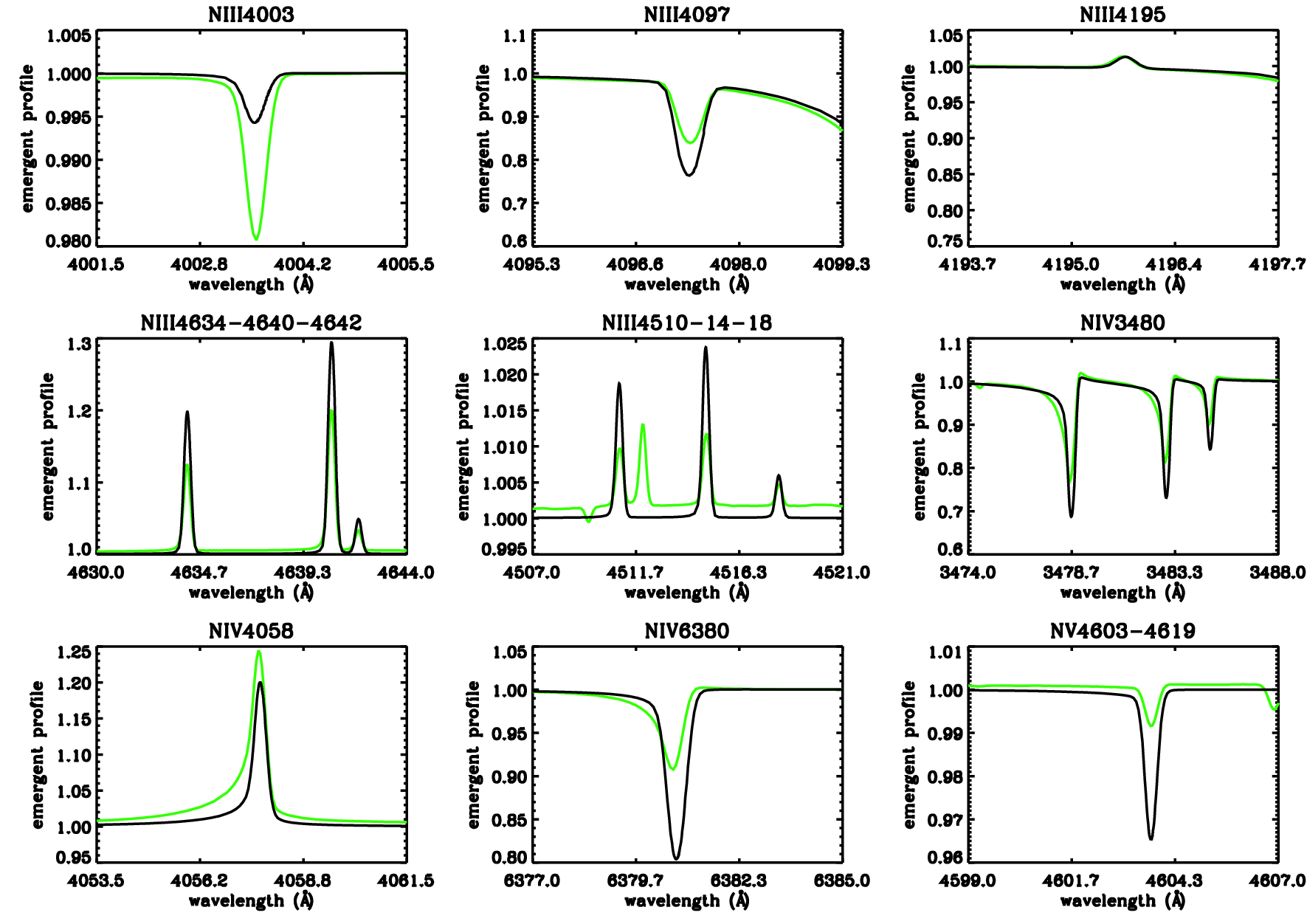}}
\end{center}
\caption{Model {\tt s4a} at [N] =7.78 (solar). Comparison of N spectra
from {\sc cmfgen} (green) and {\sc fastwind}, at the closest
grid-model (black: \Teff~=~39~kK, \logg\ = 3.5, $\log Q = -12.10$, [N] =
7.78). The H/He spectra remain as in Fig.~\ref{s4a_8.78_hhen}.} 
\label{s4a_7.78_n}
\end{figure*}

\begin{figure*}
\begin{center}
  {\includegraphics[width=15cm]{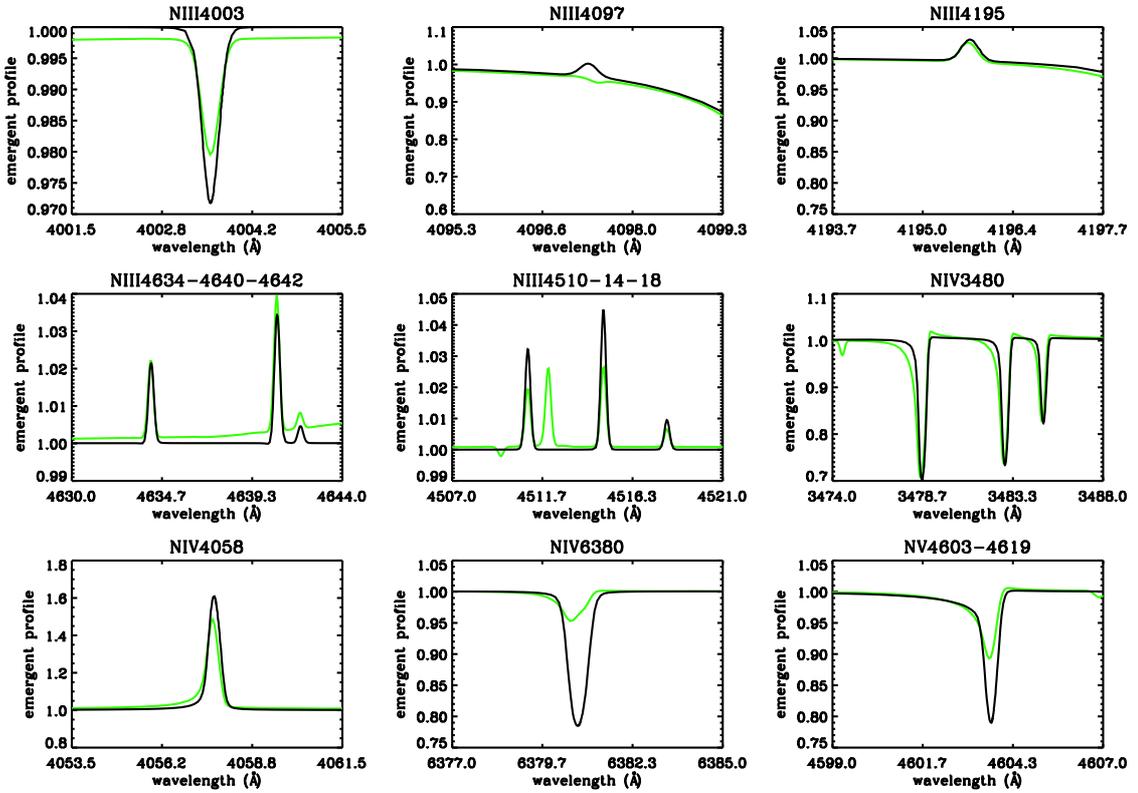}}
\end{center}
\caption{Model {\tt s2a} at [N] =7.78 (solar). Comparison of N spectra
from {\sc cmfgen} (green) and {\sc fastwind}, at the (almost) closest
grid-model (black: \Teff\ = 46~kK, \logg\ = 3.8, $\log Q = -12.10$, [N] =
7.78). The H/He spectra remain as in Fig.~\ref{s2a_8.78_hhen}.} 
\label{s2a_7.78_n}
\end{figure*}

\begin{figure*}
\begin{center}
  {\includegraphics[width=16cm]{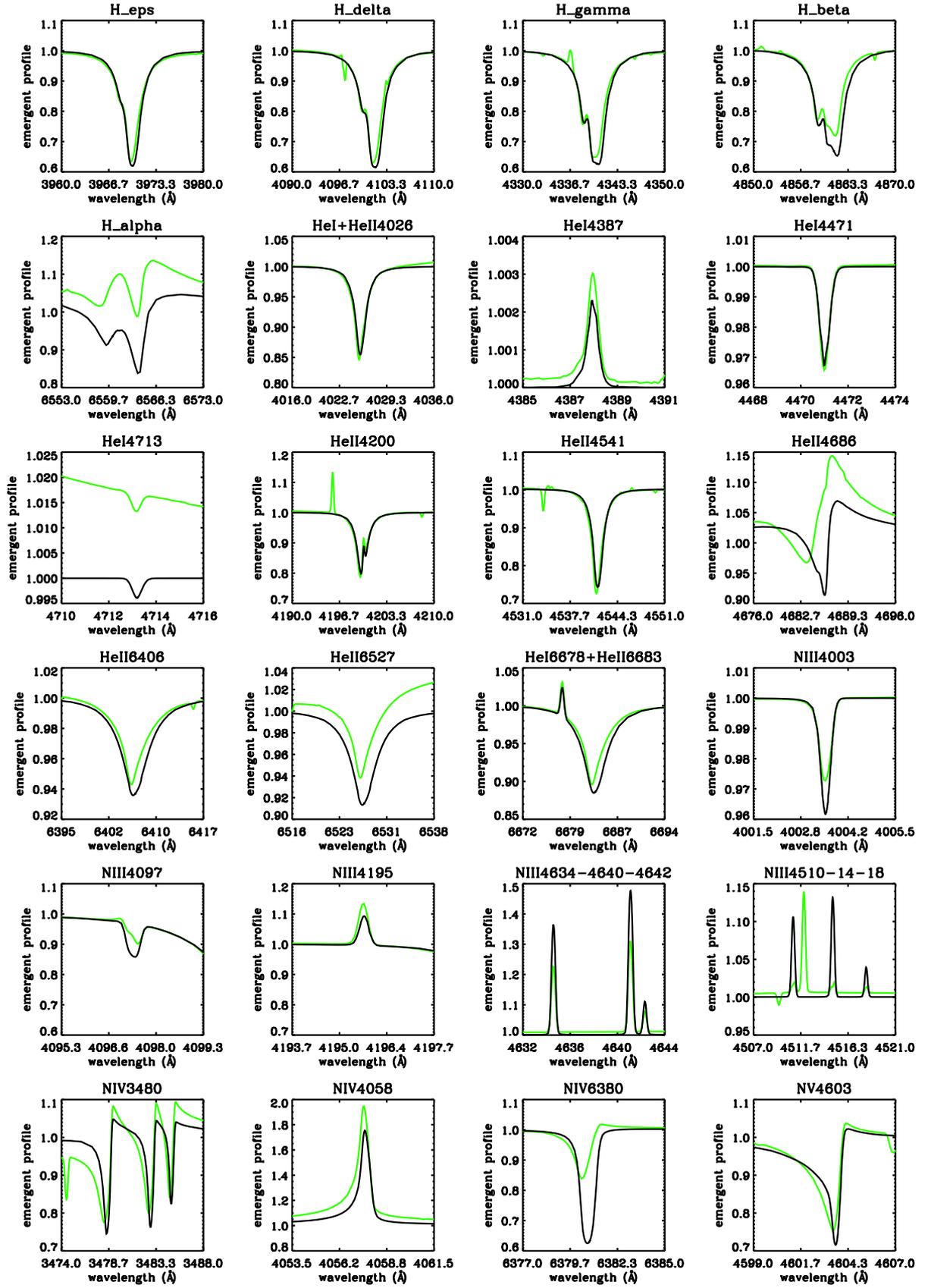}}
\end{center}
\caption{Model {\tt s2a} at [N] =8.78. Comparison of H/He/N spectra
from {\sc cmfgen} (green) and {\sc fastwind}, at the (almost) closest
grid-model (black: \Teff\ = 46~kK, \logg\ = 3.8, $\log Q = -12.10$,
[N] = 8.78).} 
\label{s2a_8.78_hhen}
\end{figure*}

Figures~\ref{d4v_8.78_hhen} to \ref{s2a_8.78_hhen} provide a detailed
comparison between H/He and \NIII/\NIV/\NV\ {\sc cmfgen} spectra for models {\tt d2v,
d4v, s2a}, and {\tt s4a} (see Table~\ref{grid-cmfgen}), and
corresponding {\sc fastwind} profiles from closest or almost
closest grid models, for a nitrogen abundance of [N]~=~8.78 and [N]~=~7.78. If
not explicitly stated else, no convolution has been applied
to the spectra. For details, see Sect.~\ref{comp_cmfgen}.

\clearpage 
\twocolumn
\section{Comments on the individual objects}
\label{comments} In the following, we give specific comments on the
individual objects, regarding peculiarities and problems found during
our analysis. We separate between galaxy membership, and sort by
luminosity class and spectral type. Line fits are displayed in
Figs.~\ref{R136-040} to~\ref{NGC346-355}, for important H/He/N lines:
\Ha, \Hb, \Hg, \Hd, \Hep, \HeI$\lambda\lambda$~4387, 4471, 4713,
\HeII(+\HeI)$\lambda$4026, \HeII$\lambda\lambda$4200, 4541, 4686,
6406, 6527, 6683, \NIII$\lambda\lambda$~4003, 4097, 4195, 4379, \trip,
and \qua, \NIV$\lambda\lambda$ 4058, 6380, and
\NV$\lambda\lambda4603/4619$. All spectra were corrected for radial
velocity shifts. 

If not explicitly stated, any comparison made in the following text
refers to the results from Mas05.

\subsection{LMC stars}
\begin{figure*}
\center
{\includegraphics[width=160mm]{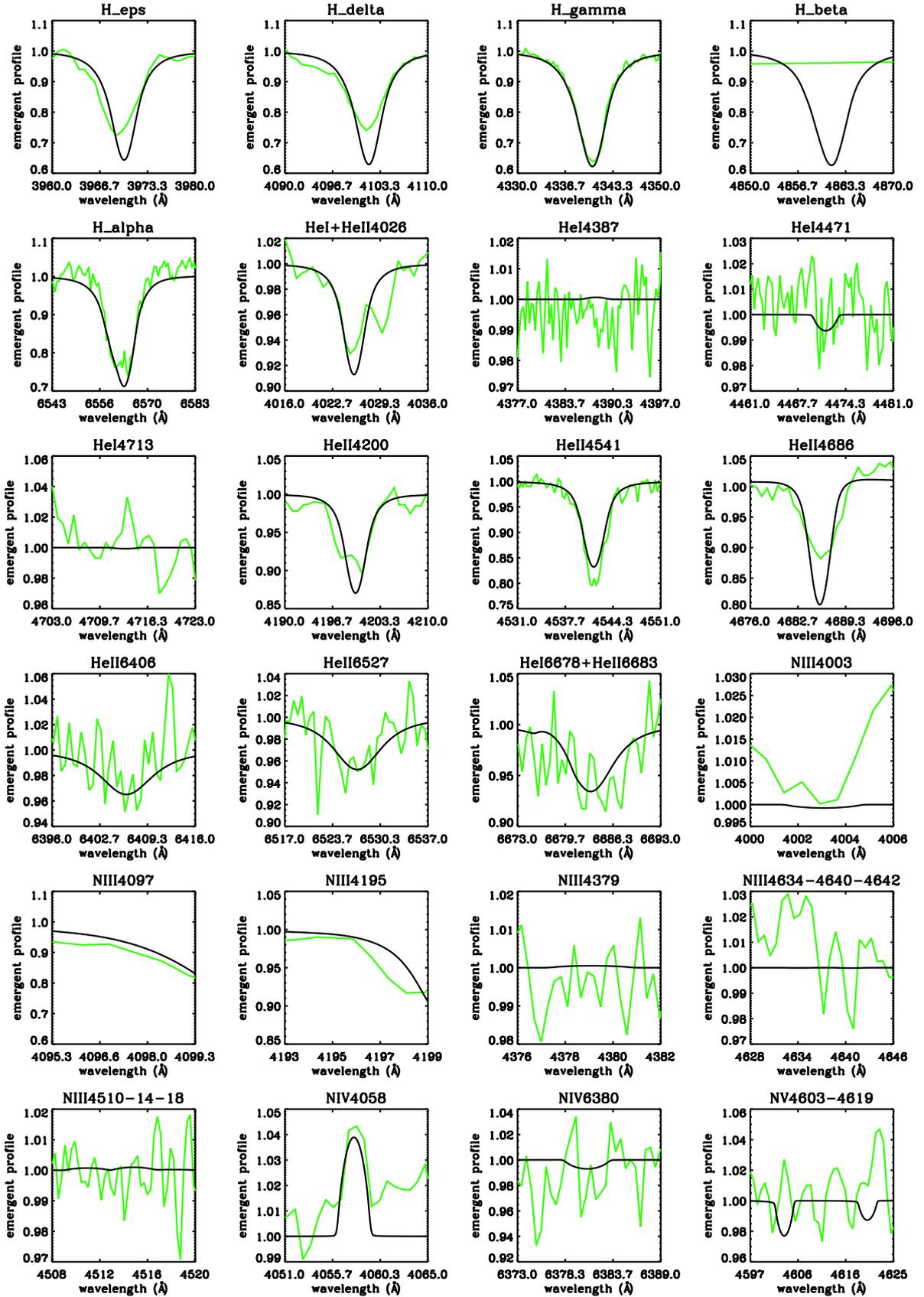}}
\caption{R136-040 - O2-3.5 V. Observed (green) and best fitting
optical H/He and N spectrum. For the R136 O-stars, observed spectra for \Ha,
\Hg, \HeI$\lambda\lambda$4387, 4471, 6678, \HeII$\lambda$4541, 6406,
6527, 6683, \NIII\qua, and \NIV\nivab\ taken from the STIS/CCD
dataset. Remaining, lower quality
spectra collected by FOS. \Hb\ was not observed for this star.}
\label{R136-040}
\end{figure*}

\paragraph{\bf R136-040 -- O2-3.5 V}(Fig.~\ref{R136-040}). This star
could not been classified by Mas05 using the scheme by
\citet{walborn02b}, because neither \NIII\trip\ nor \NIV\nivem\ were
visible in the spectra. As outlined in Sect.~\ref{obs}, for the R136
stars we have spectra from both STIS and FOS available. Taking
advantage of the better quality of the STIS data and using \Ha, \Hg,
and \HeII$\lambda$4541, we derived a similar lower limit (no \HeI\ and
no nitrogen lines)! on \Teff\ as Mas04, but a substantially larger (by
0.2 dex) gravity, which agrees better with its dwarf designation. Our
analysis also resulted in a low helium content, \YHe\ = 0.08 (Mas04:
\YHe\ = 0.1). Note the discrepancy in the cores of \Hd, \Hep, and
\HeII$\lambda\lambda$4026, 4200 when comparing with the FOS data. Such
a discrepancy was also found for the remaining H/\HeII\ lines in the
FOS data, when used instead of the STIS spectra.

The missing nitrogen lines imply an upper limit for the nitrogen
content corresponding to the LMC baseline abundance [N]~=~6.90, which
also agrees quite well with the low He content.

\begin{figure*}
\center
{\includegraphics[width=160mm]{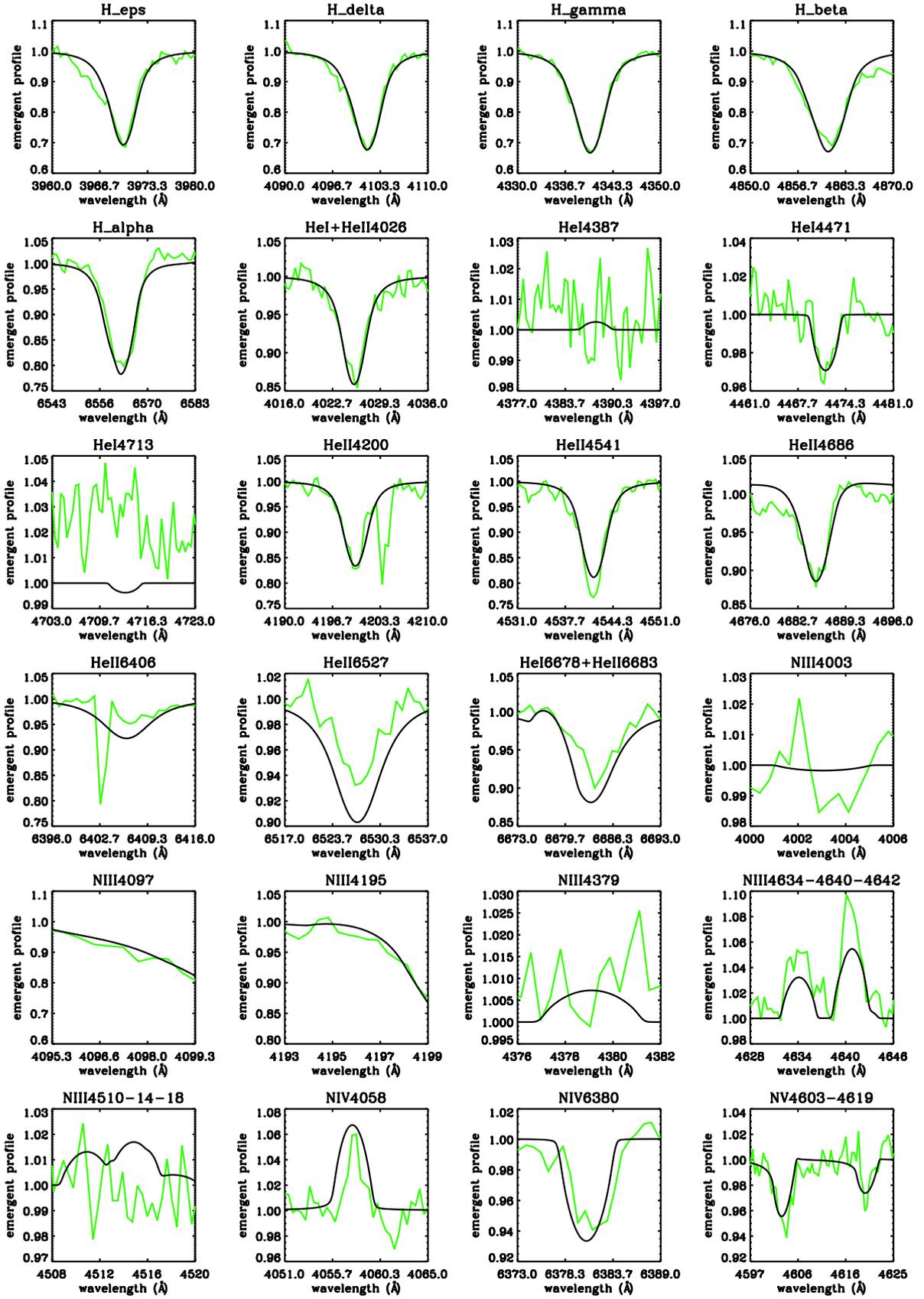}}
\caption{LH 81:W28-23 - O3.5 V((f+)).} 
\label{LH81W23}
\end{figure*}

\paragraph{\bf LH 81:W28-23 -- O3.5 V((f+))}(Fig.~\ref{LH81W23}).
The modeling of this star was straightforward, and we obtained similar
results as Mas05. However, we considered a larger $\beta$~=~1.0 to
better reproduce the marginal P-Cygni profile at \HeII$\lambda$4686,
which might indicate a luminosity class III object (see \citealt{walborn02b}).
To preserve the fit of \Ha, we needed to reduce \mdot. All nitrogen
lines are consistent with the temperature derived from the helium
ionization equilibrium. The quite large nitrogen abundance
([N]~=~8.40) agrees well with the helium abundance \YHe~=~0.25, indicating
an evolved nature of this object.\\

\begin{figure*}
\center
{\includegraphics[width=160mm]{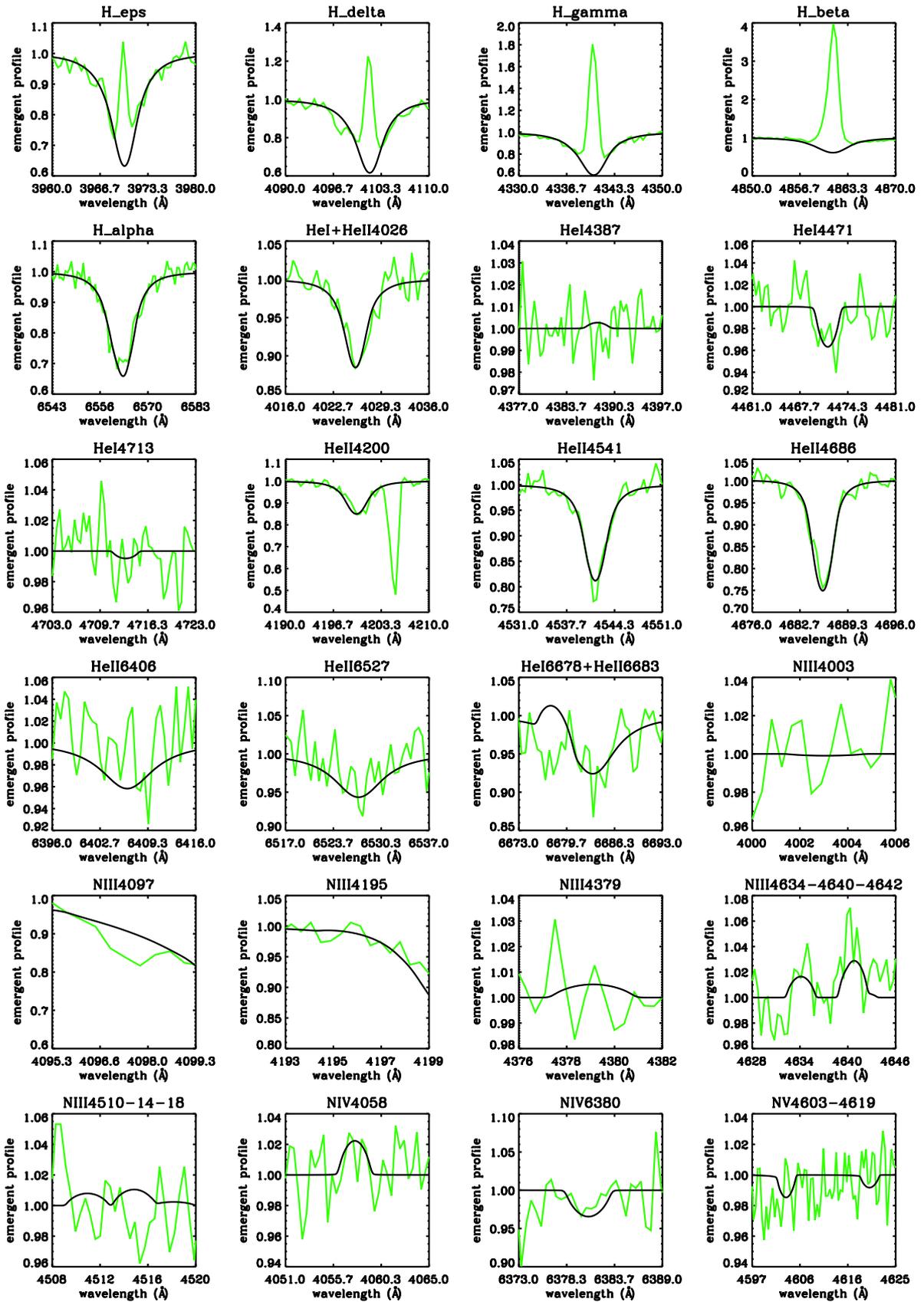}}
\caption{LH 101:W3-24 - O3.5 V((f+)).} 
\label{LH101W24}
\end{figure*}

\paragraph{\bf LH 101:W3-24 -- O3.5 V((f+))}(Fig.~\ref{LH101W24}).
We derived a somewhat cooler \Teff\ (by 1~kK) together with a lower
helium content, \YHe~=~0.10, which was consistent for all helium
lines. Nitrogen lines are barely visible, because the spectrum of this
star displays more noise than the bulk of our sample stars, caused by
the use of a narrow extraction aperture to reduce effects from nebular
emission for the ground-based observations 
(cf. Sect~\ref{obs}). Due to the noisy spectrum, we were only able to
infer an upper limit for the nitrogen abundance, [N] $\le$ 7.78. 

\begin{figure*}
\center
{\includegraphics[width=160mm]{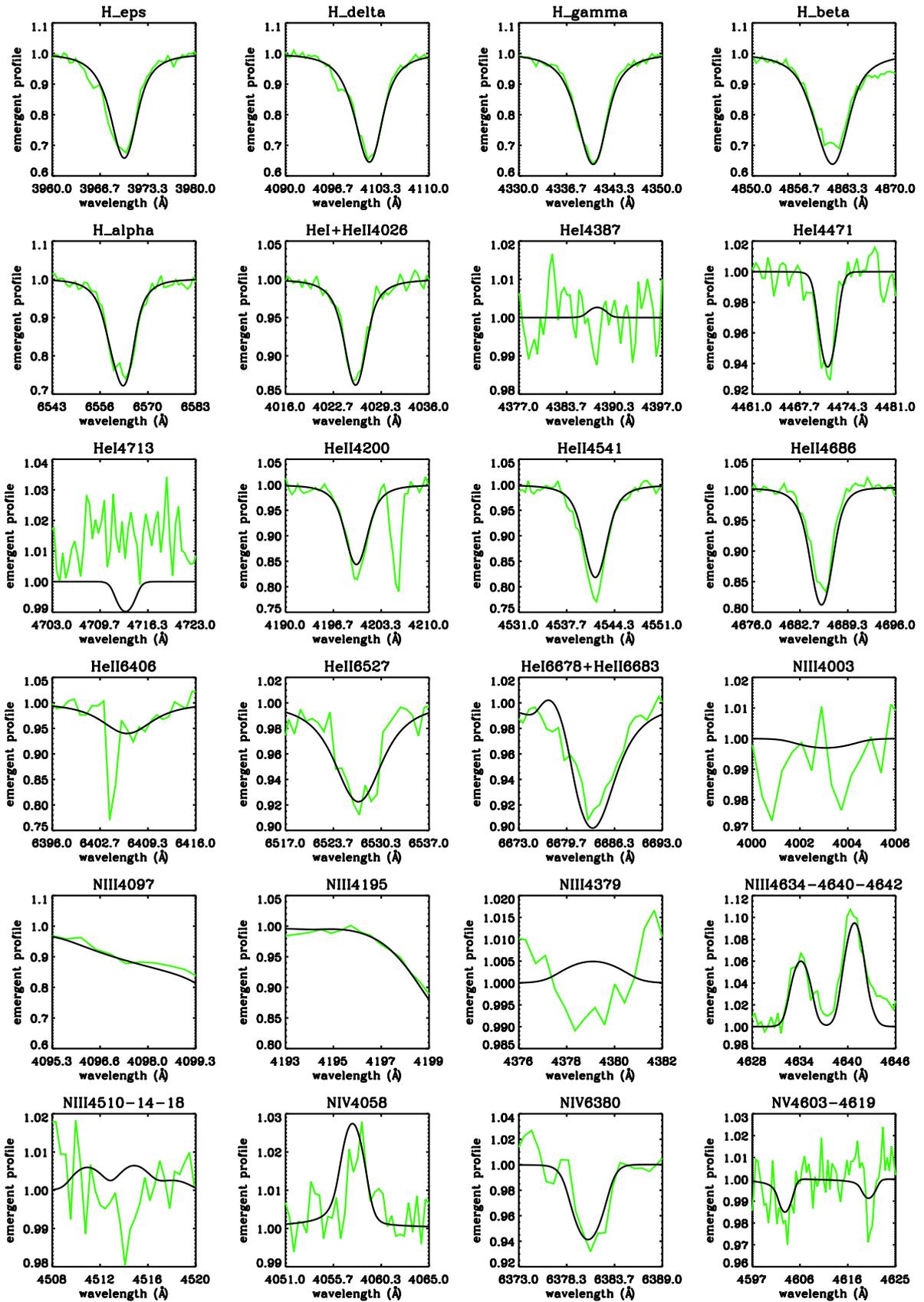}}
\caption{LH 81:W28-5 - O4 V((f+)).} 
\label{LH81W5}
\end{figure*}

\paragraph{\bf LH 81:W28-5 -- O4 V((f+))}(Fig.~\ref{LH81W5}). This is
one of the standards used by~\citet{walborn02b} for defining the O4
V((f+)) class. A consistent analysis of the helium and nitrogen
ionization equilibrium yielded a cooler temperature, \Teff~=~44~kK,
which required a helium abundance of \YHe~=~0.15 to reproduce 
HeI$\lambda$4471. An excellent fit to most nitrogen lines from all
three ionization stages was achieved for this star, indicating 
a significant enrichment, [N]~=~8.38.

\begin{figure*}
\center
{\includegraphics[width=160mm]{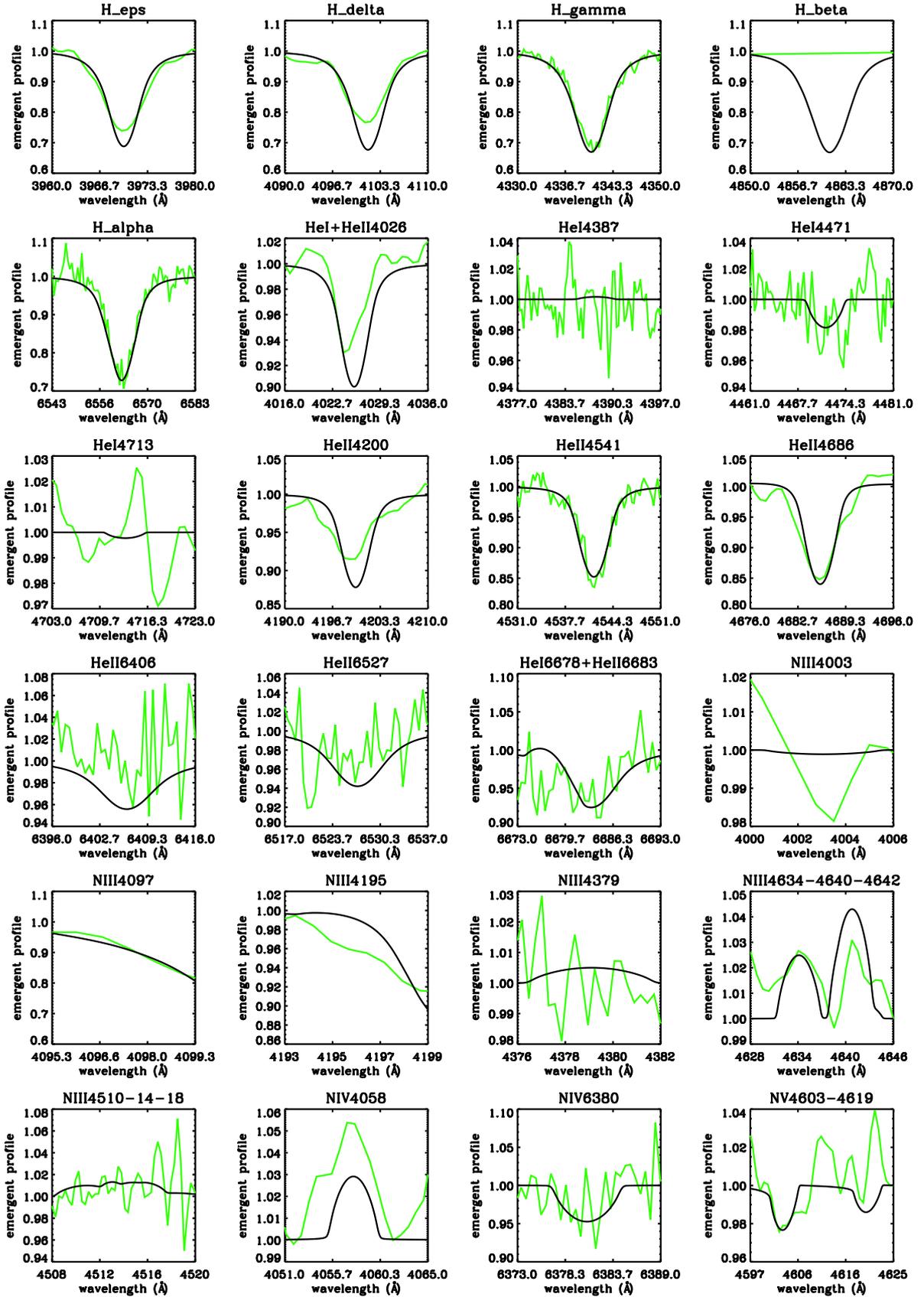}}
\caption{R136-018 - O3 III(f$^*$). Observations as for R136-040
(Fig.~\ref{R136-040}).} 
\label{R136018}
\end{figure*}

\paragraph{\bf R136-018 -- O3 III(f$^*$)}(Fig.~\ref{R136018}). Also
for this O3 giant we have used data from both STIS and FOS. A
consistent analysis of the H/He lines from STIS and nitrogen lines
(mostly from FOS) suggested a hotter \Teff, by 2~kK, as well as a
higher surface gravity, by $\sim$0.1 dex. Again, we found
discrepancies in the cores of the H/\HeII\ lines from the FOS spectra,
except for \HeII$\lambda$4686. An acceptable fit for \NIII/\NIV/\NV\
using [N]~=~8.18 was possible, where only \NIV\nivem\, was slightly
underpredicted. Even though the nitrogen lines contained in the STIS
dataset, \NIII\qua\ and \NIV\nivab, are diluted in the continuum, they
support our analysis.

\begin{figure*}
\center
{\includegraphics[width=160mm]{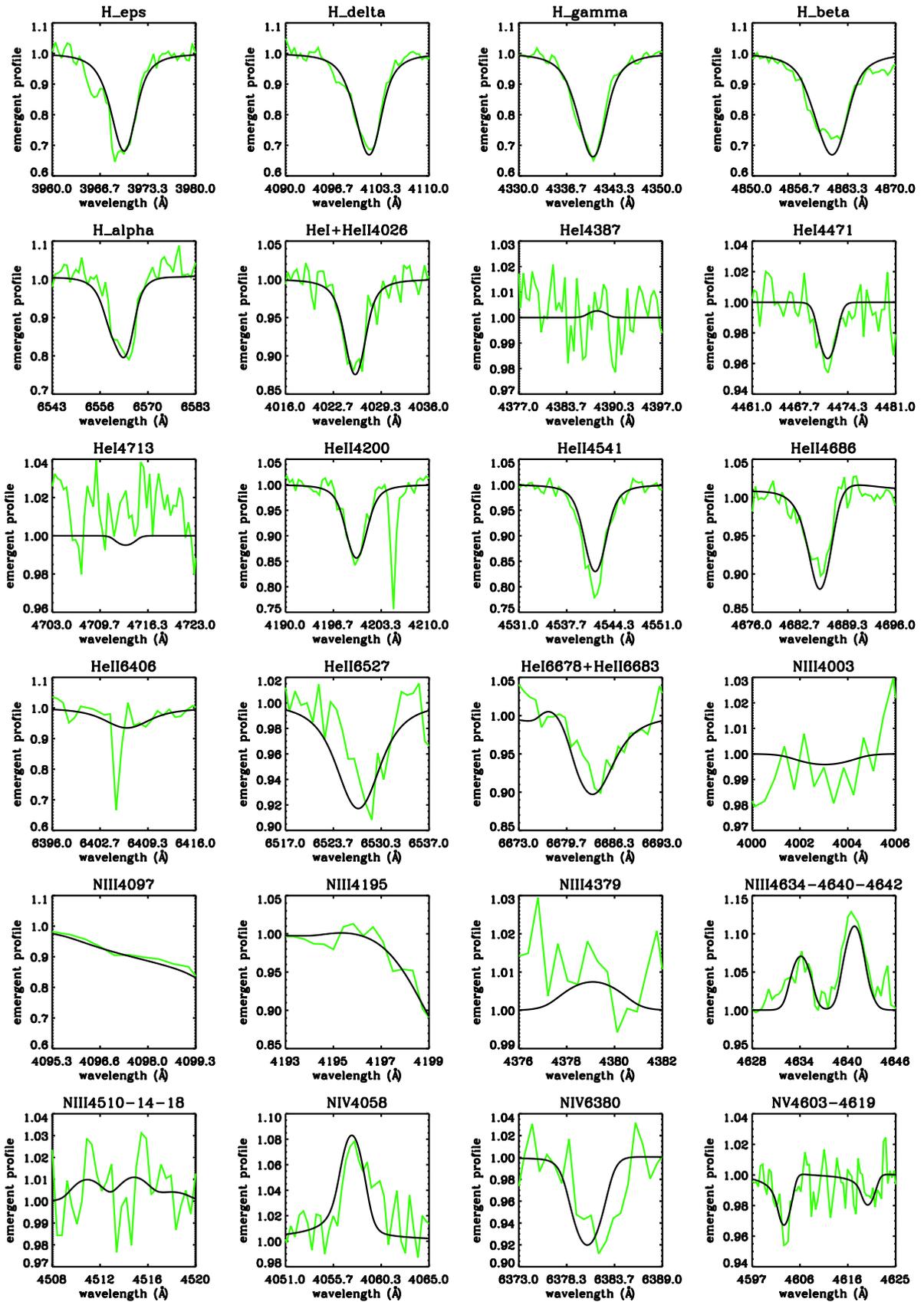}}
\caption{LH 90:ST 2-22 - O3.5 III(f+).} 
\label{LH90ST2-22}
\end{figure*}

\paragraph{\bf LH 90:ST 2-22 -- O3.5 III(f+)}(Fig.~\ref{LH90ST2-22}).
An unproblematic analysis provided the same results as obtained by
Mas05, except that we opted for a lower helium
abundance, \YHe~=~0.15. Again, a remarkable fit to the nitrogen lines
was possible, at [N]~=~8.58, indicating an extreme enrichment.

\begin{figure*}
\center
{\includegraphics[width=160mm]{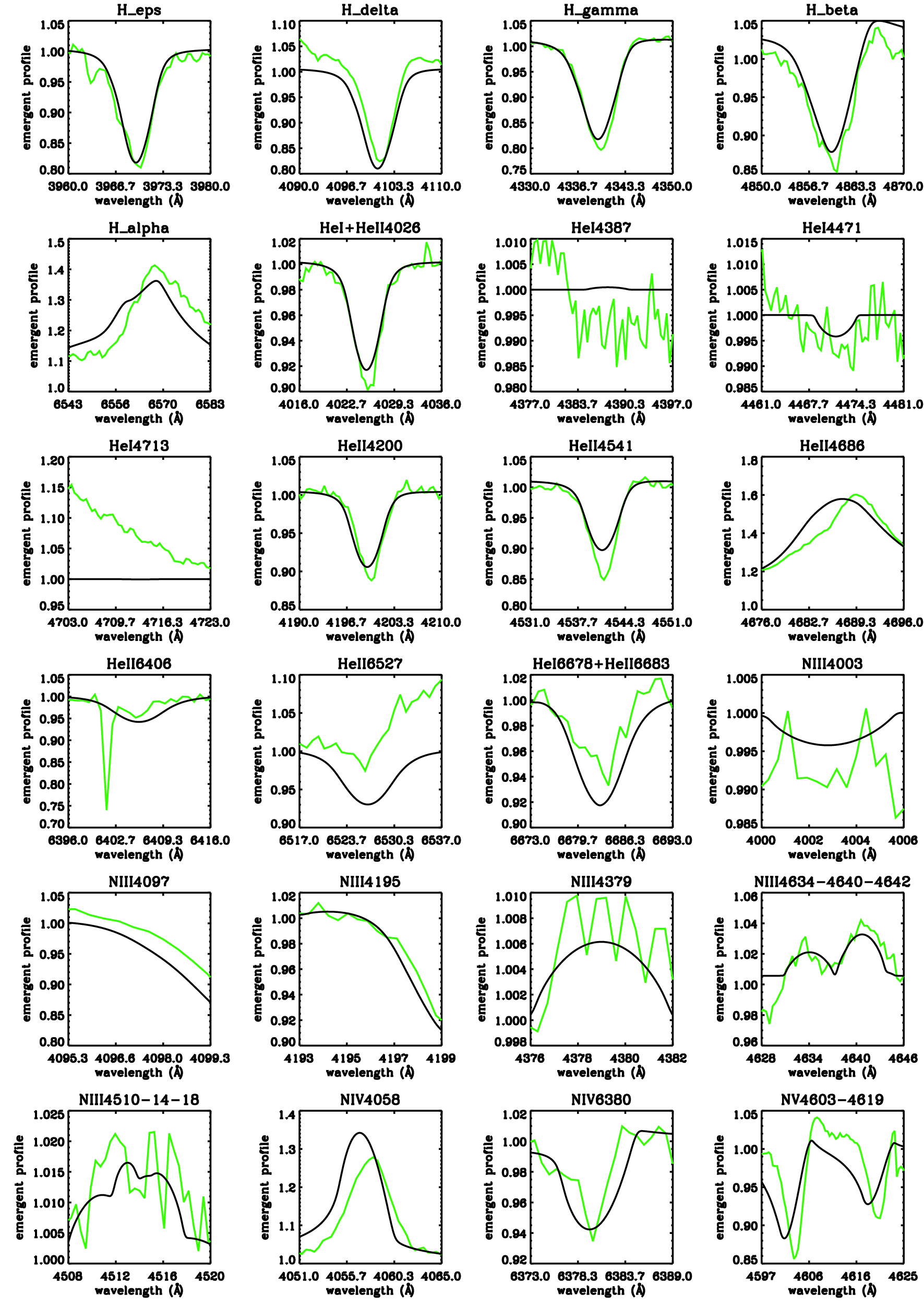}}
\caption{Sk--67$^{\circ}$ 22 - O2 If$^*$/WN5.} 
\label{Sk67-22}
\end{figure*}

\paragraph{\bf Sk--67$^{\circ}$ 22 -- O2 If$^*$/WN5} (Fig.~\ref{Sk67-22}). 
This star was re-classi\-fied\footnote{Mas05: O2 If$^*$} as O2
If$^*$/WN 5 by~\citet{crowther11} using their updated classification
scheme, because of the \Hb\ P-Cygni profile.  Using lines from
\NIII/\NIV/\NV, we inferred \Teff~=~46~kK which is hotter than the
lower limit (from very weak \HeI$\lambda$4471) quoted by Mas05. Our
fit seems to slightly overpredict the emission in \NIV\nivem\ and to
underpredict the \NV\ doublet. An extreme nitrogen abundance,
[N]~=~8.78, was required, the largest one found in our sample. Such an
abundance would be certainly too large when comparing even with
strongly nitrogen-enhanced O-stars, and also with predictions from
evolutionary calculations tailored for the LMC (\citealt{brott11b} and
paper~II), thus supporting a rather evolved nature of this object and
its `slash-star' designation. This star was also analyzed
by~\citet{doran11} using \NIV/\NV\ lines (without discussion of \HeI\
and \NIII), only providing a \Teff~=~49.3~kK for this
object. Such hotter temperature would improve our fits for
\NIV\nivem\ and the \NV\ doublet, but is inconsistent with the observed
strength of \NIII.

\begin{figure*}
\center
{\includegraphics[width=160mm]{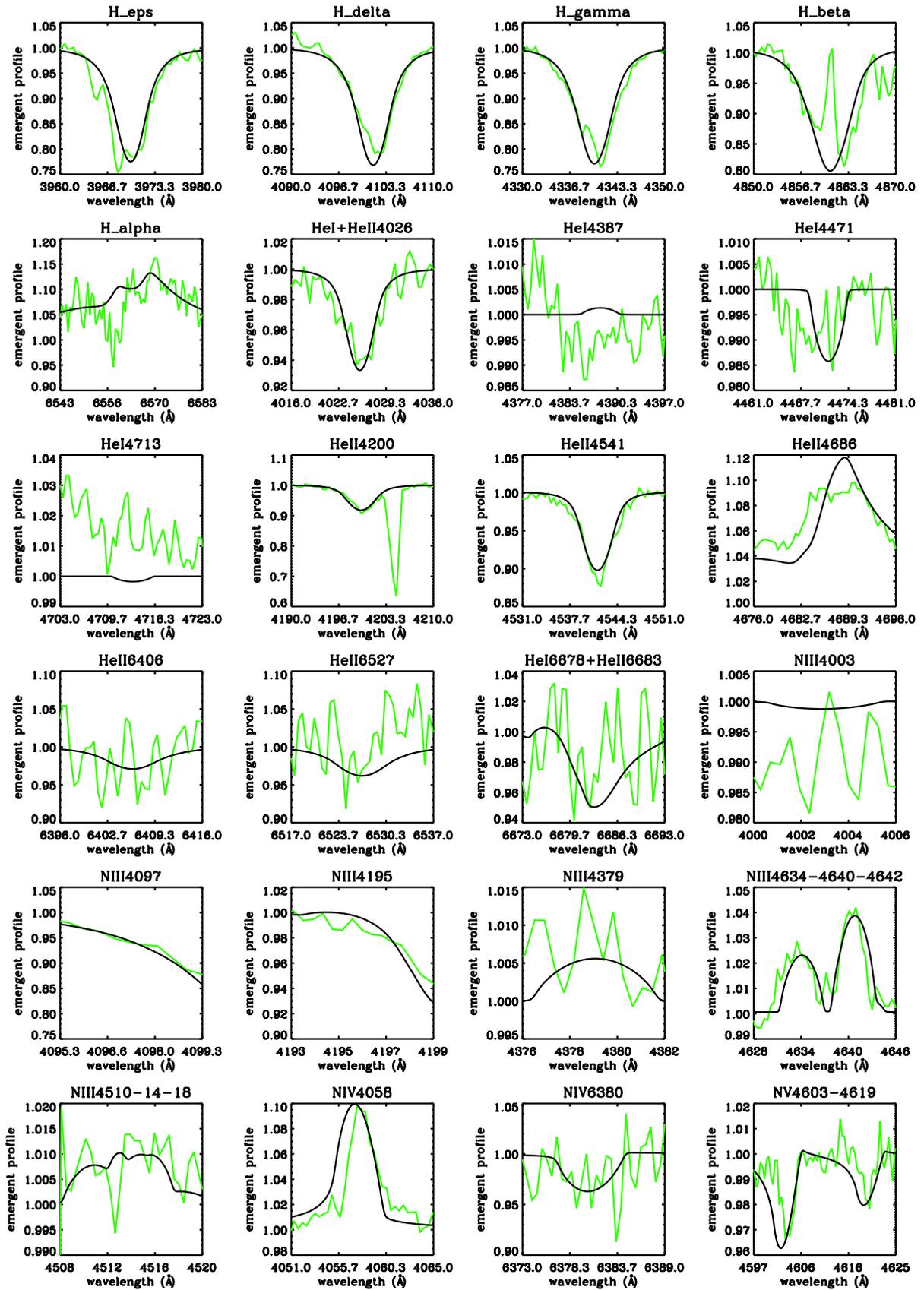}}
\caption{LH 101:W3-19 - O2 If$^*$.} 
\label{LH101W19}
\end{figure*}

\paragraph{\bf LH 101:W3-19 -- O2 If$^*$}(Fig.~\ref{LH101W19}). For this
supergiant, a consistent He/N analysis allowed us to derive
\Teff~=~44~kK, hotter than the lower limit (marginal HeI$\lambda$4471)
assigned by Mas05. Using \NIII/\NIV/\NV\ in parallel, we achieved an
almost excellent fit for the nitrogen lines at [N]~=~8.18.

\begin{figure*}
\center
{\includegraphics[width=160mm]{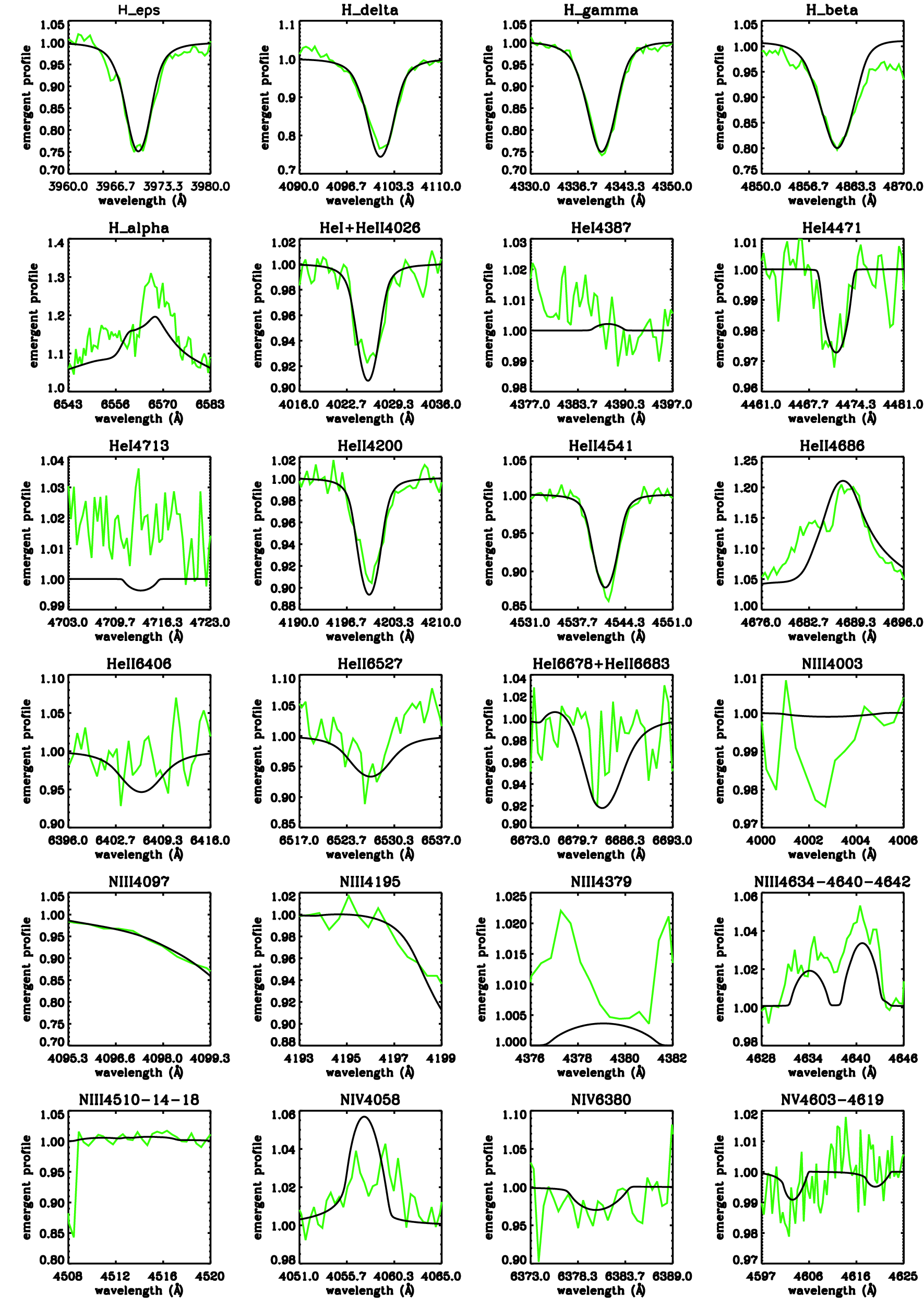}}
\caption{Sk--65$^{\circ}$ 47 - O4 If.} 
\label{Sk65-47}
\end{figure*}

\paragraph{\bf Sk--65$^{\circ}$ 47 -- O4 If}(Fig.~\ref{Sk65-47}). The
parameter set derived for this star using H/He/N lines is quite
similar to the results from Mas05, with somewhat larger
\YHe~=~0.12, A potential discrepancy provides \NIV\nivem, where we
might slightly overpredict the observed emission. 

\subsection{SMC stars}

\begin{figure*}
\center
{\includegraphics[width=160mm]{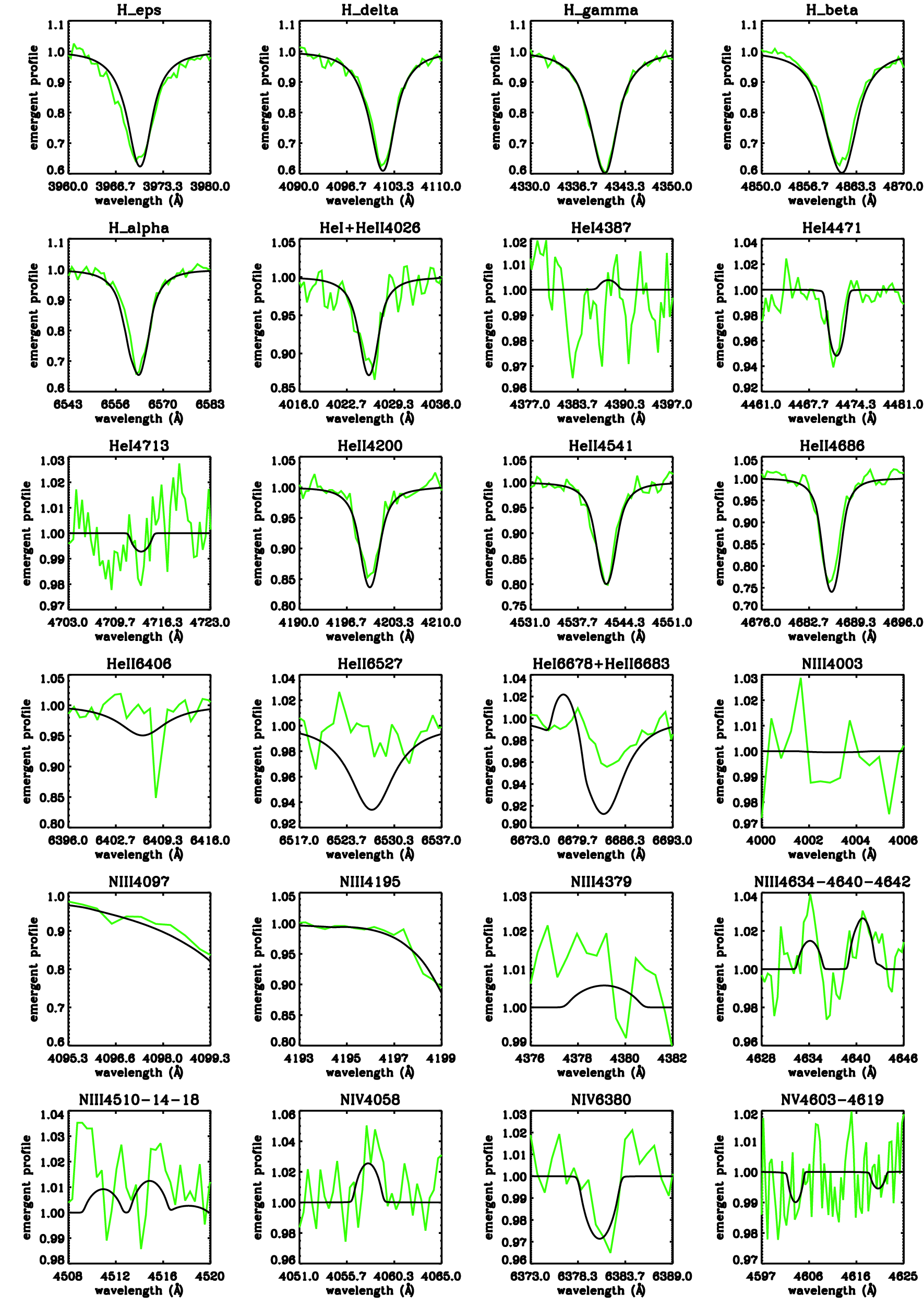}}
\caption{AV 435 - O3 V((f$^*$)).} 
\label{AV435}
\end{figure*}

\paragraph{\bf AV 435 -- O3 V((f$^*$))}(Fig.~\ref{AV435}). The only
discrepancies found during our analysis correspond to an
overprediction of \HeII$\lambda\lambda$6406, 6527, 6683. Both the
\HeI/\HeII\ and the \NIII/\NIV\ ionization equilibrium suggest a
hotter temperature than quoted by Mas05, \Teff~=~46~kK. This
temperature seems to be somewhat cool for its spectral type O3 V
assigned by Mas05 because of \NIV$\lambda$4058 $\ga$ \NIII\niiir, but
quite consistent with our predictions for the derived wind-strength
and nitrogen content, [N]~=~7.58 (cf. Figs.~\ref{iso_ratio_wind} and 
\ref{iso_ratio_abun}).

\begin{figure*}
\center
{\includegraphics[width=160mm]{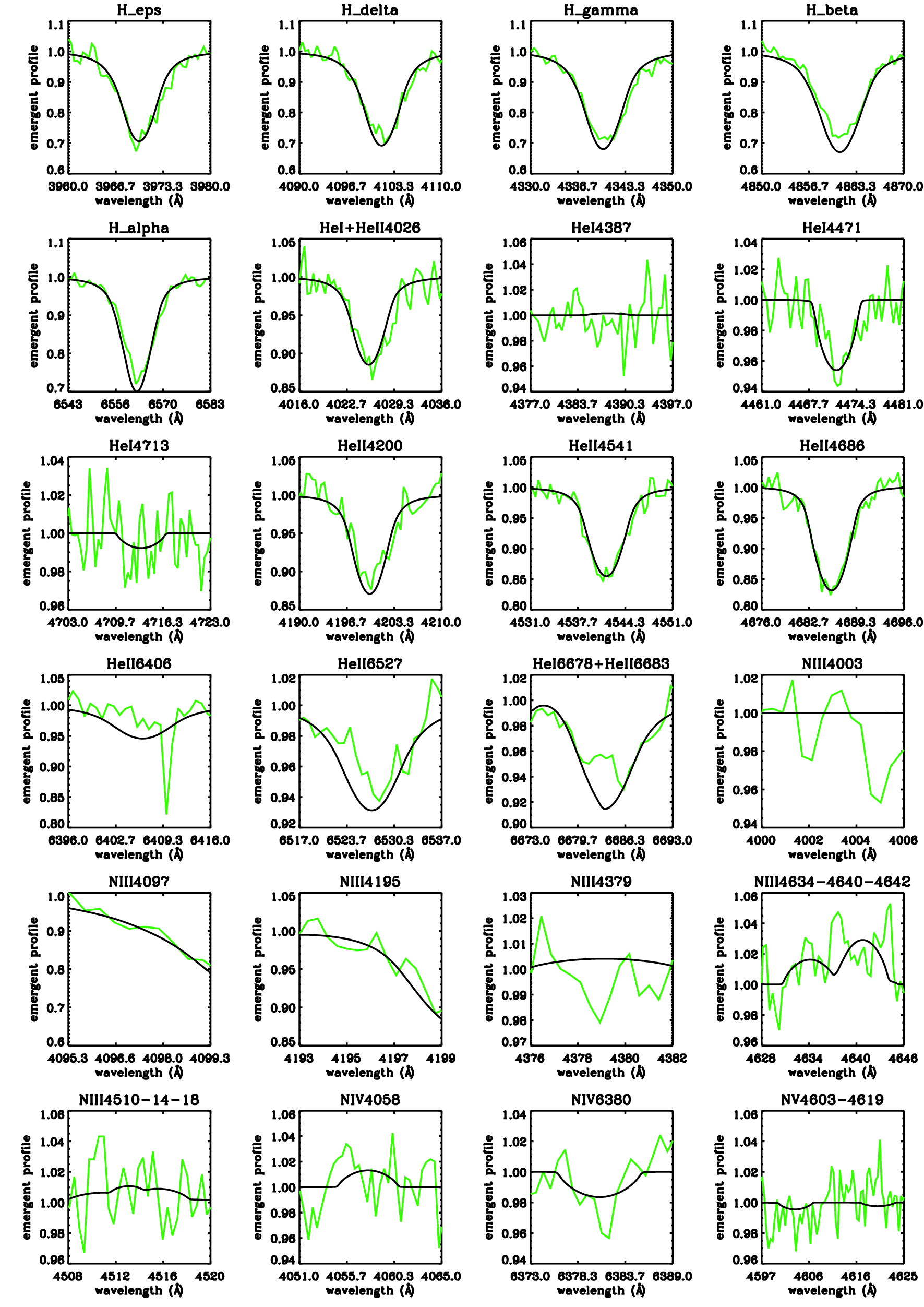}}
\caption{AV 177 - O4 V((f)).} 
\label{AV177}
\end{figure*}

\paragraph{\bf AV 177 -- O4 V((f))}(Fig.~\ref{AV177}). The H/He
analysis of this star produced similar parameters as found by Mas05. Owing to
a high rotation, \vsini~=~220~\kms, nitrogen lines are barely visible 
in the spectrum. Weak traces of emission at \NIII\trip\ and \NIII\qua\ 
together with weak absorption at \NIV\nivab\ are fitted consistently 
at [N]~=~7.78.

\begin{figure*}
\center
{\includegraphics[width=160mm]{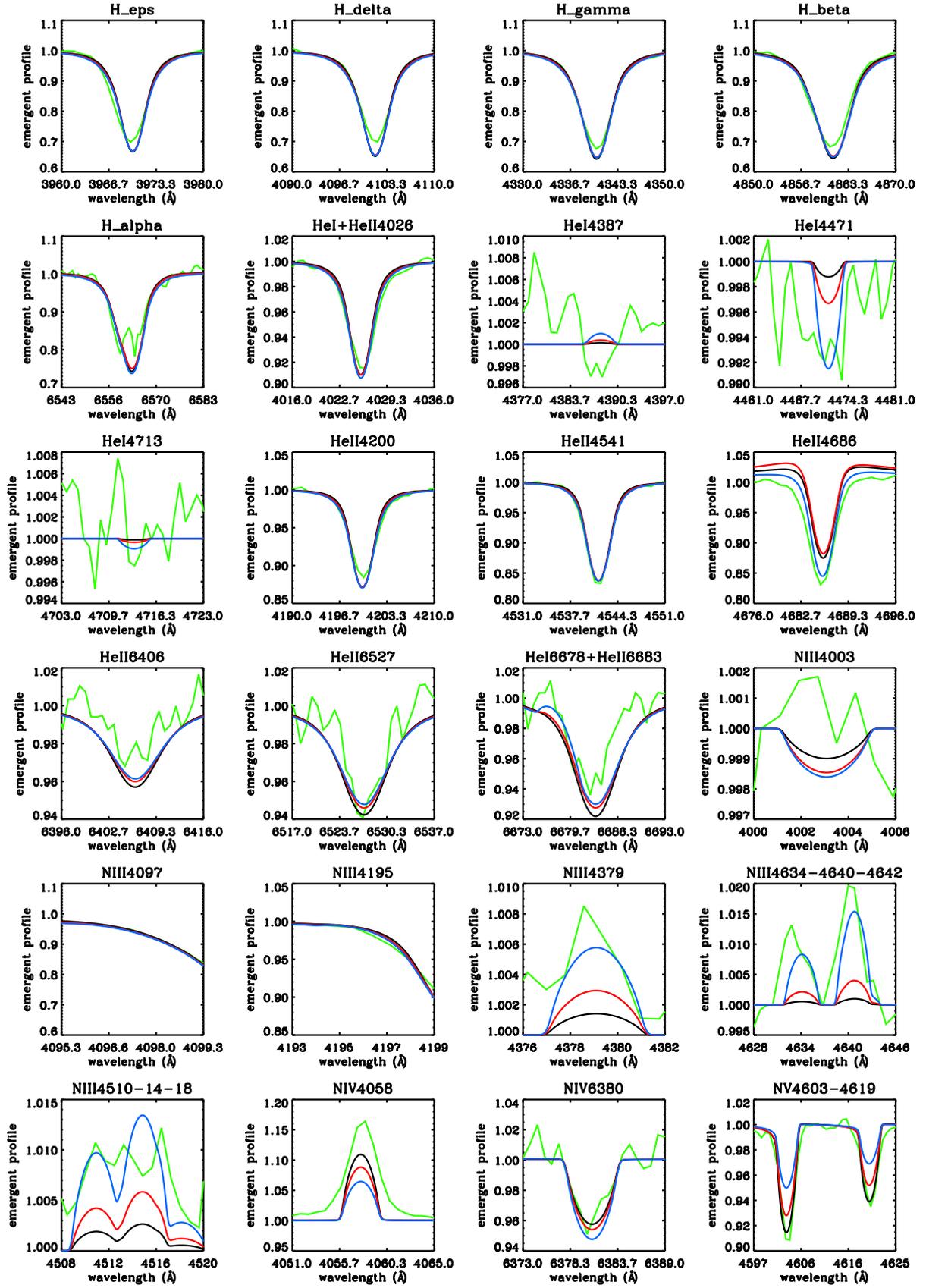}}
\caption{NGC 346-355 - ON2 III(f$^*$). Black: hotter solution (\Teff\
= 55 kK), supported by \NIV/\NV\ lines. Blue: cooler solution (\Teff\ = 51 kK),
mostly supported by \NIII\ (together with \NIV$\lambda$6380). Red: 'average'
solution (\Teff\ = 53 kK) used in Sect.~\ref{discussion}.}
\label{NGC346-355}
\end{figure*}

\paragraph{\bf NGC 346-355 -- ON2 III(f$^*$)}(Fig.~\ref{NGC346-355}).
This star was considered as a standard for the O2 III(f$^*$) category
by~\citet{walborn02b}, and later on updated to ON2 III(f$^*$)
by~\citet{walborn04}. As for N11-031 (same type!) analyzed in
paper~II, we found problems to fit all \NIII/\NIV/\NV\ lines in
parallel, but to a lesser extent. The basic difference is related to
\HeI$\lambda$4471, which is not as clearly visible as for N11-031.
During our analysis, we considered two possible parameter sets: a
cooler solution with \Teff~=~51~kK (red) and a hotter one with 
\Teff~=~55~kK (black), using either the \NIII/\NIV\ or the \NIV/\NV\
ionization equilibrium. By inspection of \HeI$\lambda$4471, we note
that both temperatures might be consistent with the very weak observed
feature. For a similar nitrogen abundance, [N]~=~7.98, we were able to
fit either \NIII\trip, \NIII\qua\, and \NIV\nivab\ for the cooler
solution, or \NIV\nivab\ and \NV$\lambda\lambda$4603-4619 for the
hotter one. 

Mas09, restricted to \HeI$\lambda$4471 as a primary temperature
indicator, derived \Teff~=~49.5~kK and \logg~=~3.9, which would agree
with our cool solution, but is insufficient for the \NIV/\NV\ lines.
The hotter solution is in better agreement with results from
\citet{bouret03} and \citet{walborn04}, who found \Teff~=~52.5~kK and
\logg~=~4.0 fitting the \NIV/\NV\ lines by means of {\sc cmfgen}. In
particular, we achieved a similar fit quality as \citet{bouret03}, for
a similar nitrogen content. Bouret et al. stated that at \Teff\ $\sim$
55~kK (identical with our hotter solution) their fit for
\HeII$\lambda$4686 would improve. Such an increase in temperature
would also improve their fit of \NV, which we are able to fit
accurately. The same stellar parameters as determined by \citet{bouret03}
and \citet{walborn04} were derived by \citet{heap06} using {\sc
tlusty}, mostly based on lines from highly ionized species, in
particular \NV\ and \NIV. Unfortunately, they did not comment on \HeI\
and \NIII, but reassuringly they derived a nitrogen
abundance very similar to ours, [N] = 7.92.

\end{document}